\documentclass[iop]{emulateapj}
\usepackage[T1]{fontenc}
\usepackage{amsmath}
\usepackage{amssymb}	
\usepackage{graphicx,aas_macros}
\usepackage{apjfonts}
\usepackage{multirow}
\usepackage{subfigure}
\usepackage[breaklinks,colorlinks,citecolor=blue]{hyperref}
\usepackage[all]{hypcap}
\usepackage{color}
\usepackage{natbib}
\usepackage{ulem}
\usepackage{xcolor}
\newcommand{\beq}{\begin{equation}}
\newcommand{\eeq}{\end{equation}}
\newcommand{\beqa}{\begin{eqnarray}}
\newcommand{\eeqa}{\end{eqnarray}}

\def\kms{{\rm km\:s^{-1}}}
\newcommand{\msun}{M_{\odot}}

\def\stacksymbols #1#2#3#4{\def\theguybelow{#2}
        \def\verticalposition{\lower#3pt}
        \def\spacingwithinsymbol{\baselineskip0pt\lineskip#4pt}
        \mathrel{\mathpalette\intermediary#1}}
\def\intermediary #1#2{\verticalposition\vbox{\spacingwithinsymbol
        \everycr={}\tabskip0pt
        \halign{$\mathsurround0pt#1\hfil##\hfil$\crcr#2\crcr
                \theguybelow\crcr}}}
\def\lta{\stacksymbols{<}{\sim}{2.5}{.2}}

\shorttitle{ALMA CO(2-1) observations of three massive elliptical galaxies}
\shortauthors{}

\begin{document}
\title{ALMA observations of molecular clouds in three group centered elliptical galaxies: \\ NGC~5846, NGC~4636, and NGC~5044}

\author{Pasquale Temi\altaffilmark{1}, Alexandre Amblard\altaffilmark{1,2}, Myriam Gitti\altaffilmark{3,4}, Fabrizio Brighenti\altaffilmark{3}, 
Massimo Gaspari\altaffilmark{5,6}, William G. Mathews\altaffilmark{7}, Laurence David\altaffilmark{8}
}

\altaffiltext{1}{Astrophysics Branch, NASA Ames Research Center, Moffett Field, CA, USA}
\altaffiltext{2}{BAER Institute, Sonoma, CA, USA}
\altaffiltext{3}{Physics and Astronomy Department, University of
  Bologna, Via Gobetti 93/2, 40129, Bologna, Italy}
\altaffiltext{4}{INAF, Istituto di Radioastronomia di Bologna, via Gobetti 101, 40129 Bologna, Italy}
\altaffiltext{5}{Department of Astrophysical Sciences, Princeton University, Princeton, NJ 08544, USA}
\altaffiltext{6}{{\it Einstein} and {\it Spitzer} Fellow}
\altaffiltext{7}{University of California Observatories/Lick Observatory, Department of Astronomy and Astrophysics, University of California, Santa Cruz, CA 95064, USA}
\altaffiltext{8}{Harvard-Smithsonian Center for Astrophysics, 60 Garden St., Cambridge, MA 02138, USA}

%

\begin{abstract}

We present new ALMA CO(2--1) observations of two well studied
group-centered elliptical galaxies: NGC~4636 and NGC~5846.  In
addition, we include a revised analysis of Cycle 0 ALMA observations
of the central galaxy in the NGC~5044 group that has been previously
published.  We find evidence that molecular gas, in the form of
off-center orbiting clouds, is a common presence in bright
group-centered galaxies (BGG). CO line widths are $\gtrsim 10$ times
broader than Galactic molecular clouds, and using the reference Milky
Way $X_{CO}$, the total molecular mass ranges from as low as
$2.6\times 10^5 M_\odot$ in NGC~4636 to $6.1\times 10^7 M_\odot$ in
NGC~5044.  With these parameters the virial parameters of the molecular
structures is $\gg 1$.
Complementary observations of NGC~5846 and NGC~4636 using
the ALMA Compact Array (ACA) do not exhibit any detection of a CO diffuse component
at the sensitivity level achieved by current exposures.
The origin of the detected molecular
features is still uncertain, but these ALMA observations suggest that they
are the end product of the hot gas cooling process and not the 
result of merger events.  
Some of the molecular clouds are associated with dust features as revealed by HST dust extinction maps suggesting that these clouds formed from dust-enhanced cooling. The global nonlinear condensation may be triggered via the chaotic turbulent field or buoyant uplift. The large virial parameter of the molecular structures and correlation with the warm ($10^3 - 10^5 K$)/hot  ($\ge10^6$)
phase velocity dispersion provide evidence that they are unbound giant molecular associations drifting in the turbulent field, consistently with numerical predictions of the chaotic cold accretion process.
Alternatively, the
observed large CO line widths may be generated by molecular gas
flowing out from cloud surfaces due to heating by the local hot gas
atmosphere.

\end{abstract}

\section{Introduction}
\noindent

Local brightest group galaxies (BGGs) are prime systems to study the physics of the hot interstellar or intergroup
medium, including its connection with the central massive black hole and the associated active galactic nuclei (AGN) feedback.
They are in many respects simple objects, where hot gas cooling and evolution can be investigated
without other complicating processes like star formation or gas-rich merging \citep[e.g.,][]{Feldmann10,o'sullivan17}.

In fact, the most massive group-centered elliptical galaxies have red and old stellar populations, with little or no
evidence of recent star formation  \citep{Ziegler, Van05, Daddi05, werner14, Gozaliasl16}.
The three galaxies which we have
selected for this study, NGC~4636, NGC~5044 and NGC~5846, 
are perfect example of passive
systems, with very old stellar populations \citep{Trager00,Annibali06, Diniz17} 
and star formation rates on the order of a few
percent M$_\odot$~yr$^{-1}$ \citep{werner14}.

Central galaxies in massive clusters have higher X-ray luminosities and
are much richer in warm/cold interstellar medium (ISM), with molecular gas masses
$M_{\rm H_2} \sim 10^9 - 10^{11}$ M$_\odot$ \citep{edge01,salome03, Pulido17}.
However, they are located at distances $\ge 100$ Mpc
and usually inhabit complex environments
-- e.g., affected by merging, stripping, star formation, and
supernovae -- which make it difficult to fully unveil the gas cooling process
and its connection to the feedback mechanism. 

The absence or dearth of star formation in BGGs is remarkable. These
galaxies usually have a massive hot gas halo radiating away
$10^{41}-10^{43}$ erg s$^{-1}$
\citep{Mathews03,Sun12,Goulding16, Anderson15}. This corresponds to a 
cooling rate $\dot M_{\rm cool}\sim 1-50$ M$_\odot$ s$^{-1}$ if
radiative losses are not balanced by some form of heating.
Evidently, a continuous gas heating or
removal mechanism is required to halt star formation. 
Since all these galaxies harbor supermassive black holes (SMBHs), it is natural to expect that the AGN feedback
process plays a pivotal role in regulating the gas cooling   \citep[e.g.,][]{croton06,mcnamara07,Gitti12}.

Indeed, {\it Chandra} and {\it XMM} observations have shown that hot gas ($T \sim 10^7$ K) 
cooling is suppressed by more than an order of magnitude with
respect to the classical cooling flow expectation (e.g., \citealt{mcnamara07,peterson03,peterson06,mcnamara12,molendi16}).
This is true at all mass scales, from
galaxies to clusters. X-ray images clearly show that the SMBH strongly
interacts with its environment, generating outflows/jets, bubbles, turbulence, and shocks 
that heat the surrounding hot atmosphere \citep[e.g.][for a review]{mcnamara07,fabian12}.

Multi-wavelength observations of massive ellipticals
show nevertheless a multiphase ISM,
suggesting that residual cooling occurs and regulates 
the triggering of the AGN and galaxy evolution.
H$\alpha$ emission is
regularly detected \citep[e.g.,][]{caon00,mcdonald11,sarzi13,werner14}
and probes photo-ionized warm $T\sim 10^4$ K gas
\citep{Haardt12}.
The warm gas
typically occupies an irregular region 
several kpc in size and displays a complex and chaotic kinematics.
Dust, associated with both the hot
and cold ISM phases is also often detected
\citep{temi09,smith12,o'sullivan15}.
Recent observations have shown that [CII] emission
is also a common
presence in massive ellipticals \citep{werner14}.
The [CII] and H$\alpha$ emitting gas are largely co-spatial, 
which suggests [CII] emission also traces the warm phase.
While a significant fraction of the
cold gas mass in low to intermediate mass early-type galaxies (ETGs)
is thought to have an external, merger-related origin \citep[e.g.,][]{davis11}, 
in the most massive ETGs, of interest here,
the cold and warm gas phases are likely generated by in-situ cooling 
\citep{davis11,david14,werner14}.

Recent theoretical developments have shown that, while AGN-generated outflows
and cavities globally inhibit cooling, they also stimulate residual cooling
in spatially extended regions $\sim\,$1\,-\,10 kpc in size,
where the ratio of the cooling time to free-fall time 
is $t_{\rm
  cool}/t_{\rm ff} \lta 10-30$
\citep{revaz08,gaspari11,gaspari17_cca,sharma12,brighenti15,valentini15,voit15a,li15,barai16,lau17}.

The multiphase condensation mechanism might be related to the feeding of SMBHs,
through the so-called chaotic cold accretion (CCA) mechanism
\citep[e.g.,][]{gaspari13, gaspari15_cca, gaspari17_cca}. 
In this scenario, warm filaments and cold clouds condense out of the hot halo,
and recurrently boost the
accretion rate up to a few orders of magnitude compared with hot (Bondi) accretion.
The AGN responds by injecting back the energy in form of entrained
massive outflows and/or relativistic jets, establishing a tight
self-regulated loop (e.g., \citealt{gaspari17_uni} for a brief
review). During CCA the gas phases are spatially and kinematically correlated,
showing comparable cooling and eddy turnover time $t_{\rm cool}/t_{\rm eddy} \approx 1$
\citep{gaspari17b}.

Currently, the coldest ISM phase (molecular or
neutral gas) in normal massive elliptical galaxies is the least studied and not well characterized yet.
Recent ALMA observations of the galaxy group NGC~5044 have revealed approximately 20 
CO-emitting clouds in the central 10 kpc of the galaxy \citep{david14}.
In this paper we present new ALMA CO(2--1) observations of two group-centered
elliptical galaxies (NGC~4636 and NGC~5846)
to seek confirmation that molecular gas is a common presence in BGGs,
and is not a result of merger events, but has cooled directly from the hot gas.
In addition, we include a revised analysis of Cycle 0  ALMA observations of NGC~5044 that have been previously
published by \cite{david14}.
The current, more reliable pipeline, is used to reduce ALMA Cycle 0 data, known
to suffer from early calibration issues. 
These three galaxies
have the most complete observational coverage, with available
deep {\it Chandra} X-ray data, SOAR and HST H$\alpha$ observations,
{\it Herschel} [CII] data, {\it Spitzer} FIR data and detailed HST dust absorption
maps. 
All three galaxies satisfy the empirical criteria for extended multiphase gas
cooling \citep[][and references therein]{gaspari13,voit15a,Voit17,brighenti15,valentini15, Pulido17}:
central entropy $\lesssim 15$ keV cm$^2$ at radii of 5-15 kpc , minimum $t_{\rm cool}/t_{\rm ff}
\lesssim 20$, $t_{\rm cool}\lesssim 3\times10^8$ yr,
and $t_{\rm cool}/t_{\rm eddy} \approx 1$.
Equally important, they all  
exhibit X-ray evidence of recent feedback, bubbles, and radio emission, and each is expected to have low-entropy regions that 
are currently cooling \citep{gaspari13,voit15a,Voit17,brighenti15,valentini15}.

\section{Galaxies sampled}
\noindent
In this section we describe in depth the two galaxies for which new
CO(2--1) data have been collected. We refer to \cite{david14}
for a description of NGC~5044. 
With the adopted distances of 31.2, 24.2, and 17.1 Mpc  for NGC~5044, NGC~5846, and NGC~4636 \citep{tonry01},  the corresponding physical scales are  151, 118 and 83 pc
per arcsec, respectively. 


\begin{table}
\begin{center}
\begin{tabular}{lcc}
  & NGC~5846 & NGC~4636 \\
  \hline
  AGN  & yes & yes \\
  AGN kinetic power (erg s$^{-1}$) & 7.5$\times$10$^{41}$& 3$\times$10$^{41}$\\
  L$_\mathrm{H\alpha+[NII]}$ (erg\,s$^{-1}$) & 2.5$\times$10$^{40}$& 6$\times$10$^{39}$ \\
  PAH lines ($\mu$m)& -- & 11.3, 17\\
  Distance (Mpc)& 24.2 & 17.1 \\
  ATLAS$^{3D}$ classification & slow-rotator & slow-rotator \\
  effective radius (arcsec, kpc) & 59, 6.9&  89, 7.4\\
  radio power (erg s$^{-1}$) & 2.5$\times$10$^{38}$ & 1.5$\times$10$^{38}$\\
  \hline
\end{tabular}
\caption{Summary of NGC~5846 and NGC~4636 properties, as measured or referenced in \cite{allen06,caon00,trinchieri02,werner14,rampazzo13,tonry01,mei07,emsellem11,cappellari11,cavagnolo10}.}
\end{center}
\end{table}

\subsection{NGC~5846}
\noindent
NGC~5846, spherical in shape, is classified as a giant elliptical E0 and
is the central and brightest galaxy in its group
\citep{mahdavi05}.


Its proximity
makes this object one
of the closest example of AGN feedback in massive ellipticals. 
This is a very well studied system in every
wavelengths other than at rad
io frequencies \citep{machacek11}.
There is good knowledge of
its hot ($\sim 10^7 K$) and ionized medium, while little is known about the content
and the dynamics of the cold ($\le 10^4 K$) phase, likely a significant component
in the central kpc or so. 

The dramatic effects of AGN outflows in NGC~5846 have been first
discovered in the X-ray band, which probes hot ($\approx 10^7$ K) gas
\citep{trinchieri02,machacek11}.
Two inner bubbles in the hot gas, at a distance
of 600 pc from the center, filled with radio emission, are clear indication
of recent AGN feedback.
A weak radio source, elongated in the NE-SW
direction, connects the inner cavities.
X-ray bright rims surround
the inner X-ray bubbles \citep{machacek11}.
Many X-ray knots are visible, suggesting cooling sites. The scenario indicated by
the {\it Chandra} observation is that of an AGN outflow, compressing and cooling the gas \citep[e.g.][]{brighenti15}
in the central $\sim\,$2 kpc (20$^{\prime \prime}$ at the distance of NGC~5846).

H$\alpha$ observations \citep{caon00,trinchieri02,werner14}
reveal the presence of warm ($T\sim10^4$ K) ionized gas in the inner 2 kpc of
NGC~5846.
Spectra of this gas indicate
irregular motion, with typical velocity of 150\,-\,200 km\,s$^{-1}$. 
The warm $10^4$ K gas traces the X-ray bright features, 
suggesting again a multiphase AGN outflow.

Using {\it Spitzer} IRS, \cite{rampazzo13} have detected
mid infrared lines (e.g., [NeII] 12.81 $\mu$m, [NeIII]
15.55 $\mu$m ), but no trace of polycyclic aromatic hydrocarbon (PAH) emission.
\cite{filho04} have identified several sources in radio
at 2.3, 5 and 15 GHz using VLBA data, these sources
are aligned in the South-North direction.

Recent {\it Herschel} PACS observations have detected the presence of 
[CII] emitting gas that
extends to a radius of $\sim 2$ kpc and is
centrally peaked. 
The [CII] emission is almost exactly co-spatial with the of $H\alpha+[NII]$ emission
and the total fluxes in [CII] and $H\alpha+[NII]$ have a ratio  $F_{H\alpha+[NII]}/F_{[CII]} = 2.5$;
a very similar flux ratio value observed in other group centered ellipticals \citep{werner14}.
Furthermore, the velocities inferred from the [CII] line are consistent
with those measured for the H$\alpha$ line \citep{caon00}.
All these evidences suggest that the [CII] line is emitted by the warm gas, and it is not necessarily  
tracing the molecular phase.

NGC~5846 has another indication that the cold gas being disturbed by an AGN outburst. 
It has in fact an excess of cold ($T\sim 30$ K) dust, approximately co-spatial
with the ionized and the molecular gas
\citep{temi07b,mathews13}. With a 70
$\mu$m luminosity of $3.5 \times 10^{41}$ erg s$^{-1}$ \citep{temi09}, 
NGC~5846 shares the same dust properties of several giant
ellipticals (e.g., NGC~4636 and NGC~5044; \citealt{temi07a}) which
are best explained with ejection of dusty gas from their center by AGN
activity that occurred $\sim 10^7$ yr ago.
This dust has to still be embedded in cold gas, otherwise it would 
be sputtered away in $\sim 10^7$ yr.

\subsection{NGC~4636}
\noindent
NGC~4636 is the central galaxy of a poor group in the outskirt of the
Virgo Cluster.
It harbors a relatively small SMBH
of 7.9$\times$10$^7$ M$_\odot$ \citep{merritt01}
as inferred from the bulge velocity dispersion.

The dramatic effects of AGN outflows in NGC~4636 have been first
discovered in the X-ray band, which probes the hot ($\approx 10^7$ K) gas
\citep{jones02,baldi09}. Large bubbles in the hot gas,
surrounded by bright rims 
are likely the results of shocks
generated by the AGN jets. A weak
radio source, elongated in the NE-SW
direction, connects the NE and SW bubbles.

Of major interest is the X-ray bright core, having a
radius of $\sim 1$ kpc.
As discussed in \cite{baldi09}, the
core shows a central cavity surrounded by a bright
edge. Interestingly, the small X-ray cavity surrounds the $\sim 1$\;kpc 
radio jet detected at 1.4 GHz \citep{allen06}, and is likely
generated by the jet. Thus, the X-ray and radio observations point to 
a scenario in which gas may be currently outflowing in the central kpc
of NGC~4636.
UV emission \citep{bregman01,bregman05} exhibits  O VI emission, 
which is a tracer of gas cooling. The measured emission indicates
a cooling rate of 0.3 $M_\odot$\,yr$^{-1}$.
\cite{rampazzo13} detected PAH emission at 11.3 and 17 $\mu$m as well as
[Ne II], [Ne III] and [S III] lines in the center of NGC~4636 (within
 $r_\mathrm{e}$/8) using {\it Spitzer}-IRS.
H$\alpha$ observations \citep{caon00,werner14} reveal
the presence of warm ($T\sim10^4$ K) ionized gas in the inner kpc of
NGC~4636.
Spectra of this gas indicate
irregular motion, with typical velocity of 150\,-\,200 km\,s$^{-1}$. 
H$\alpha$ maps of the galaxy core show the presence of a cavity in the distribution of the ionized gas,
encircled by a dense shell located at a distance of $\sim 400$ pc from the
center. Again, the most plausible explanation is gas expansion
caused by the AGN activity.

In NGC~4636 the [CII] emission extends to a radius of $\sim 1$ kpc and is
centrally peaked.  
The velocities inferred from the [CII] line are consistent
with those measured for the H$\alpha$ line.
Finally, NGC~4636 has an excess of cold dust, approximately cospatial
with the ionized and the molecular gas
\citep{temi07a,mathews13}.
As above, this dust is expected be embedded in cold gas, to be protected against rapid sputtering. 
In \cite{temi07a} we showed that the
extended dust distribution originates from the ejection of cold gas by the
AGN activity occurred 10 Myr ago.
 
\section{Observations and data reduction}

\begin{table*}
\begin{center}
\begin{tabular}{lcccc}
  & \multicolumn{2}{c}{NGC~5846} & \multicolumn{2}{c}{NGC~4636} \\
  Observation center (RA,DEC) & \multicolumn{2}{c}{(15h 6m 29.253s, 1$^\circ$ 36' 20.290")} & \multicolumn{2}{c}{(12h 42m 49.867s, 2$^\circ$ 41' 16.010")} \\
  \hline
  & \multicolumn{4}{c}{ALMA 12m} \\
  Observation dates (yyyy/mm/dd) & \multicolumn{2}{c}{2016/05/13} & 2016/05/02 & 2016/05/03 \\
  \hline
  Observation duration (min) & \multicolumn{2}{c}{71.15}  & 49.21 & 48.81 \\
  On-target duration (min) & \multicolumn{2}{c}{40.82} & 29.74 & 29.74 \\
  PWV (mm) & \multicolumn{2}{c}{1.08} & 1.12 & 2.02 \\
  T$_\mathrm{sys}$ at 230 GHz (K) & \multicolumn{2}{c}{75} & 74 & 90 \\
  Calibrators (flux, bandpass, phase) & \multicolumn{2}{l}{J1517-2422,  J1550+0527, J1505+0326} & \multicolumn{2}{c}{J1229+0203}\\
    Line central wavelength (GHz) & \multicolumn{2}{c}{229.23}& \multicolumn{2}{c}{229.83} \\
    Line bandwidth \& resolution (MHz, km s$^{-1}$) & \multicolumn{4}{c}{(937.5, 1200) \& (0.5, 0.6)} \\
    Continuum central wavelengths (GHz) & \multicolumn{2}{c}{216.45, 218.17, 230.61} & \multicolumn{2}{c}{215.62, 217.33, 229.91} \\
    Continuum bandwidth \& resolution (MHz, km s$^{-1}$) & \multicolumn{4}{c}{(1850, 2400) \& (31.25, 42)}\\
    \hline
  & \multicolumn{4}{c}{ ALMA Compact Array (ACA)} \\
    Observation dates (yyyy/mm/dd) & 2016/12/21 & 2017/04/29 & 2016/12/28 & 2017/03/26 \\
    \hline
      Observation duration (min) & 77.13 & 77.31  & 91.62 & 91.55 \\
  On-target duration (min) &  \multicolumn{2}{c}{34.27} & \multicolumn{2}{c}{49.90} \\
  T$_\mathrm{sys}$ at 230 GHz (K) & 86 & 68 & 61 & 64 \\
  Calibrators (flux, bandpass, phase) & \multicolumn{2}{l}{
    Callisto, J1256-0547, J1516+0015} & \multicolumn{2}{l}{J1256-0547, J1229+0203, J1229+0203}\\
    Line central wavelength (GHz) & \multicolumn{2}{c}{229.23}& \multicolumn{2}{c}{229.83} \\
    Line bandwidth \& resolution (MHz, km s$^{-1}$) & \multicolumn{4}{c}{(937.5, 1200) \& (0.5, 0.6)} \\
    Continuum central wavelengths (GHz) & \multicolumn{2}{c}{214.02, 215.71, 228.02} & \multicolumn{2}{c}{214.33, 216.02, 228.52} \\
    Continuum bandwidth \& resolution (MHz, km s$^{-1}$) & \multicolumn{4}{c}{(1850, 2400) \& (31.25, 42)}\\
    \hline
\end{tabular}
\caption{Summary of ALMA observations of NGC~5846 and NGC~4636}
\end{center}
\label{tab:obs}
\end{table*}

\subsection{New Cycle 3-4 observations of NGC~5846 and NGC~4636}

We observed NGC~5846 and NGC~4636 with ALMA during Cycle 3 (Project
code: 2015.1.00860.S, PI: Temi).
For both galaxies, the ALMA interferometer was configured such that its longest baseline was about 640 m and its shortest baseline about 15 m.  This configuration
resulted in an angular resolution of about 0.6$^{\prime \prime}$ and a maximum
recoverable scale of 5.4$^{\prime \prime}$.  Assuming a distance of 17.1 Mpc for
NGC~4636 and 24.2 Mpc for NGC~5846, 0.6$^{\prime \prime}$ and 5.4$^{\prime \prime}$ correspond to
50 and 450 pc for NGC~4636 and
72 and 644 pc for NGC~5846, respectively.
All data were taken in the ALMA band 6, one spectral window (spw) was centered around the CO (2-1) line and threo other spws measured the continuum. A detailed description of the observations is presented in table \ref{tab:obs}.

%
%

The data were reduced using the CASA software \citep[version
4.5.3][]{mcmullin07}.  As a first step we carefully checked the
results of the quality assurance (QA) processes provided by the
Science Pipeline team, focusing in particular on the calibration
process.  For NGC~5846, the pipeline calibration appears to be
reasonable, therefore we simply produced the calibrated dataset by
running the original pipeline reduction scripts.  We further attempted
the self-calibration, but the image quality did not improve.  For
NGC~4636, a bad antenna (DV02) partially corrupted the second
execution block (EB), therefore we flagged it and performed a new
manual calibration following the steps in the script provided to the
PI, and then combined the two EBs to produce the final calibrated
dataset. Self-calibration was attempted as well, but no good solutions
were found.

The continuum-subtracted image cubes were produced using the CLEAN
algorithm provided in the CASA package, and several velocity
resolutions, Briggs weightings and threshold values were explored to
determine the optimum setup for each image and to investigate the
goodness of the detection. In the end, we used in this paper
  two thresholds to produce maps that will be used to detect the
  potential signals. One threshold corresponds to about 1.5 times
  the noise rms of the treated data, it is a more aggressive
  ``cleaning'' version. A lower threshold creates a lot
  of sources that most likely are just noise. The other threshold corresponds
  to about 4 times the noise rms of our data, it is a more conservative ``cleaning''. Using a much larger threshold would correspond to making a final image which is identical to the initial dirty map, since the signal-to-noise in our data is not very large and the algorithm would not find any signal to iterate on. Using thresholds between these two values does not produce significantly different results with our data, table \ref{tab:AAAA} indicate with which threshold a source has been detected. CO maps in figures have been produced using the more aggressive threshold, since it returns a better S/N on maps.

In this paper, we present images and spectra obtained with 10 ${\rm km\,s^{-1}}$
resolution image cubes and natural weighting (robustness parameter
$=2$), with the CLEAN algorithm running in non-interactive mode
(threshold = 0.5 mJy, that roughly corresponds about 1.5 times the rms noise). This provided a synthesized beam of
0.79$^{\prime \prime}\times$0.70$^{\prime \prime}$  with a position angle (P.A.) of 79$^{\circ}$ for NGC~5846 observations and a synthesized beam of 0.74$^{\prime \prime}\times$0.67$^{\prime \prime}$ with a P.A. of 116$^{\circ}$ for NGC~4636.  For the reasons explained above, the
final images presented here are not self-calibrated.  The rms noise in
the line-free channels was 0.4 and 0.3 mJy beam$^{-1}$ for NGC~5846 and NGC~4636, respectively.
Images of the continuum emission were also produced by averaging
channels free of any line emission.  An unresolved central continuum
source is detected in both NGC~5846 and NGC~4636, with flux of 
$10.63 \pm 0.03$ mJy at 220.997 GHz and $0.4 \pm 0.1$ mJy at 221.390 GHz, respectively.
The position of the continuum is in good agreement with radio and optical images  with an offset of 0.10$\pm$0.01 and 0.15$\pm$0.01 arcsec for NGC~4636 and NGC~5846 respectively.

In addition to the ALMA 12-m array observations, we obtained
complementary data from the ALMA Compact Array (ACA) for NGC~5846 and
NGC~4636 in Cycle 4 (Project code: 2016.1.00843.S, PI: Temi).
As for the 12-m observations, all data were taken in the ALMA band 6 with a similar spectral setup. These observations are presented in more details in table \ref{tab:obs}.\\
%
%
%
%
We used the products provided by the ALMA pipeline for ACA data. The
pipeline used CASA 4.7.2 and CASA 4.7.38335 for NGC~5846 and NGC~4636,
respectively.  Using the Briggs weighting scheme with a robust
parameter of 0.5, the resulting beam size was 7.0$^{\prime \prime}\times$4.5$^{\prime \prime}$ and
6.8$^{\prime \prime}\times$4.4$^{\prime \prime}$ for NGC~5846 and NGC~4636, respectively. The
final estimates of the rms noise values are 4.02 and 2.45 mJy beam$^{-1}$
for NGC~5846 and NGC~4636 with a velocity bin of 10 ${\rm km\,s^{-1}}$.


\subsection{Re-analysis of Cycle 0 observations of  NGC~5044}

NGC~5044 was observed during Cycle 0 (Project code: 2011.0.00735.SSB,
PI: Lim) and is described in details in \cite{david14}, we only
summarize here the main characteristics of these observations. 
NGC~5044 was observed with an angular resolution of 2.0$^{\prime \prime}\times$1.4$^{\prime \prime}$ for
29 minutes with a spectral resolution of 0.64 ${\rm km\,s^{-1}}$ around the CO(2-1)
emission line.

Given the uncertainties associated with the early phase of ALMA data reduction and calibration,
we decided to reduce again the NGC~5044 observations with the latest release of the CASA reduction and calibration software package.  
In particular, we followed the CASA
guides \footnote{https://casaguides.nrao.edu/index.php/Guide\_To\_Redo\_Calibration\_For\_\\
ALMA\_Cycle\_0;
  https://casaguides.nrao.edu/index.php/Updating\_a\_script\_to\_\\
  work\_with\_CASA\_4.2}
to modify the packaged calibration script in order to apply the proper
amplitude calibration scale (which has changed since Cycle 0) and also
remove some atmospheric lines in the calibrators. The channel
containing these lines were further removed from the calibrated target
in order to perform a correct continuum subtraction.

Similarly to what was done for NGC~5846 and NGC~4636, we present here the
images and spectra obtained with 10 ${\rm km\,s^{-1}}$ resolution image cubes and
natural weighting, with the CLEAN algorithm running in non-interactive
mode (threshold = 1.6 mJy, that roughly corresponds to 1.5 time the rms noise).  This provided a synthesized
beam of 1.99$^{\prime \prime}\times$1.41$^{\prime \prime}$ with a position angle (P.A.) of 148$^\circ$.
The rms noise in the line-free channels was 1.0 mJy beam$^{-1}$.

\renewcommand{\arraystretch}{1.5}
\begin{table*}
\begin{center}
\begin{tabular}{lcccccccc}
\hline
 \hline
ID & (RA,DEC) offset & $<{v} >$ & $\sigma$ & ${S_{CO}\Delta v}$ & ${M_{mol}}$ & Ang. Size & Phys. Size & Methods\\
 & ($^{\prime \prime}$),(kpc) & (km~s$^{-1}$) & (km~s$^{-1}$) & (Jy~km~s$^{-1}$)  & ($10^5$~M$_{\odot}$) & FWHM ($^{\prime \prime}$) & (pc) & \\
\hline
\hline
\multicolumn{9}{c}{NGC~5846}\\
1 &(-2.3,-1.5),(-0.3,-0.2)& -230.7 $\pm$  1.6 & 23.3 $\pm$  1.6 &  0.32 $\pm$  0.02 & 6.4 $\pm$  0.4 & $^{\;\;\,1.2\pm0.2}_{\times0.2\pm0.2}$ & 143$\times$24 & I,F,i,f\\ 
2 &(-0.8,-18.0),(-0.1,-2.1)& -155.4 $\pm$  3.0 & 14.6 $\pm$  3.0 &  0.24 $\pm$  0.04 & 4.8 $\pm$  0.8  & < 0.7 &  < 82 & i,f\\ 
3 &(8.2,3.2),(1.0,0.4)& 110.6 $\pm$  1.6 & 19.9 $\pm$  1.6 &  0.45 $\pm$  0.04 & 8.9 $\pm$  0.7  & $^{\;\;\,2.9\pm0.3}_{\times0.7\pm0.04^\dagger}$  & 346$\times$83 & I,F,i,f\\ 
\hline
  \multicolumn{9}{c}{NGC~4636}\\
1 &(-19.0,-0.3),(-1.6,-0.0)& 140.3 $\pm$  8.4 & 26.4 $\pm$  8.4 &  0.20 $\pm$  0.06 &  1.9 $\pm$  0.5 & < 0.7 & < 50 & i,f\\ 
2 &(2.0,0.3),(0.2,0.0)& 209.5 $\pm$  3.9 & 25.8 $\pm$  3.9 &  0.07 $\pm$  0.01 &  0.7 $\pm$  0.1 & < 0.7 &  < 50 & I,F,i,f\\ 
  \hline
  \multicolumn{9}{c}{NGC~5044}\\
1 &(-0.1,-2.0),(-0.0,-0.3)& -556.9 $\pm$ 11.9 & 67.0 $\pm$ 10.7 & 0.8 $\pm$  0.1 & 24.4 $\pm$ 3.8  & < 2.0 & < 300 & I,F,i,f\\
2 &(0.0,2.8),(0.0,0.4)& -313.1 $\pm$ 9.5 & 36.6 $\pm$ 9.4 & 0.4 $\pm$  0.1 & 12.2 $\pm$ 2.5  & < 2.0 & < 300 & I,F,i,f\\
3 &(-2.2,-1.4),(-0.3,-0.2)& -274.4 $\pm$ 4.5 & 28.4 $\pm$ 4.4 & 0.4 $\pm$  0.1 & 12.8 $\pm$ 1.8  & < 2.0 & < 300 & F,i,f\\
5 &(-12.0,-5.2),(-1.8,-0.8)& -192.8 $\pm$ 6.6 & 30.5 $\pm$ 8.5 & 0.7 $\pm$  0.1 & 23.2 $\pm$ 3.8  & < 2.0 & < 300 & F,i,f\\
6 &(-1.9,1.8),(-0.3,0.3)& -226.8 $\pm$ 4.3 & 20.4 $\pm$ 4.3 & 0.3 $\pm$  0.1 & 9.5 $\pm$ 1.8  & < 2.0 & < 300 & I,F,i,f\\
7 &(-1.3,6.4),(-0.2,1.0)& -207.0 $\pm$ 6.9 & 31.1 $\pm$ 6.4 & 0.4 $\pm$  0.1 & 12.6 $\pm$ 2.5  & < 2.0 & < 300 & F,i,f\\
8 &(-0.4,-0.4),(-0.1,-0.1)& -148.8 $\pm$ 8.2 & 76.2 $\pm$ 7.2 & 1.1 $\pm$  0.1 & 33.9 $\pm$ 3.1  &  $^{\;\;\,3.9\pm1.3}_{\times1.2\pm0.6}$ &  590$\times$182 & I,F,i,f\\
11 &(3.0,-2.6),(0.5,-0.4)& -95.8 $\pm$ 6.0 & 41.0 $\pm$ 4.9 & 0.5 $\pm$  0.1 & 16.5 $\pm$ 2.4  & < 2.0 & < 300 & F,i,f\\
12 &(-8.6,3.4),(-1.3,0.5)& -132.9 $\pm$ 3.9 & 18.4 $\pm$ 3.9 & 0.4 $\pm$  0.1 & 13.6 $\pm$ 2.5  & < 2.0 & < 300 & F,i,f\\
13 &(-1.2,-2.4),(-0.2,-0.4)& -80.9 $\pm$ 6.9 & 43.1 $\pm$ 8.1 & 0.8 $\pm$  0.1 & 26.0 $\pm$ 2.6  &  $^{\;\;\,3.8\pm1.0}_{\times0.7\pm0.7}$ &  575$\times$106 & I,F,i,f\\
14 &(-14.4,-5.8),(-2.2,-0.9)& -46.8 $\pm$ 10.6 & 39.0 $\pm$ 8.2 & 1.0 $\pm$  0.2 & 30.7 $\pm$ 7.2  &  $^{\;\;\,3.1\pm1.3}_{\times1.0\pm0.6}$ &  469$\times$151 & F,i,f\\
18 &(0.8,1.2),(0.1,0.2)& 27.8 $\pm$ 1.7 & 37.4 $\pm$ 1.7 & 3.0 $\pm$  0.1 & 93.8 $\pm$ 3.7  & $^{\;\;\,1.9\pm0.2}_{\times0.8\pm0.5}$ & 287$\times$121 & I,F,i,f\\
20 &(-15.0,-7.8),(-2.3,-1.2)& 29.2 $\pm$ 12.8 & 73.0 $\pm$ 12.8 & 1.2 $\pm$  0.2 & 37.6 $\pm$ 6.0  & < 2.0 & < 300 & F,i,f\\
21 &(-8.4,11.0),(-1.3,1.7)& 41.8 $\pm$ 9.3 & 46.8 $\pm$ 8.4 & 1.0 $\pm$  0.2 & 32.0 $\pm$ 5.5  & < 2.0 & < 300 & F,i,f\\
22 &(1.2,14.4),(0.2,2.2)& -18.3 $\pm$ 26.1 & 125.6 $\pm$ 24.6 & 1.7 $\pm$  0.3 & 53.2 $\pm$ 9.0  & < 2.0 & < 300 & F,i,f\\
25 &(17.3,10.4),(2.6,1.6)& -573.7 $\pm$ 3.2 & 10.3 $\pm$ 2.1 & 0.7 $\pm$  0.2 & 22.8 $\pm$ 6.6  & < 2.0 & < 300 & i,f\\
26 &(15.8,-12.8),(2.4,-1.9)& -108.1 $\pm$ 3.3 & 10.9 $\pm$ 3.4 & 0.6 $\pm$  0.2 & 19.2 $\pm$ 5.4  & < 2.0 & < 300 & i,f\\
\hline
\end{tabular}
\caption{Cloud candidates for NGC~5846 (top), NGC~4636 (center) and NGC~5044 (bottom) with their location with respect to the galaxy center (offset along RA,DEC axis in arcsec and kpc),
  their average velocity, their velocity dispersion, their total CO(2-1) flux (corrected from the primary beam effect), the corresponding
  molecular mass calculated using Eq.~\ref{eq:mmol}, the dimension of the cloud (in arcsec and pc) and the methods with which the clump was detected (i when detected in the image, f when detected in the spectrum with a 1.5 noise rms cleaning threshold, I and F when detected with a 4 noise rms cleaning threshold). 
To be consistent with cloud labeling in \citet{david14}, the original cloud number sequence for NGC~5044 has been maintained. 
The spectral properties have been
  calculated with a Gaussian fitted on the sum of the emission from pixels with a signal-to-noise
  larger than 4. The dimension of the cloud were calculated with the {\it imfit} function of CASA and correspond
  to the deconvolved FWHM $^\dagger$ : for cloud \#3 of NGC~5846, the dimension of the cloud is not deconvolved from the beam, given the minor axis length is too close to the beam size.}
\end{center}
\label{tab:AAAA}
\end{table*}

\section{Results}
\noindent
Using Cycle 0 ALMA observations,
\cite{david14} discovered over 20
CO(2-1) emitting clouds in the brightest X-ray galaxy/group NGC~5044.
With our new ALMA Cycle 3 observations, we seek confirmation that CO molecular clouds are common features 
in group-centered elliptical galaxies that are, in many respects, similar to NGC~5044.

In order to identify CO clouds in our data, we used the cubes
described previously and 3 other sets of cubes for each galaxy.  One
set was produced by increasing the threshold of the cleaning algorithm
to 1.5 mJy for NGC~5846 and NGC~4636 and to 4 mJy for NGC~5044 (roughly 4 times the rms noise)  from 0.5 and 1.6 mJy.  Two
extra sets were produced by refining the velocity resolution to 3 ${\rm km\,s^{-1}}$
and setting the CLEAN threshold to 0.9 and 2.5 mJy for NGC~5846 and
NGC~4636, and to 1.6 and 7 mJy for NGC~5044, respectively (thresholds were scaled with the bin size, $\sqrt{10/3}$). This thinner spectral resolution is neeeded to perform the detection of the lines in the spectral domain. Our detection techniqure requires to convolve the signal by a Gaussian kernel as thin as 9 ${\rm km\,s^{-1}}$ wide and the 3 ${\rm km\,s^{-1}}$ provides an adequate sampling to do so.\\
With each 10 ${\rm km\,s^{-1}}$ cubes (2 per galaxy), we identified clouds in each
images by detecting pixels with a flux larger than five times the rms
noise. These sets of pixel-channel are then clustered together
(spatially and between velocity channels) to form a list of cloud
candidates.
\\
With each 3 ${\rm km\,s^{-1}}$ cubes (2 per galaxy), 
we whitened the noise by multiplying each pixel frequency stream by a function
in Fourier mode that approximates well the noise power spectrum of these time streams; 
this operation reduces the effect of the Hanning filter introduced by the ALMA processing.
We then convolved each pixel spectra by Gaussian kernels of various width
($\sigma$ = 9, 18, 30, 45, 60 ${\rm km\,s^{-1}}$) and selected pixels with signal
larger than 5 times the rms noise.
Again, we clustered these lists of pixel-velocity to
obtain a list of cloud candidates.
Both of these techniques are tuned to measure lines larger than about 10 ${\rm km\,s^{-1}}$ given one is using a 10 ${\rm km\,s^{-1}}$ bin size and the other has a smallest kernel of 9 ${\rm km\,s^{-1}}$. Using thinner spectral resolution is very challenging with the data collected that do not have a large signal-to-noise ratio. Furthermore, given the angular resolution of these observations it is not expected that we could resolve individual GMCs with small velocity dispersion but more likely some GMAs with a typical velocity dispersion of a few tenths of ${\rm km\,s^{-1}}$.\\
We produced a combined list of cloud candidates from these 4 clouds
candidate lists by requiring at least 1 detection in the image and 1
detection in the spectrum. The final list of cloud candidates
presented in this paper (Table \ref{tab:AAAA}) was obtained by
requiring that each cloud line had a minimal velocity dispersion of 6
${\rm km\,s^{-1}}$ and a total signal-to-noise
ratio of 6. We increased the signal-to-noise ratio criteria to 6 in order to remove potential false detections, given the data sample size is of a few millions (depending on the velocity resolution). A signal-to-noise of 5 could still return a few false detections (1 per 3 million sample on average).
  We note however that the velocity dispersion criteria and the requirement to detect the line in a pixel spectrum as well as in the image is more stringent and that a signal-to-noise of 5 return the same cloud candidate for our data.
  The minimum velocity dispersion of 6 ${\rm km\,s^{-1}}$ allows to sample properly the lines, given the data have been analyzed at velocity resolution of at most 3 ${\rm km\,s^{-1}}$.

\begin{figure}[ht!]
\includegraphics[width=8.5cm]{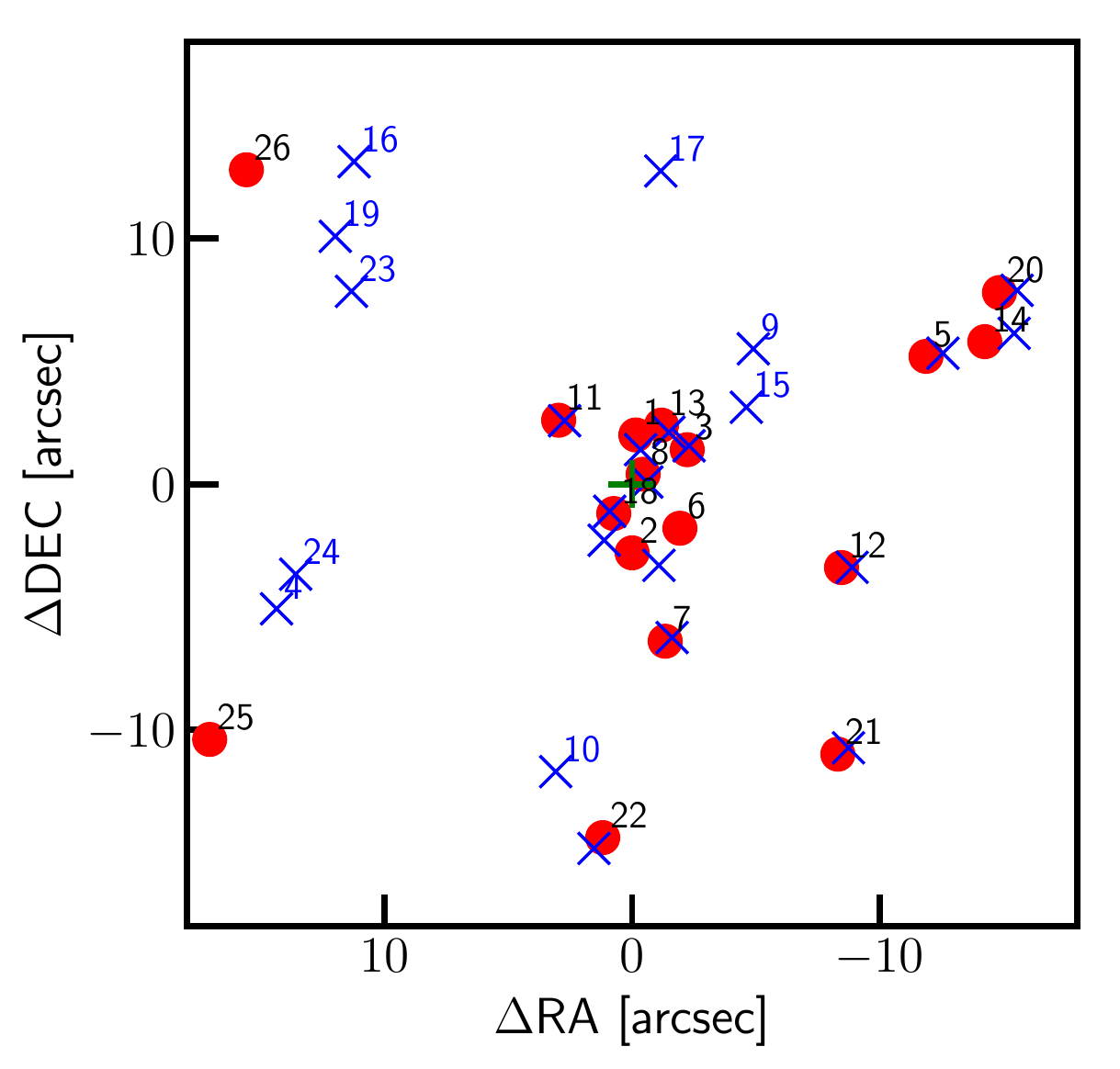}
 \caption{CO clouds detected in NGC~5044. Blue crosses and red dots show the location of CO clouds detected by \citet{david14} and with our updated reduction, respectively. For consistency with cloud labeling in \citet{david14}, the original cloud number sequence has been maintained. Our detection criteria did not confirm 9 clouds previously listed by  \citet{david14}, while two new clouds, 25 and 26, are now apparent in the field.
 }
\label{fg:cloudsD}
\end{figure}
\renewcommand{\thefigure}{\arabic{figure}}

The rather conservative approach used in selecting cloud candidates 
makes us worry about losing real clouds that could have been marginally detected.
Indeed, by relaxing the selection criteria to a signal-to-noise ratio greater than 3 and allowing a detection in 
either the spatial or velocity space, a number a clouds candidates become apparent
in NGC~5846 and NGC~4636.
These additional clouds seem to follow closely the filaments and ridges evident in the 
H$\alpha$+[NII] image, and are not randomly distributed in the field as expected from spurious signals.
Although there are hints that these clouds may be real with a low-level detection, 
we decided not to include them in our analysis because 
often their extent in the velocity domain was very limited and the detection was not 
confirmed in the spatial plane.
Deeper ALMA observations would be required to confirm a positive detection of these clouds. 

\begin{figure*}[ht!]
\hskip-0.0cm
  \includegraphics[width=9.35cm]{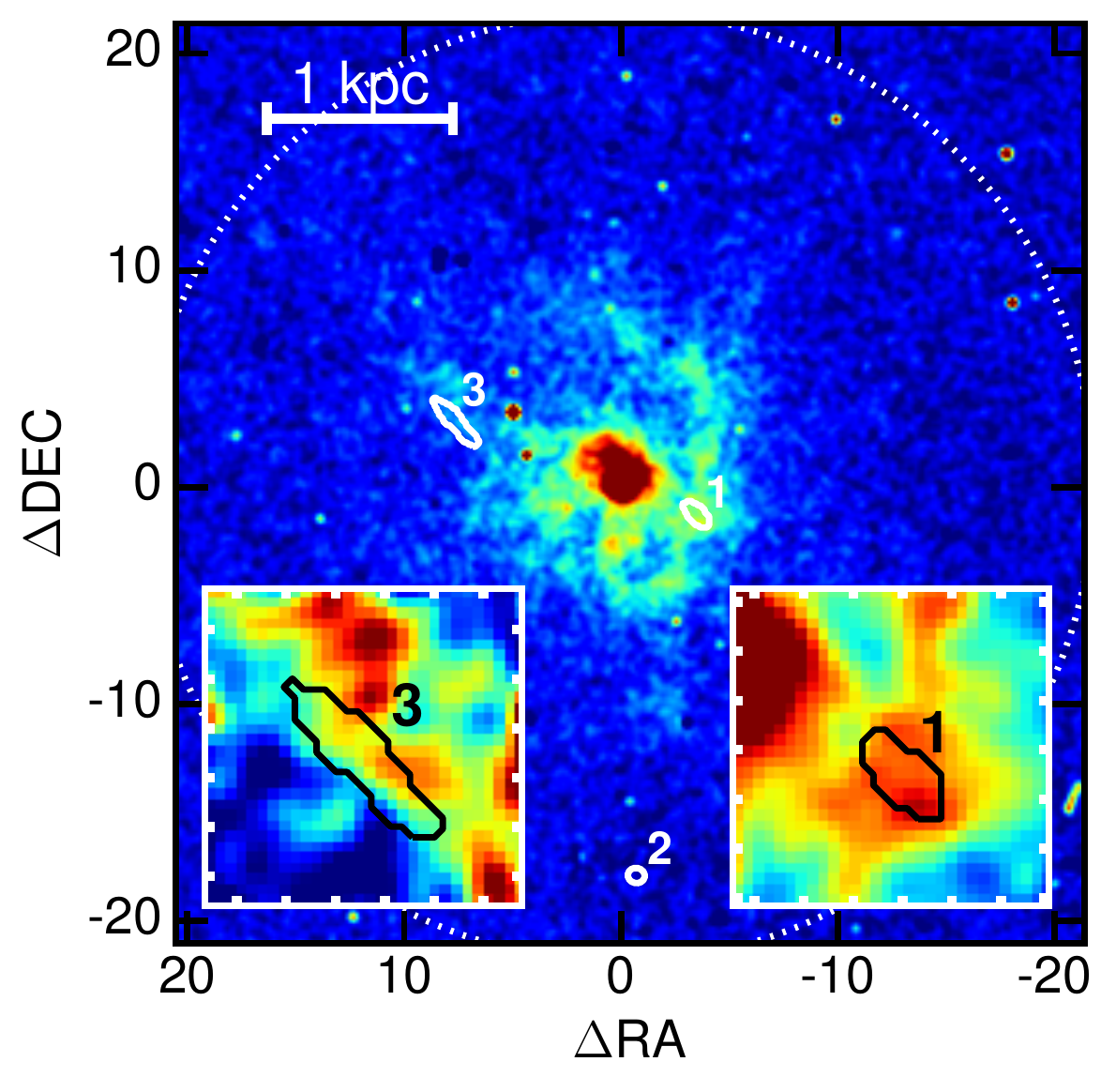}
  \includegraphics[width=8.1cm]{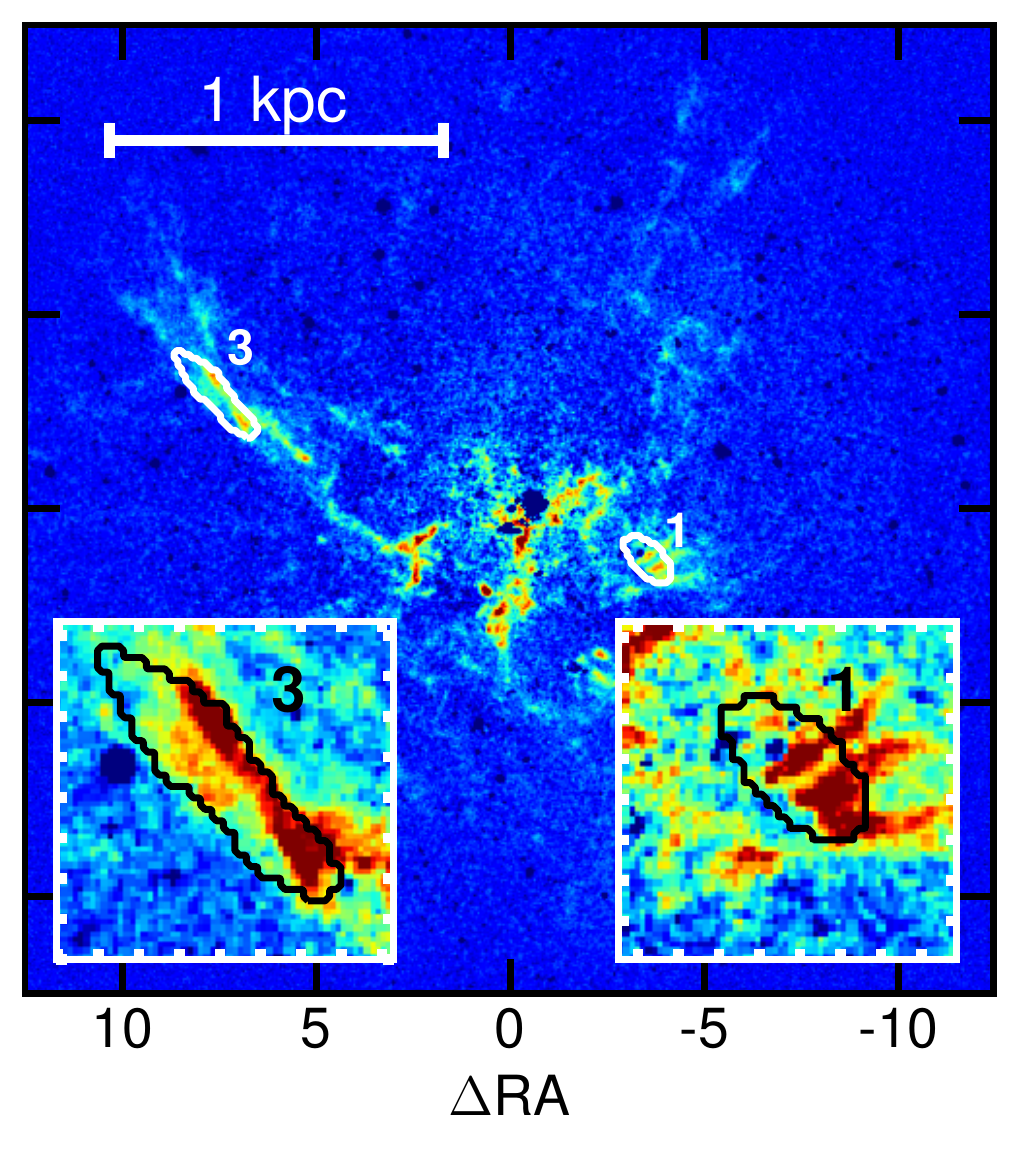}
  \caption{
Images of NGC~5846 showing detected CO(2-1) clouds projected
against an optical H$\alpha$+N[II] emission (left) and a dust starlight extinction image (right).
Color sequence (blue $\rightarrow$ green $\rightarrow$ yellow
$\rightarrow$ red) indicates increasing H$\alpha$+[NII] 
and dust.
CO clouds, indicated with black (in the insert) and white contours, are defined as the area where the emission
  line signal-to-noise is greater than 4. Clouds \#1 and \#3 are somewhat resolved and extend about 1.2$^{\prime \prime}$ 
and 2.9$^{\prime \prime}$.
Note that cloud  \#3 (see enlarged insert) is aligned almost exactly along a 
dust filament, and coincides with knots and filamentary structures in the H$\alpha$+[NII] emission.
Cloud \#1 also coincides with H$\alpha$+[NII] emission and dust extinction, but 
other similar dusty regions were not detected in CO(2-1).
Cloud \#2 is not associated with detectable optical emission
and out of dust extinction map FOV.  The registration of the H$\alpha$+[NII] and CO images is correct within an uncertainty of
about 0.1$^{\prime \prime}$- 0.2$^{\prime \prime}$  in the astrometry in the SOAR data.
The effective radius of NGC~5846 in the K-band is $35.75^{\prime \prime}$. The white dashed circle in the H$\alpha$+N[II] emission map identifies the field of view of the ALMA primary beam.
}
\label{fg:clloc1}
\end{figure*}

The final list of clouds contains 3 detections for NGC~5846, 2 for NGC~4636 and 17 for NGC~5044. 
Our new data reduction and cloud selection criteria for NGC~5044 yield a lower number of clouds detection (17) when compared to the earlier published list (24) by \cite{david14}.
Figure \ref{fg:cloudsD} shows the location of CO clouds detected by \citet{david14} and with our updated reduction as blue crosses and red dots, respectively. To be consistent with cloud labeling in \citet{david14}, the original cloud number sequence has been maintained. 
There are 9 unconfirmed clouds previously listed in \citet{david14}, and two new clouds, 25 and 26, that passed our detection criteria. The vast majority of clouds in the central region are confirmed as well as is the ridge of clouds extending to the north-west part of the sky.
The reason for the discrepancy in CO cloud detection in NGC~5044, may be caused by the selection criteria that we used to detect clouds. 
Indeed, \cite{david14} used a lower signal-to-noise ratio of 4 as threshold for cloud
detection in the earlier NGC~5044 reduction.
To avoid false-positive detections we set our criteria to be quite restrictive and conservative, requiring a robust signal-to-noise ratio threshold to be met.
This is supported by the fact that most of the missing clouds resides toward the edge of the primary beam extension where the instrumental noise is higher.
Also, some of the confirmed clouds are displaced by $\sim 1^{\prime \prime}$ when compared with the cloud location of our new reduction. This may be due to updated housekeeping astrometry in the CASA software.
The physical properties of the confirmed clouds in NGC~5044 will be discussed in the following sections, but in general terms they are in good agreement with  results  previously published by  \citet{david14}.

\begin{figure*}[ht]
\hskip-0.0cm
     \includegraphics[width=9.3cm]{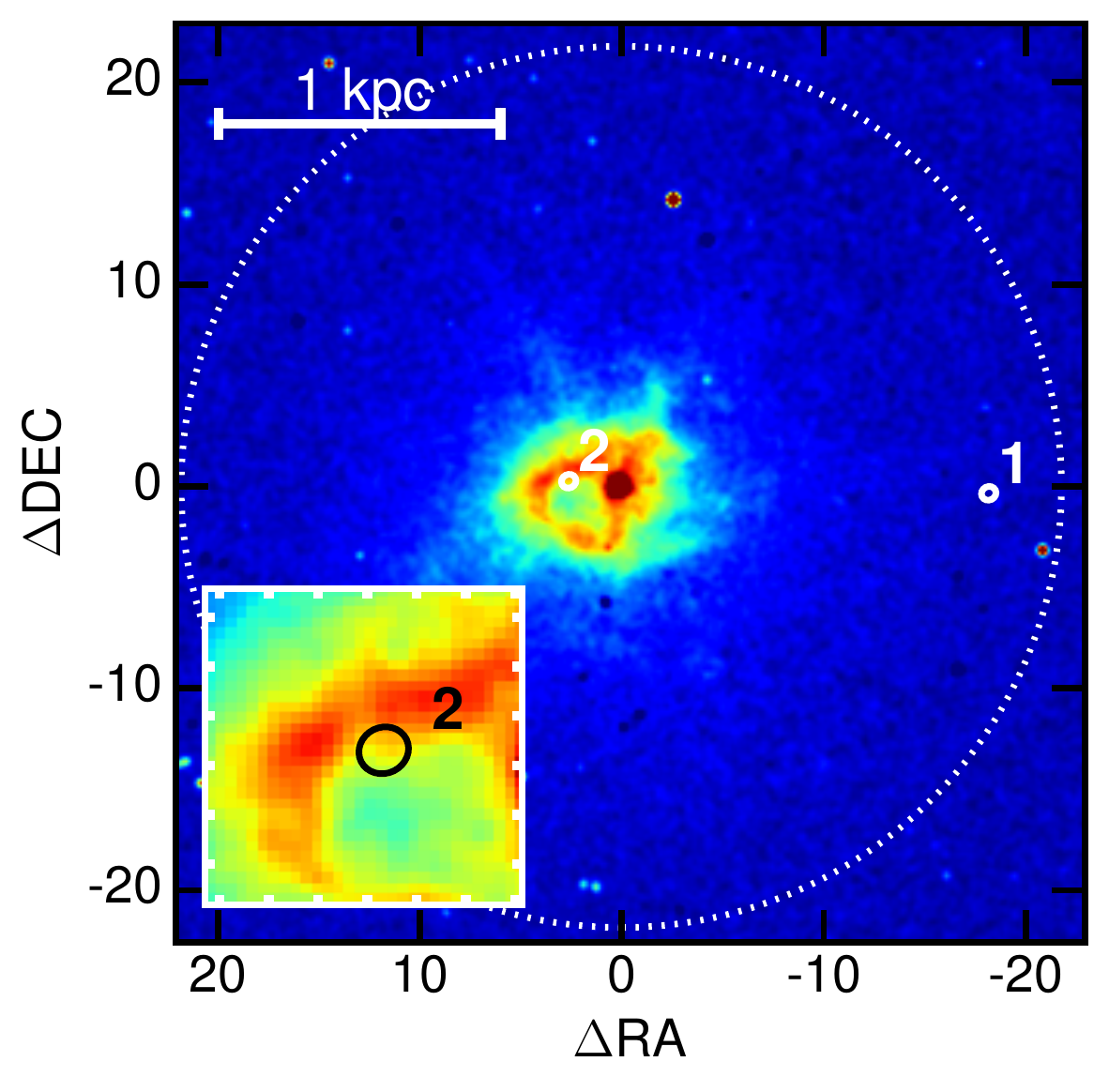}
  \includegraphics[width=8.1cm]{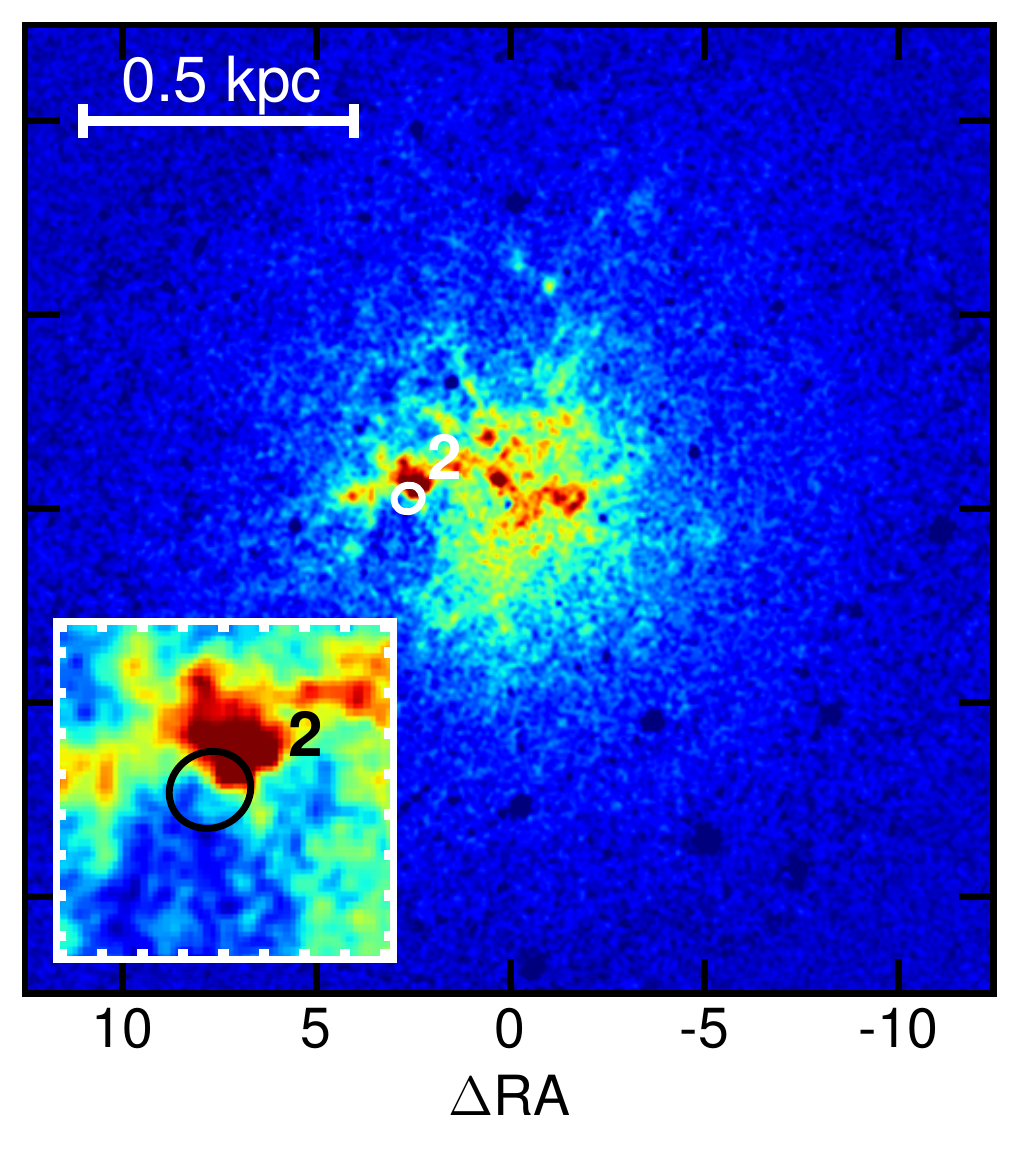}
  \caption{
 CO(2-1) clouds in NGC~4636 projected against the galaxy images (same as in Figure  \ref{fg:clloc1}).
Cloud \#1 is not associated with detectable optical emission
and out of the dust extinction map FOV, while cloud \#2 is centered on a dust absorption knot and also aligned with a ridge in the optical line emission map.
 None of the clouds in NGC~4636 are resolved. 
 The effective radius of NGC~4636 in the K-band is $56.29^{\prime \prime}$.
The white dashed circle in the H$\alpha$+N[II] emission map identifies the field of view of the ALMA primary beam.
}
\label{fg:clloc2}
\end{figure*}

\begin{figure*}[ht!]
  \centerline{\includegraphics[width=6cm]{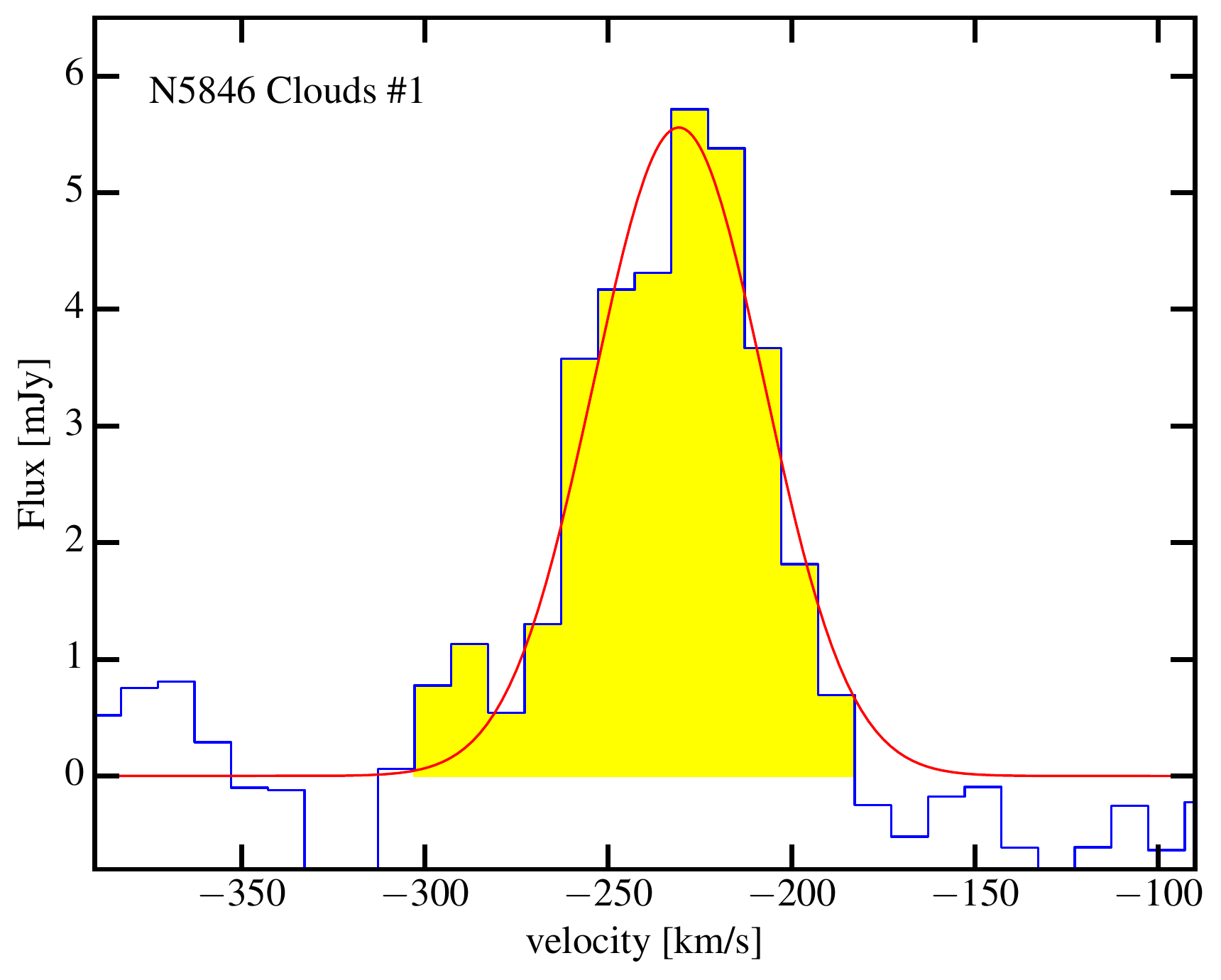}
  \includegraphics[width=6cm]{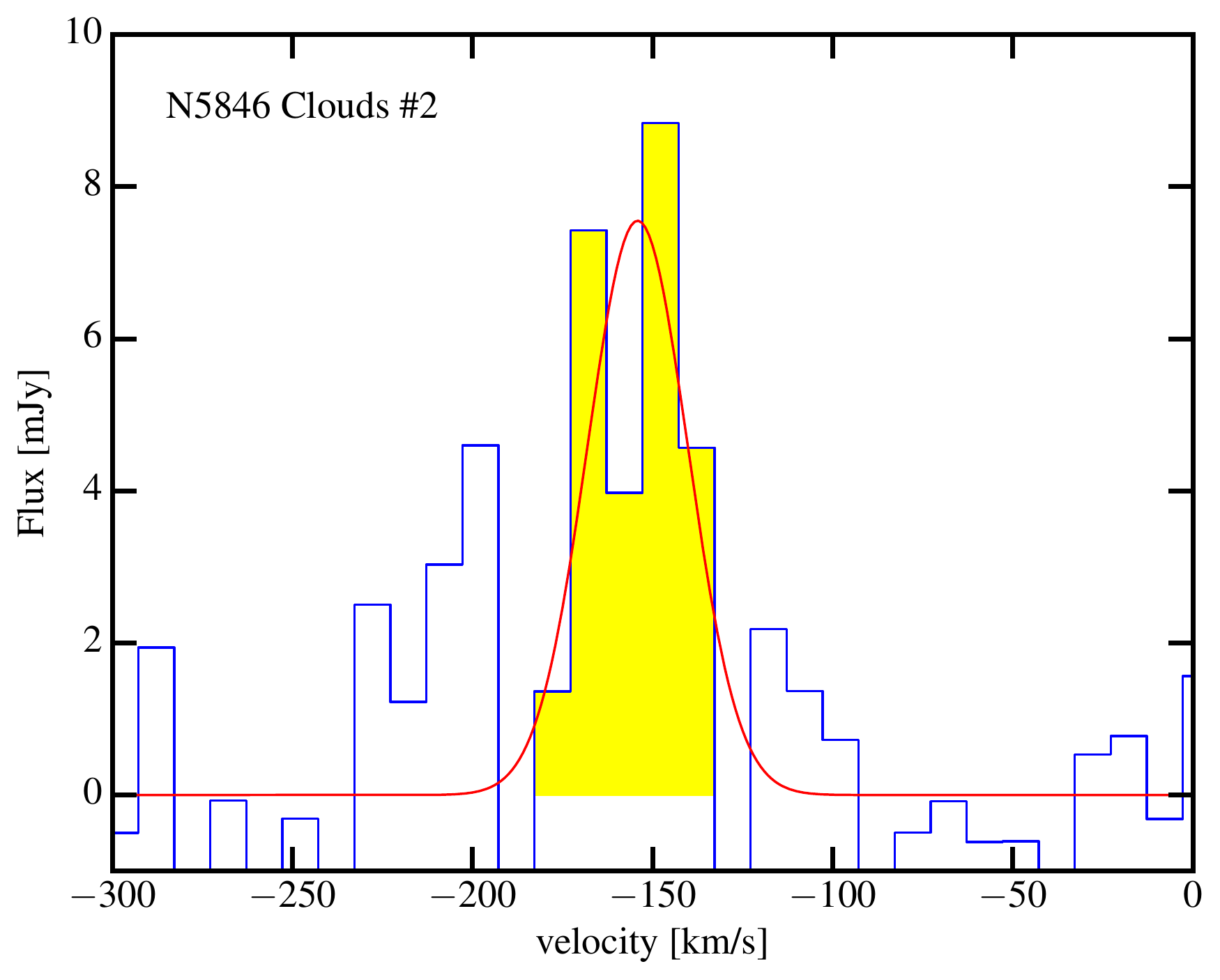}
  \includegraphics[width=6cm]{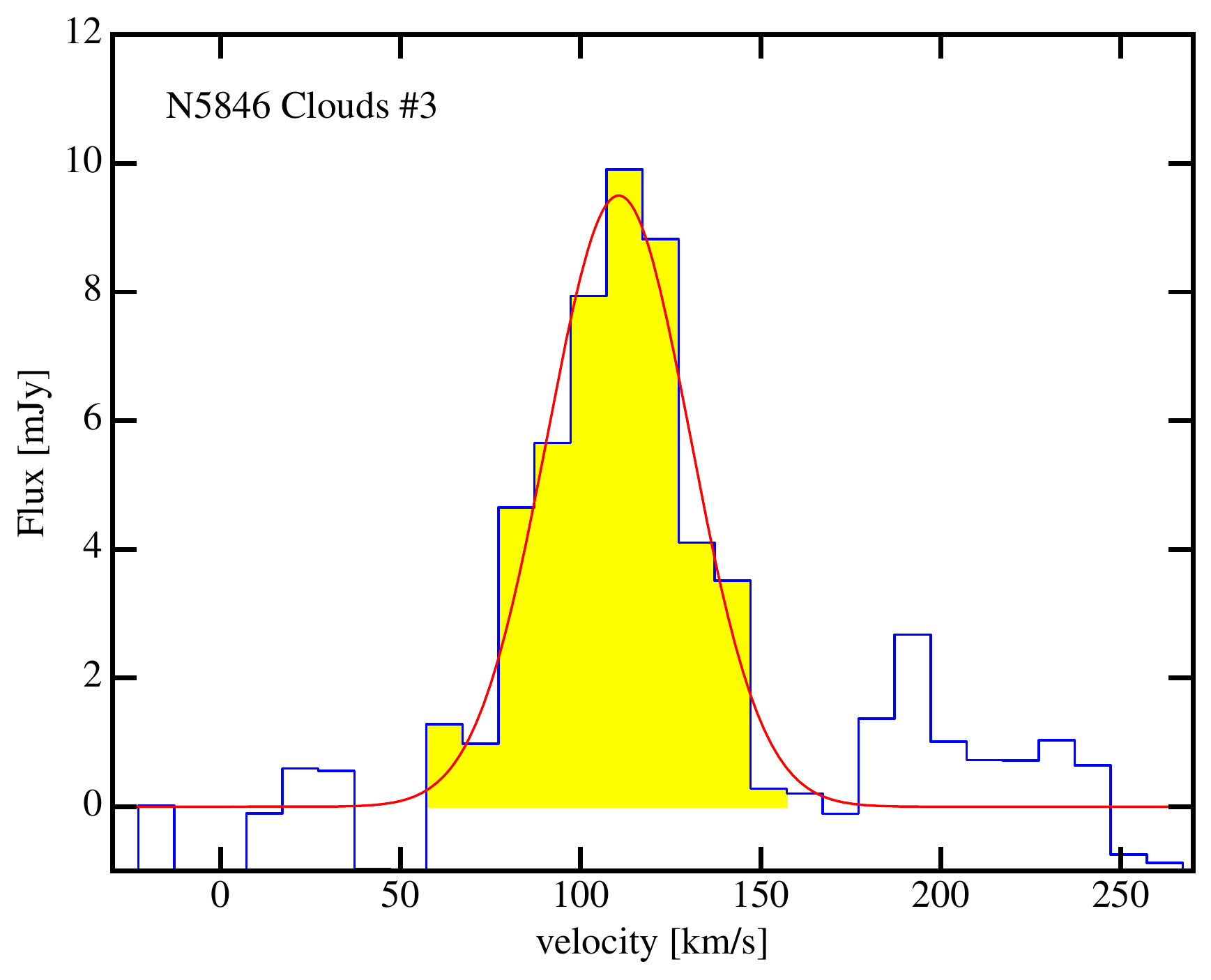}}
  \centerline{
  \includegraphics[width=6cm]{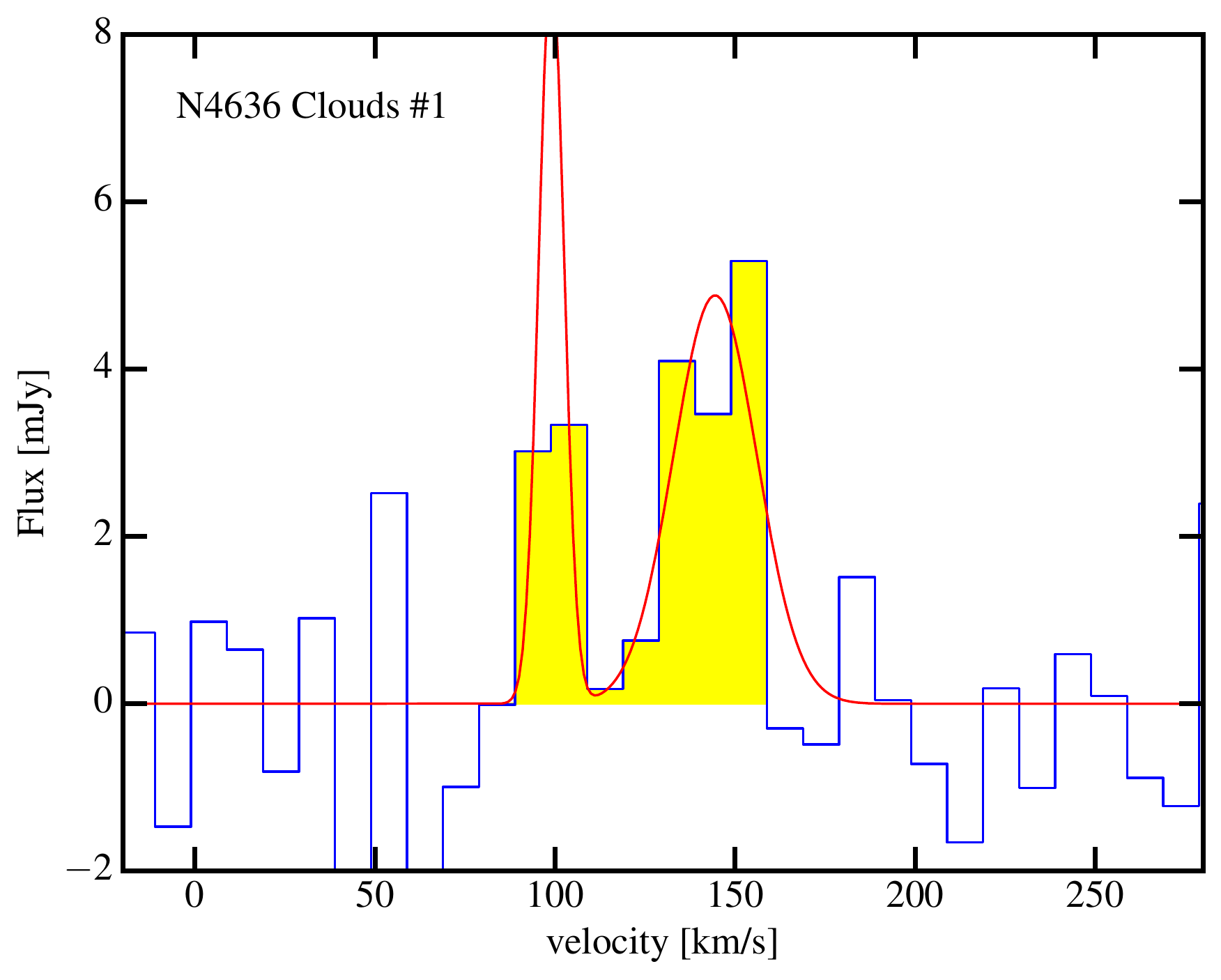}
  \includegraphics[width=6cm]{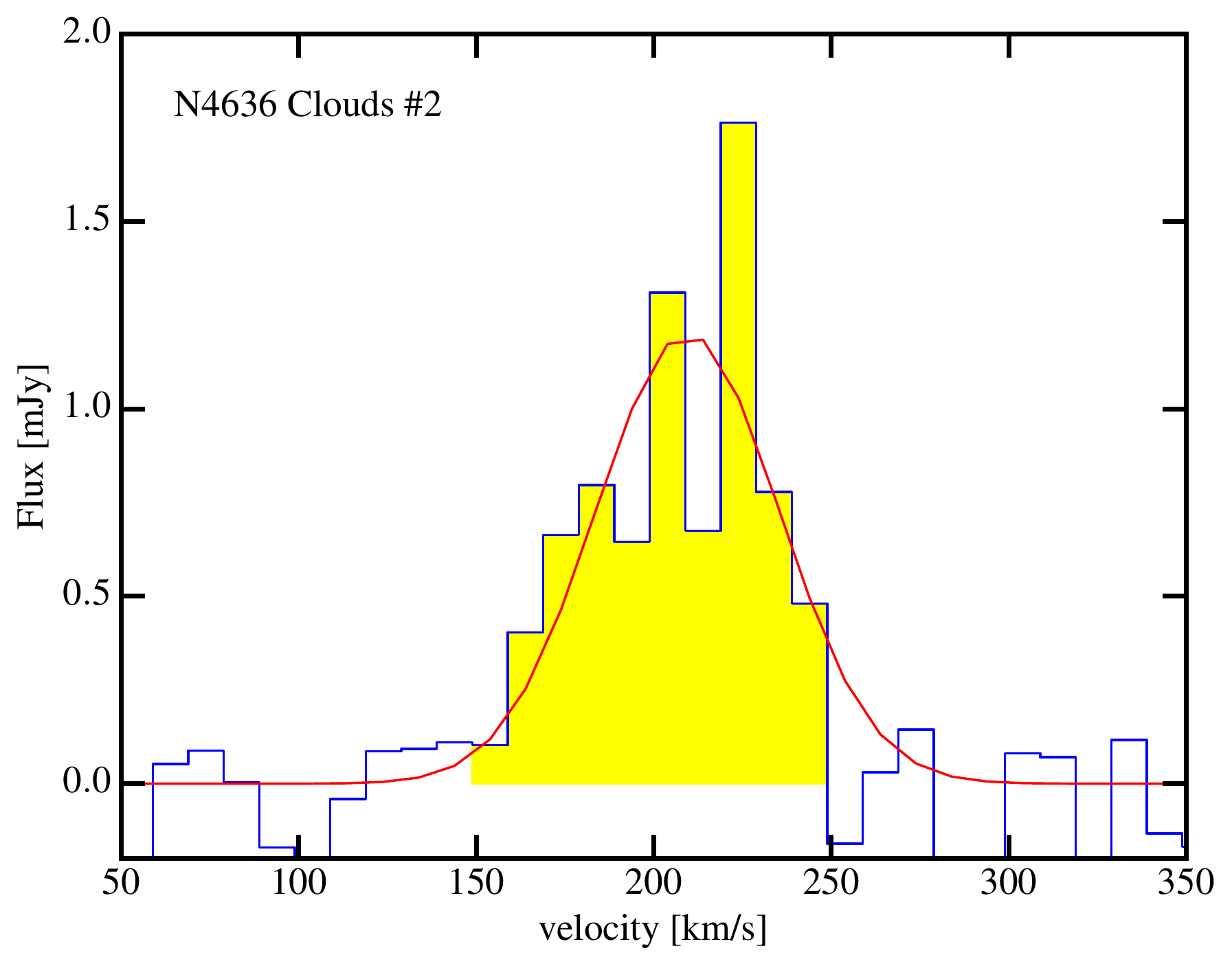}}
  \caption{Spectra of the 5 clouds detected in ALMA CO(2-1) images of NGC~5846 (top) and NGC~4636 (bottom),
  with a Gaussian fit to the emission line (red). Cloud \#1 and \#3 spectra of NGC~5846 were computed respectively over 144 and 221 pixels covering a solid angle of  1.4 and 2.2 arcsec$^2$ (contiguous pixels with signal-to-noise ratio greater than 3), whereas since the other clouds are unresolved, the spectra correspond to the pixel with the largest signal-to-noise.}
\label{fg:clspc}
\end{figure*}

\begin{figure*}[!ht]
  \centerline{\includegraphics[width=9.3cm]{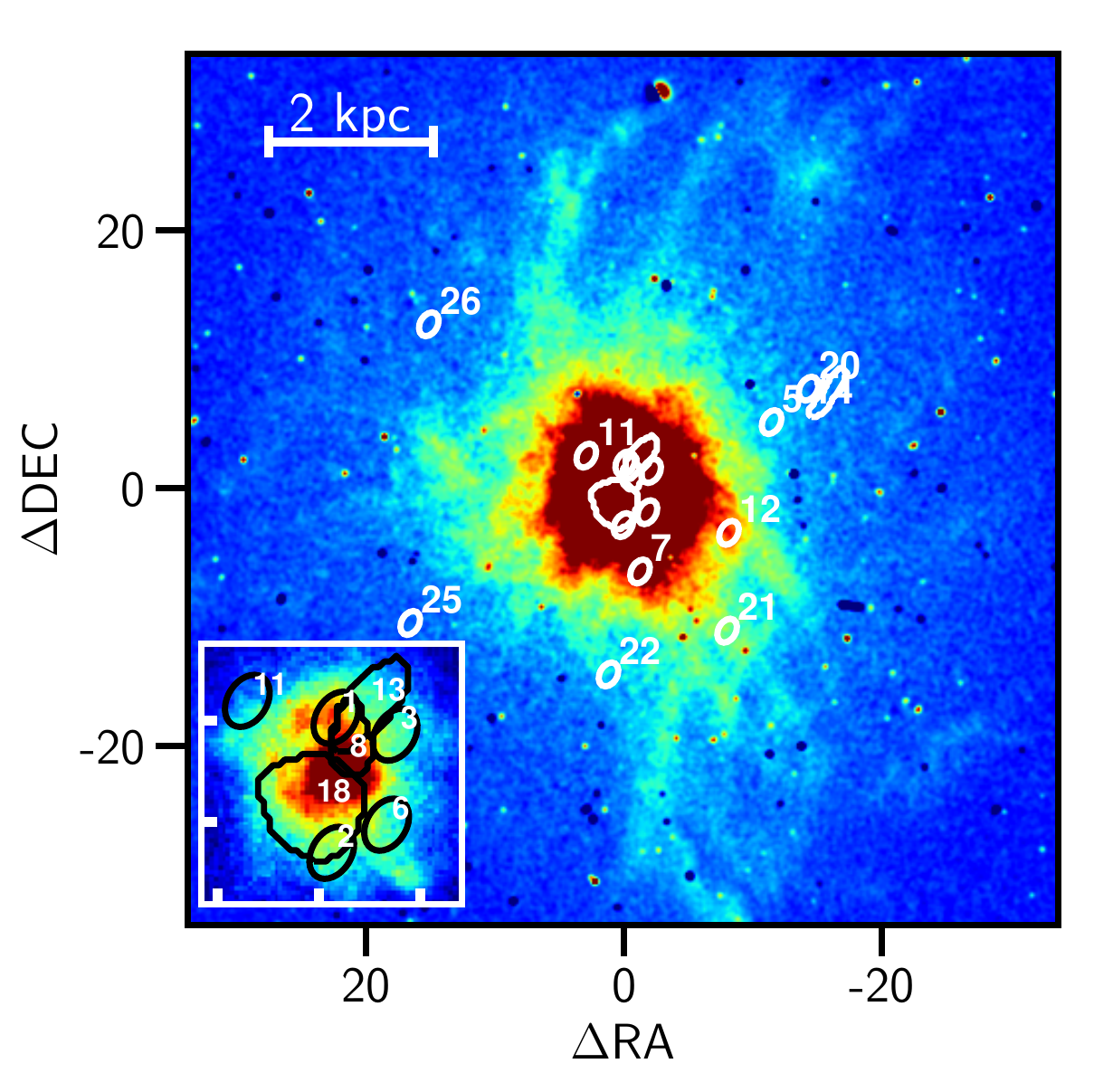}
  \includegraphics[width=8.2cm]{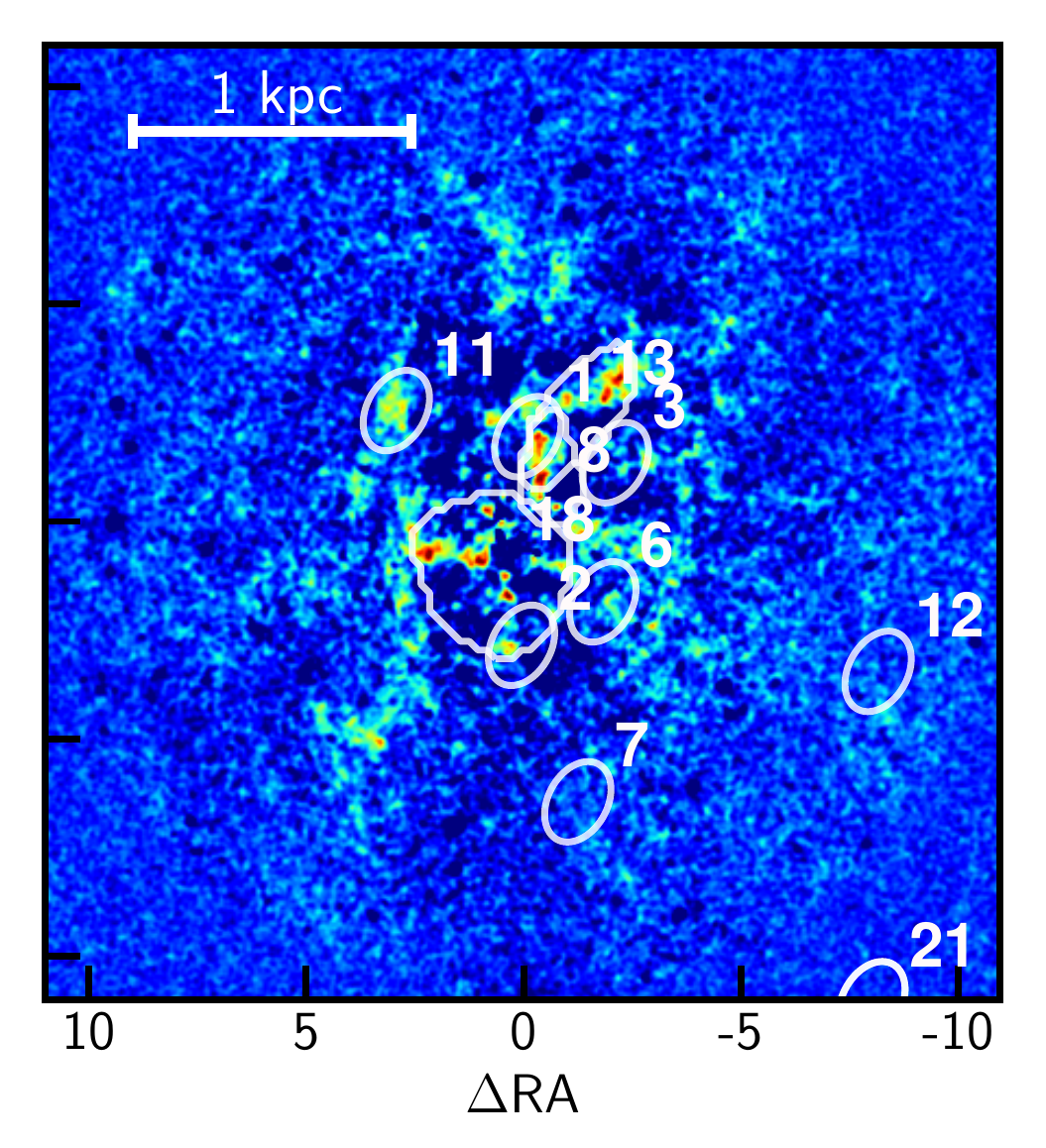}
}
  \caption{Images of NGC~5044 showing detected CO(2-1) clouds projected
against an optical H$\alpha$+N[II] emission (left) and a dust starlight extinction image (right).
Color scheme of the background images and the cloud contours are as described in \ref{fg:clloc1}).
The revised data reduction accounts for 17 CO clouds in the central $\sim \ 6 \ kpc$). Cloud labeling reflects the
original sequence presented in \cite{david14}. The combination of newer pipeline software and a more stringent set of criteria for clouds detection did not confirm few CO clouds originally listed in \cite{david14}.
Clouds \#8, \#13, \#14, and \#18 are resolved.
Three of the four resolved clouds reside in the $\sim \ 2 \ kpc$ (see enlarged insert)  and overlap with several knot in the dust extinction maps as well as with the H$\alpha$+[NII] emission in the galaxy center. 
The registration of the H$\alpha$+[NII] and CO images is correct within an uncertainty of
about 0.1$^{\prime \prime}$- 0.2$^{\prime \prime}$ in the astrometry in the SOAR data.
}
\label{fg:clouds3}
\end{figure*}

\begin{figure*}[ht!]
 \centerline{\includegraphics[width=4.4cm]{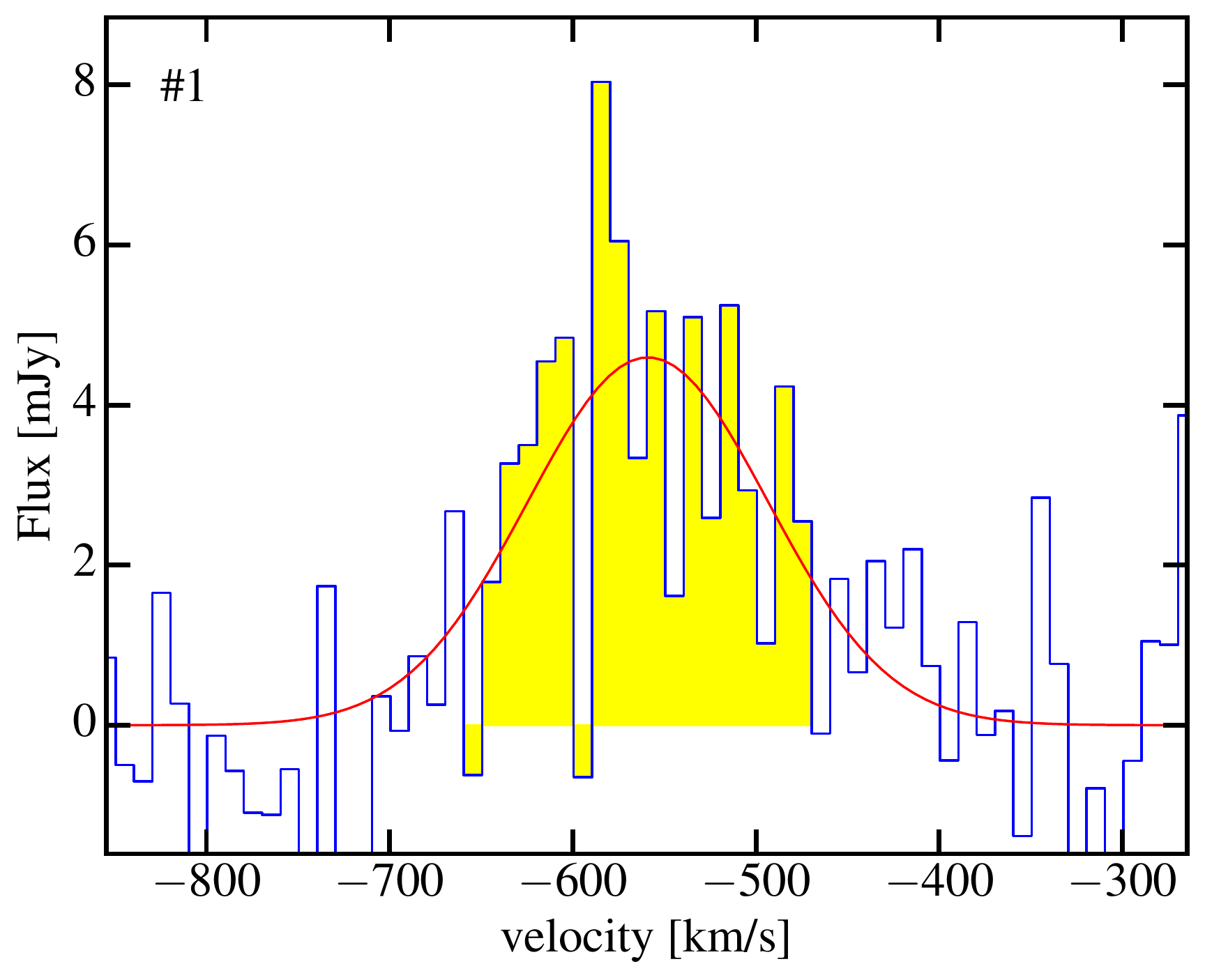}
 \includegraphics[width=4.35cm]{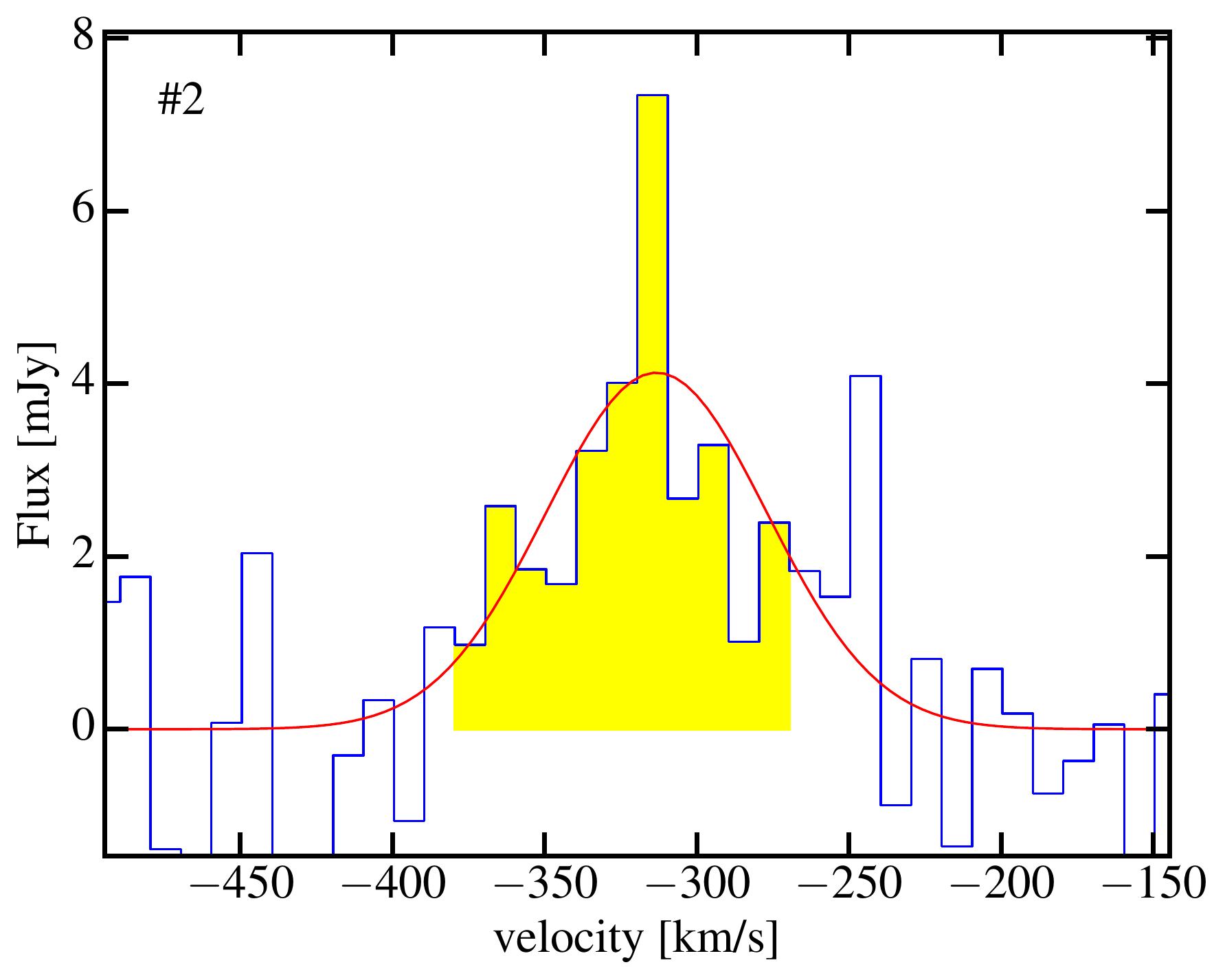}
 \includegraphics[width=4.30cm]{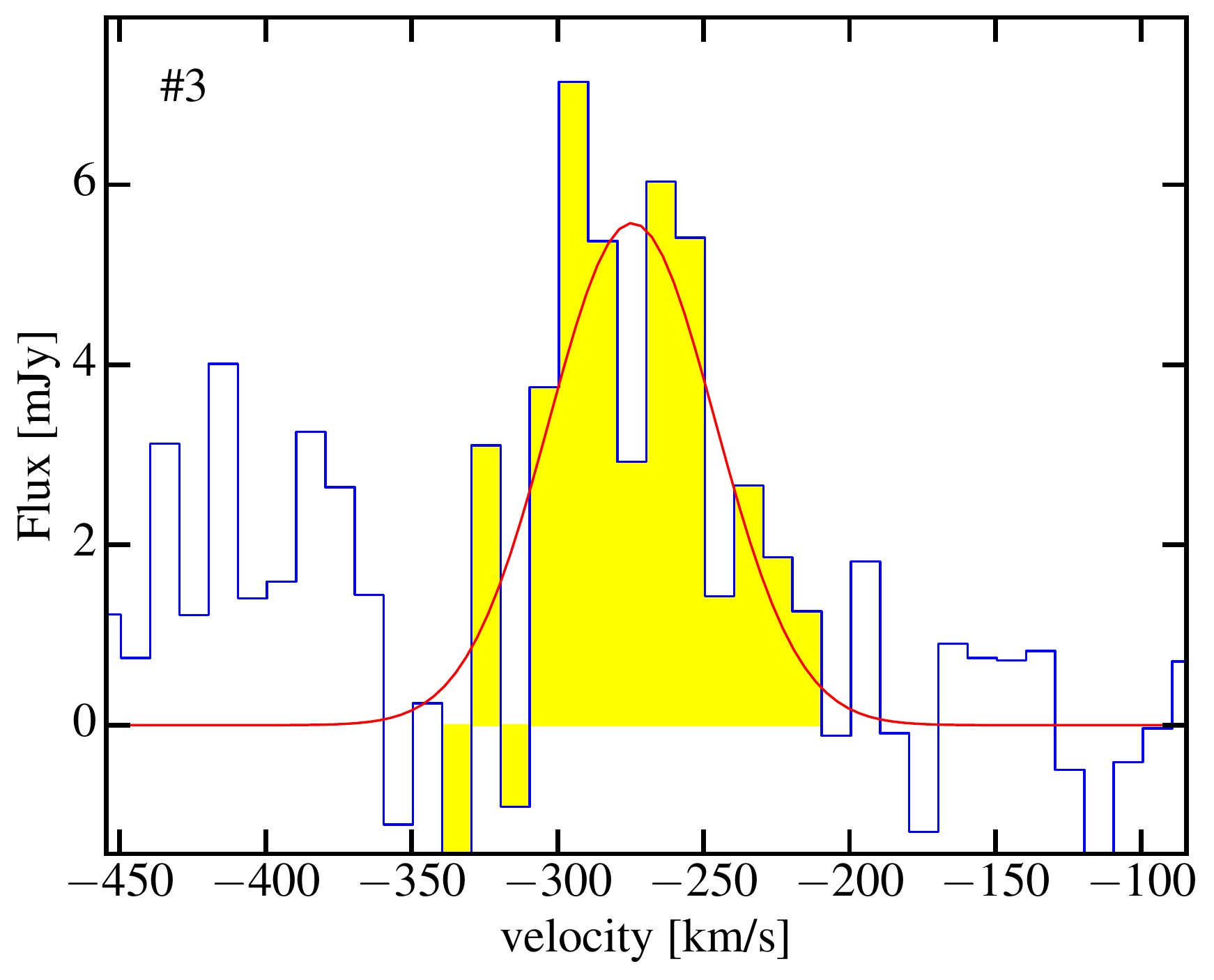}
 \includegraphics[width=4.35cm]{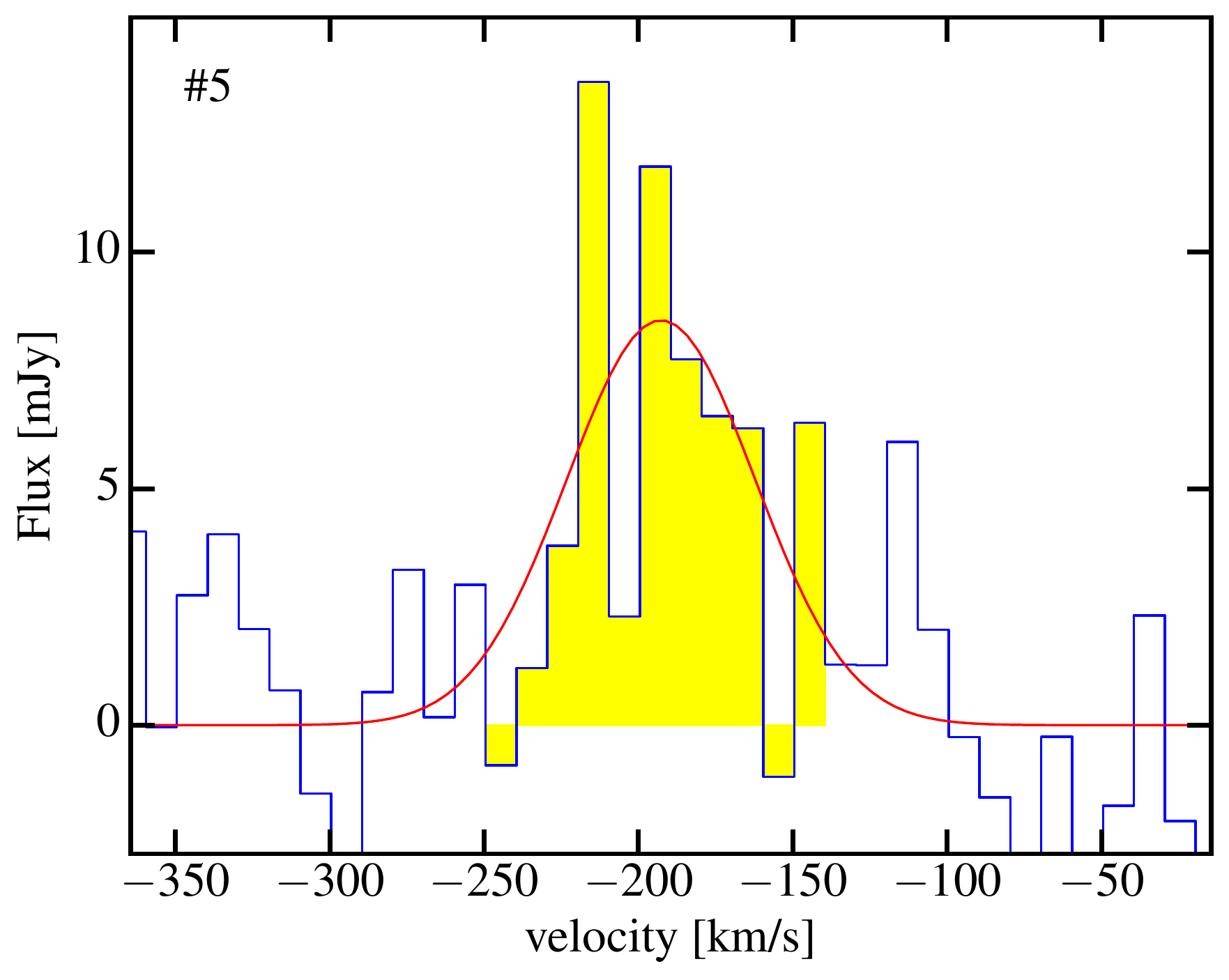}}
 \centerline{
 \includegraphics[width=4.3cm]{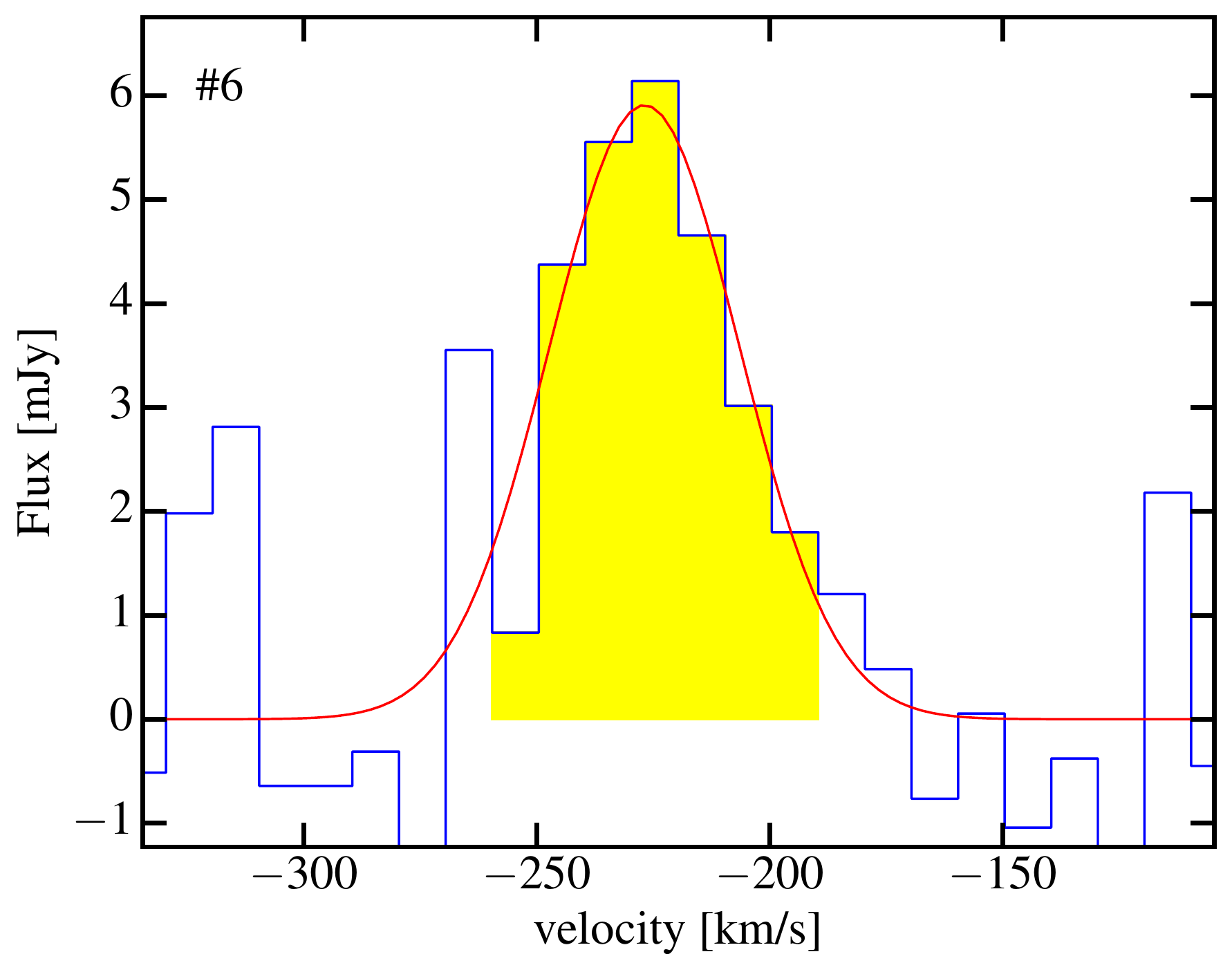}
 \includegraphics[width=4.3cm]{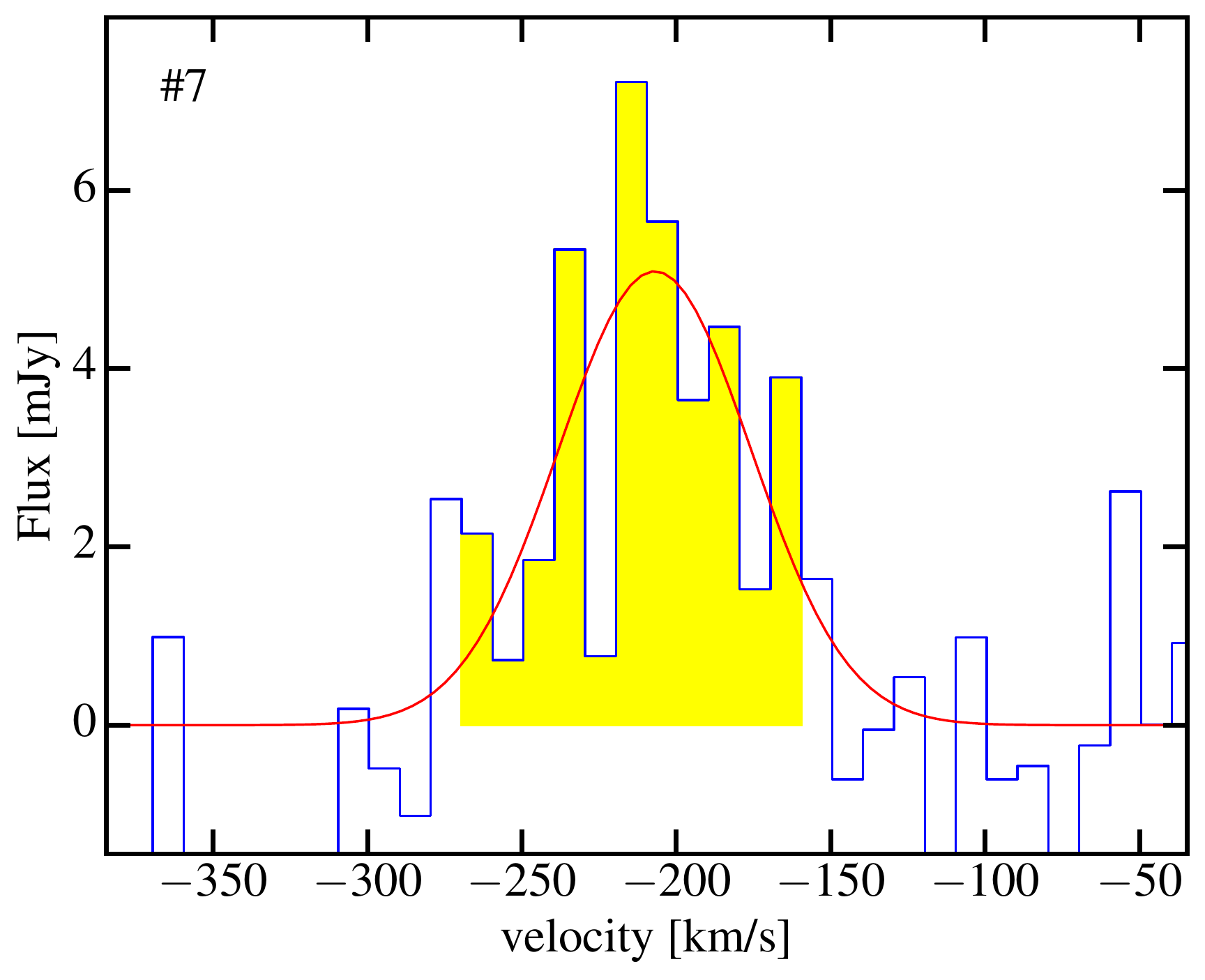}
\includegraphics[width=4.3cm]{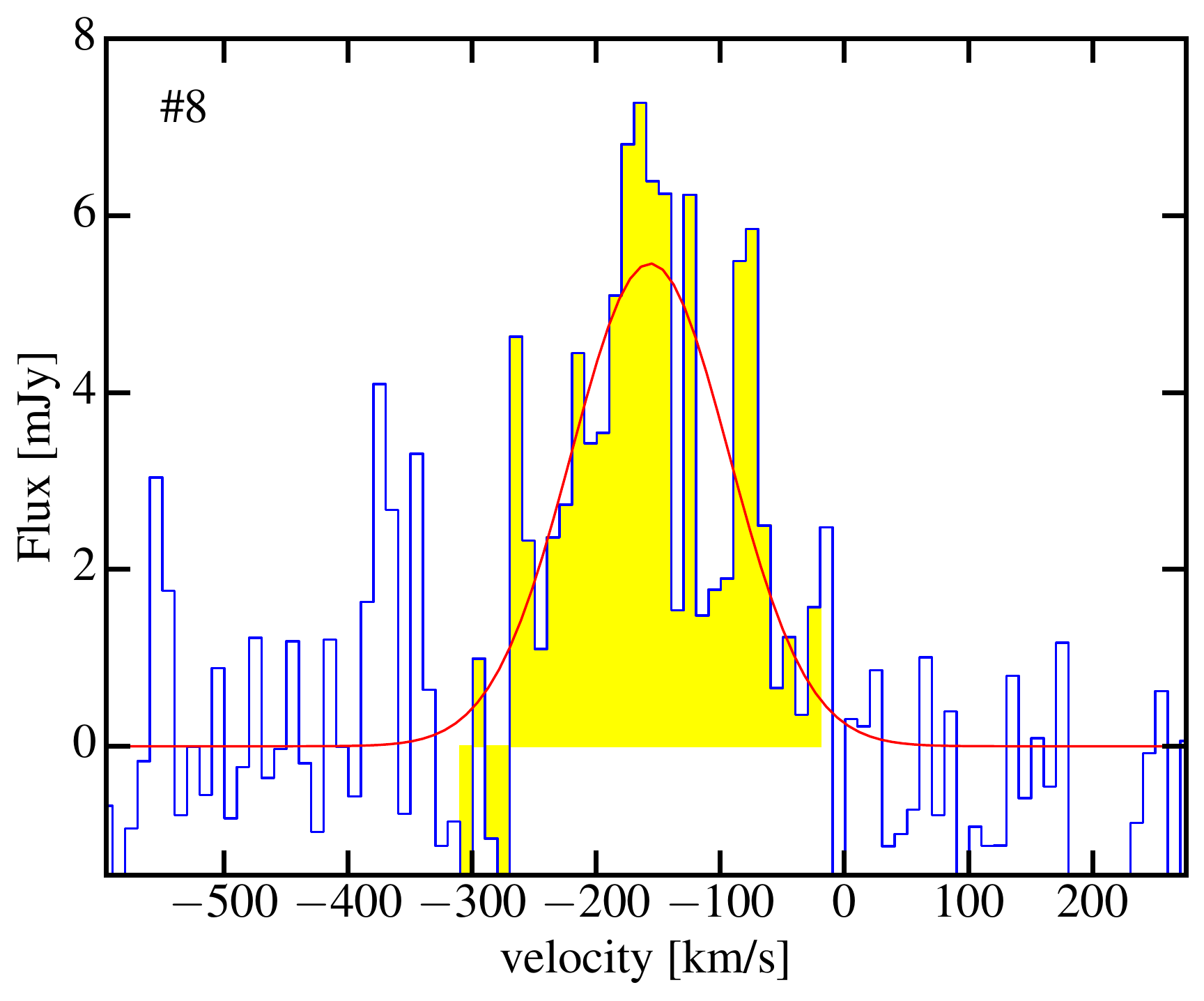}
 \includegraphics[width=4.25cm]{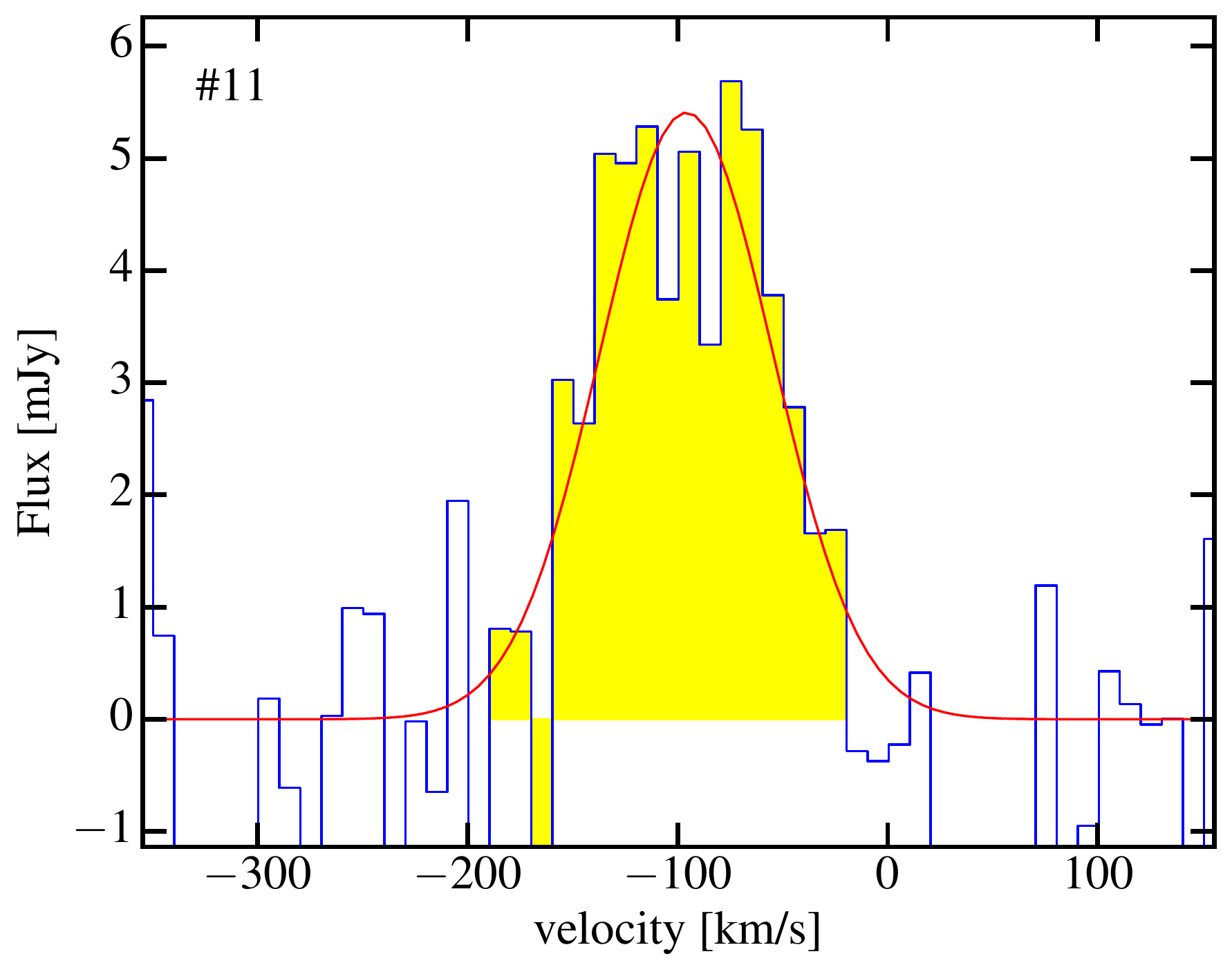}}
 \centerline{
 \includegraphics[width=4.3cm]{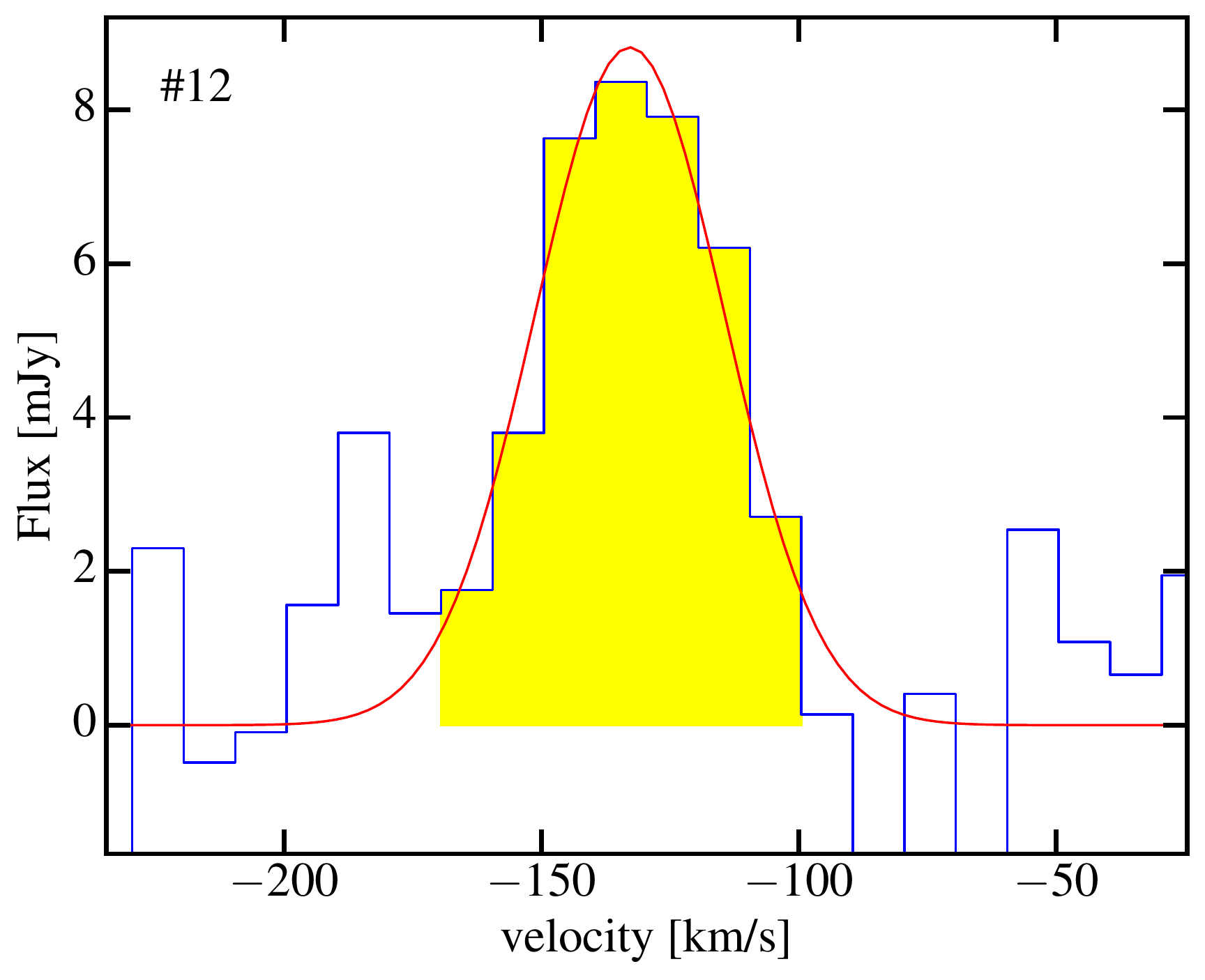}
 \includegraphics[width=4.3cm]{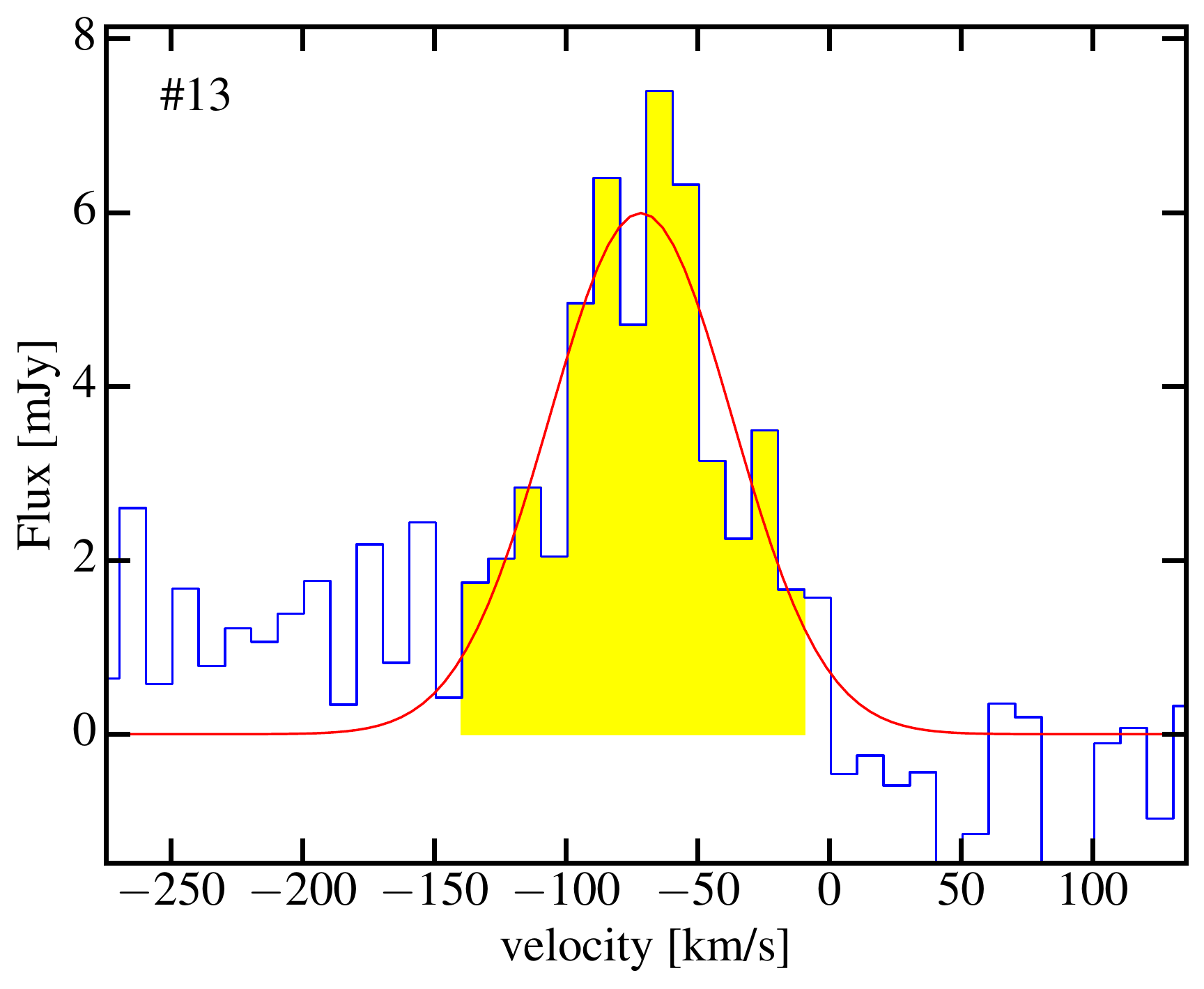}
 \includegraphics[width=4.3cm]{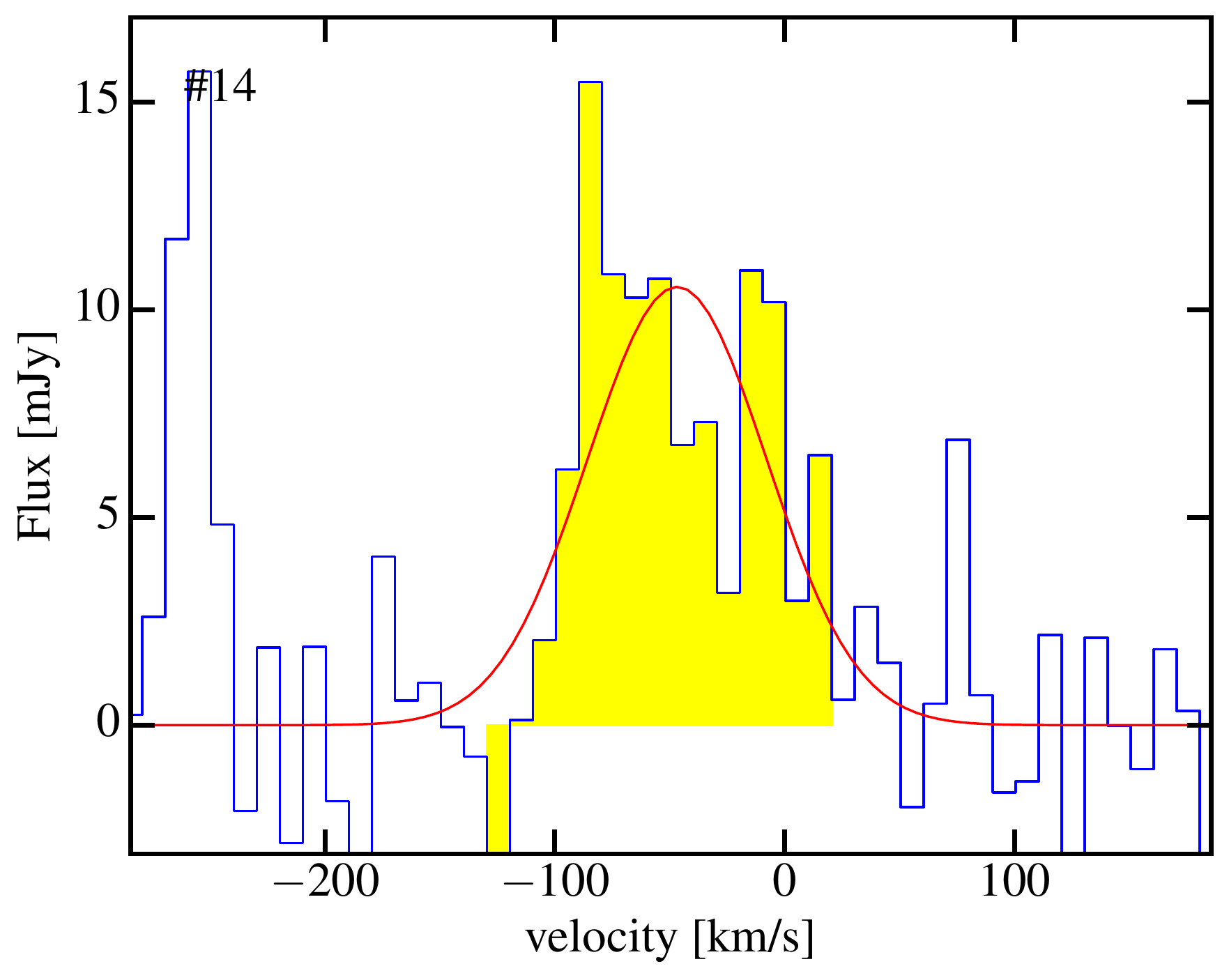}
 \includegraphics[width=4.3cm]{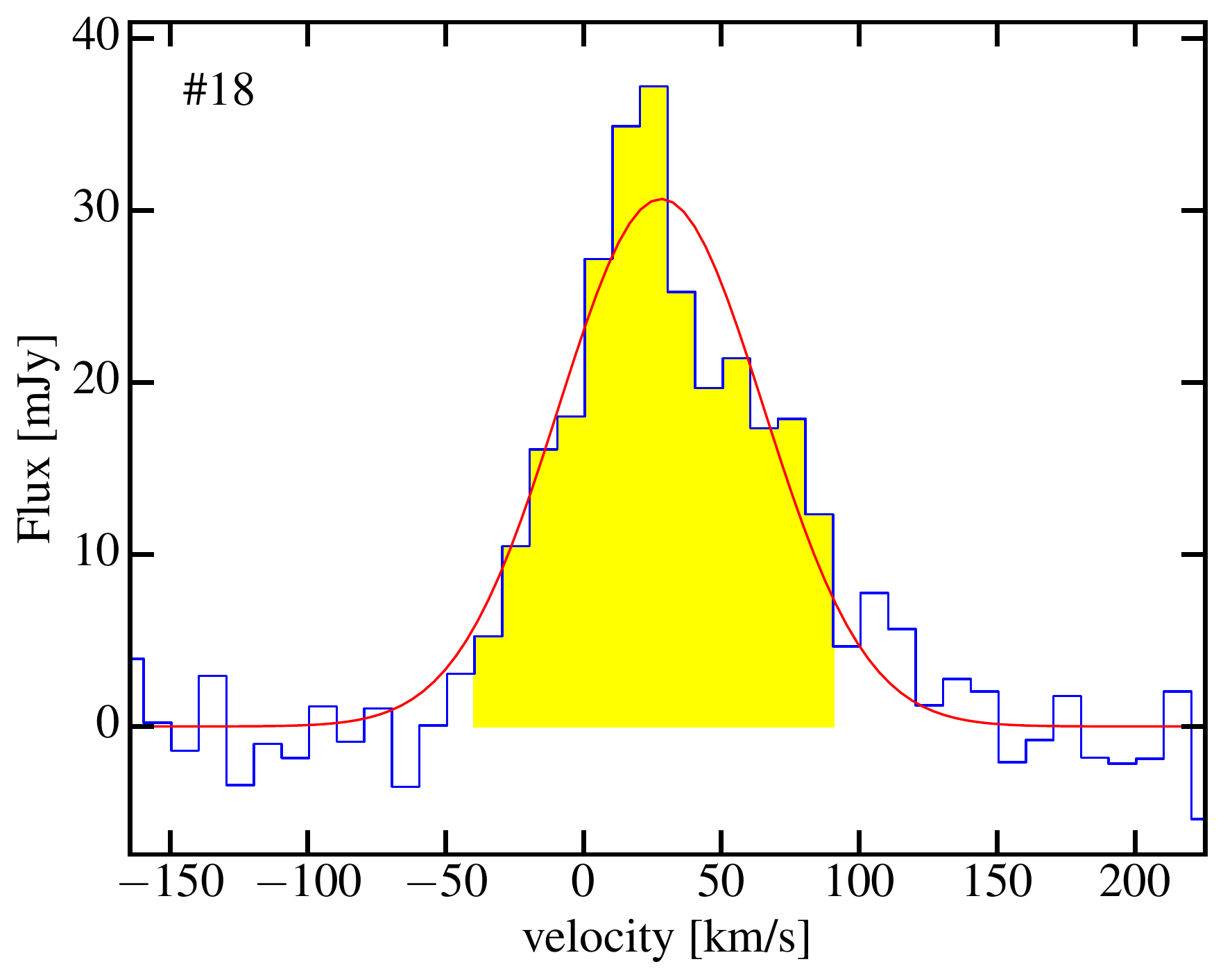}}
 \centerline{
\includegraphics[width=4.32cm]{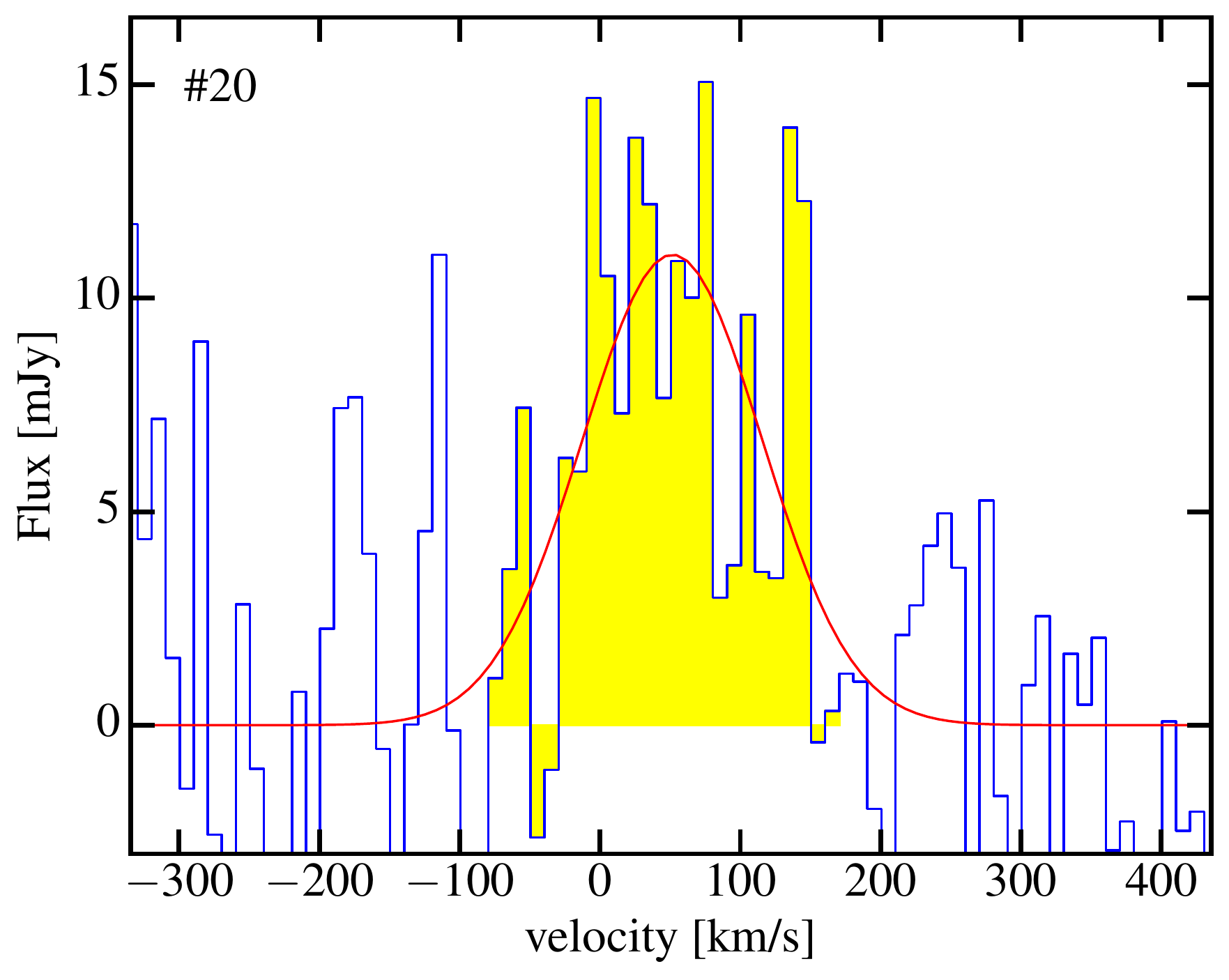}
 \includegraphics[width=4.5cm]{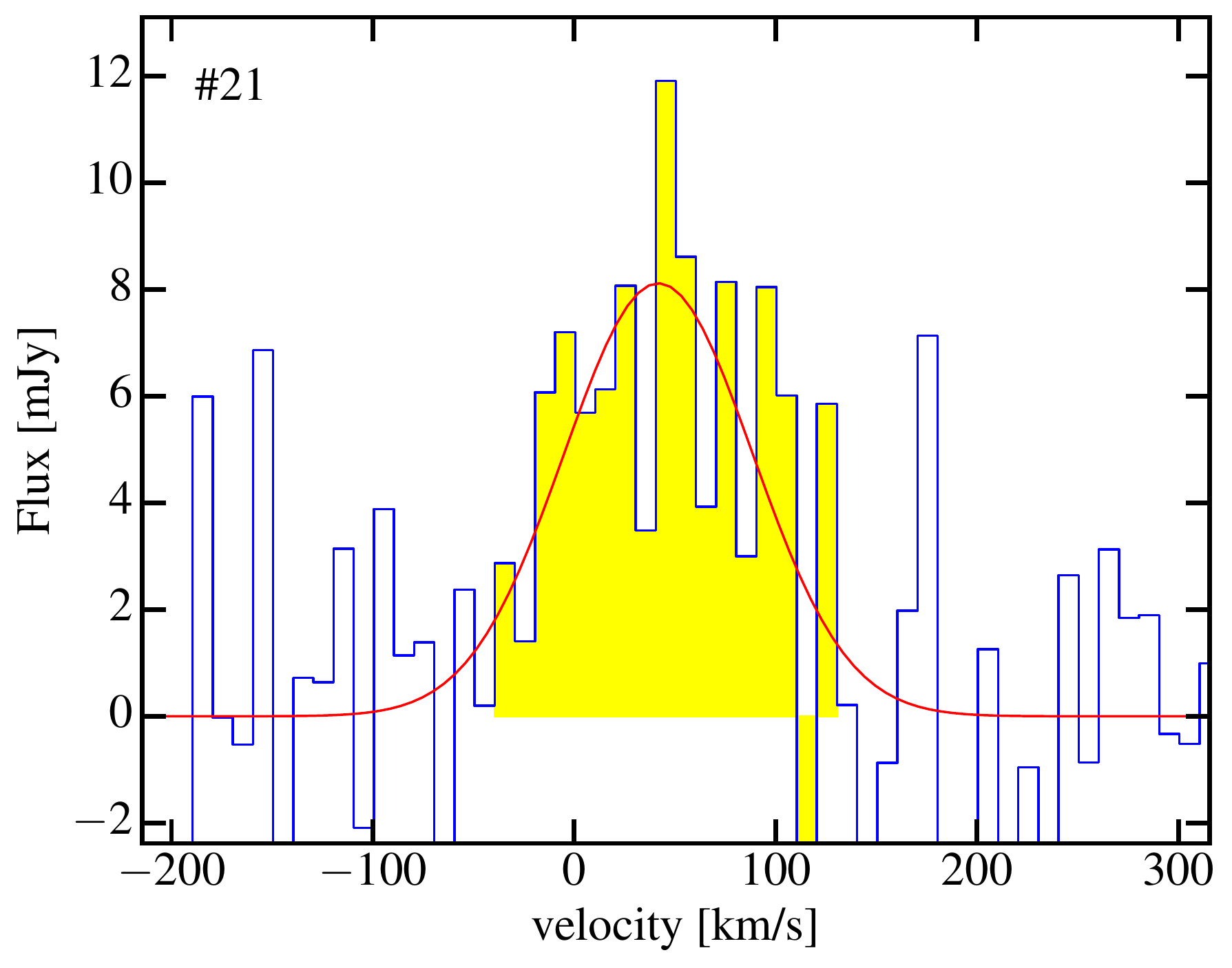}
 \includegraphics[width=4.22cm]{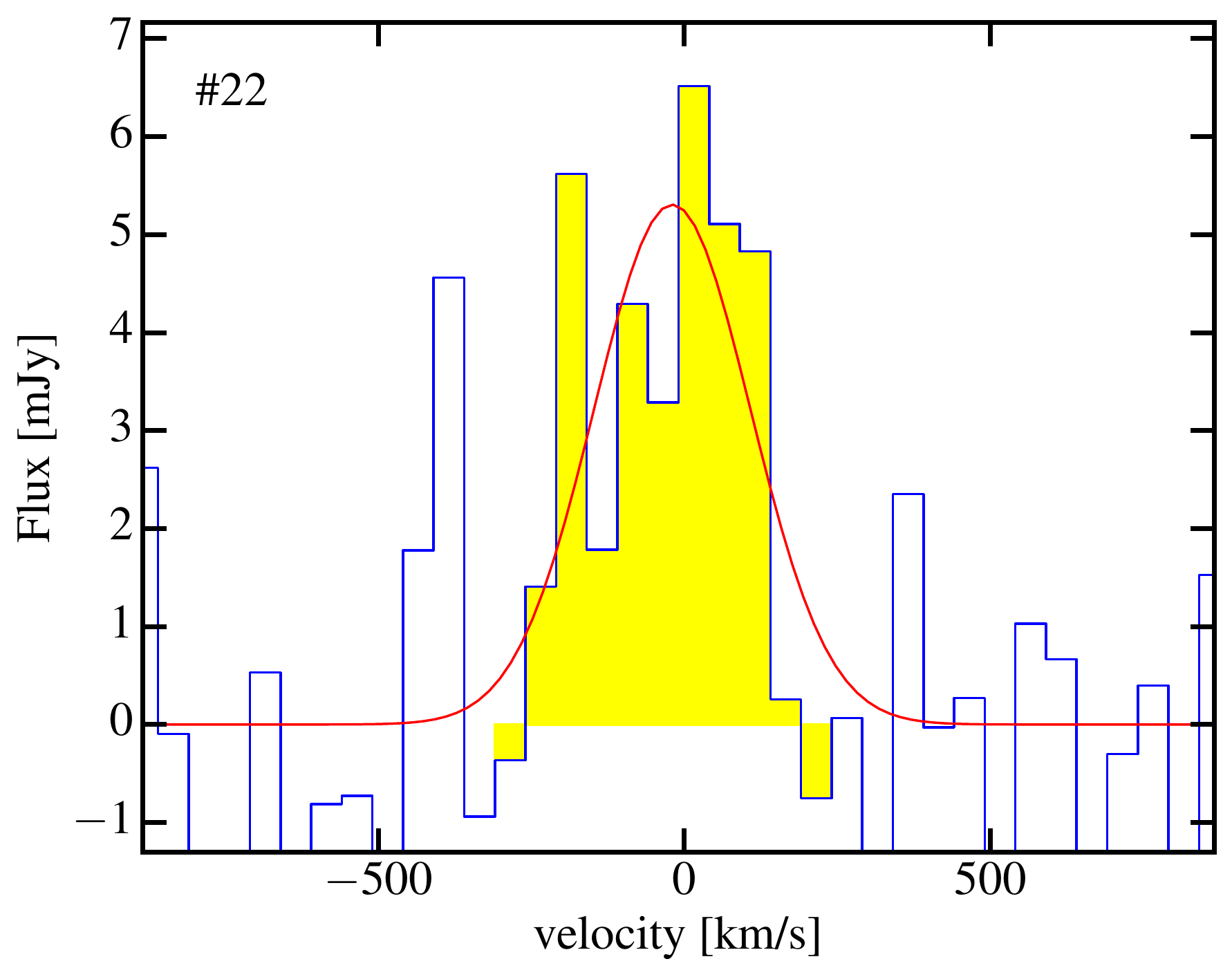}
 \includegraphics[width=4.3cm]{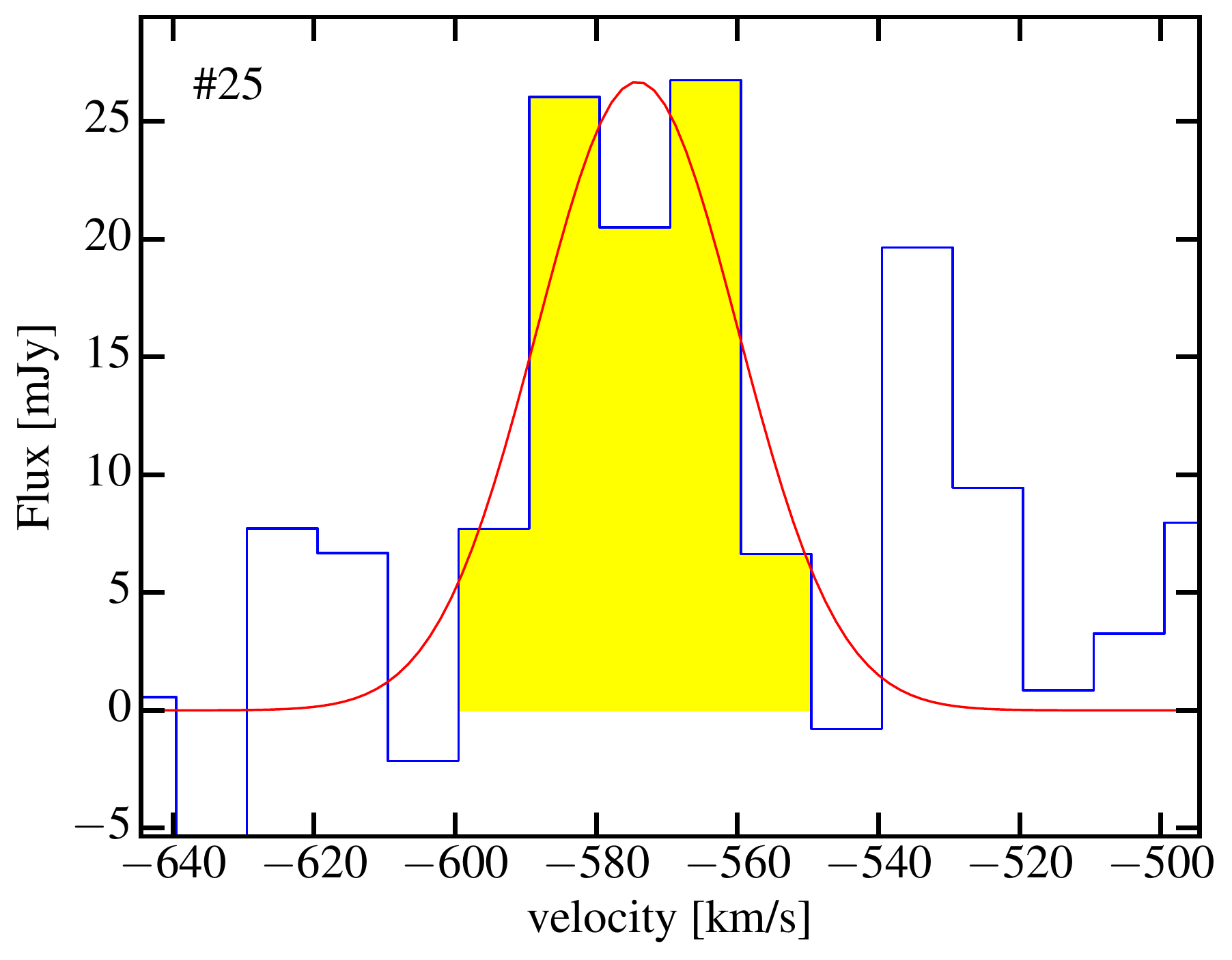}}
\centerline{
 \includegraphics[width=4.3cm]{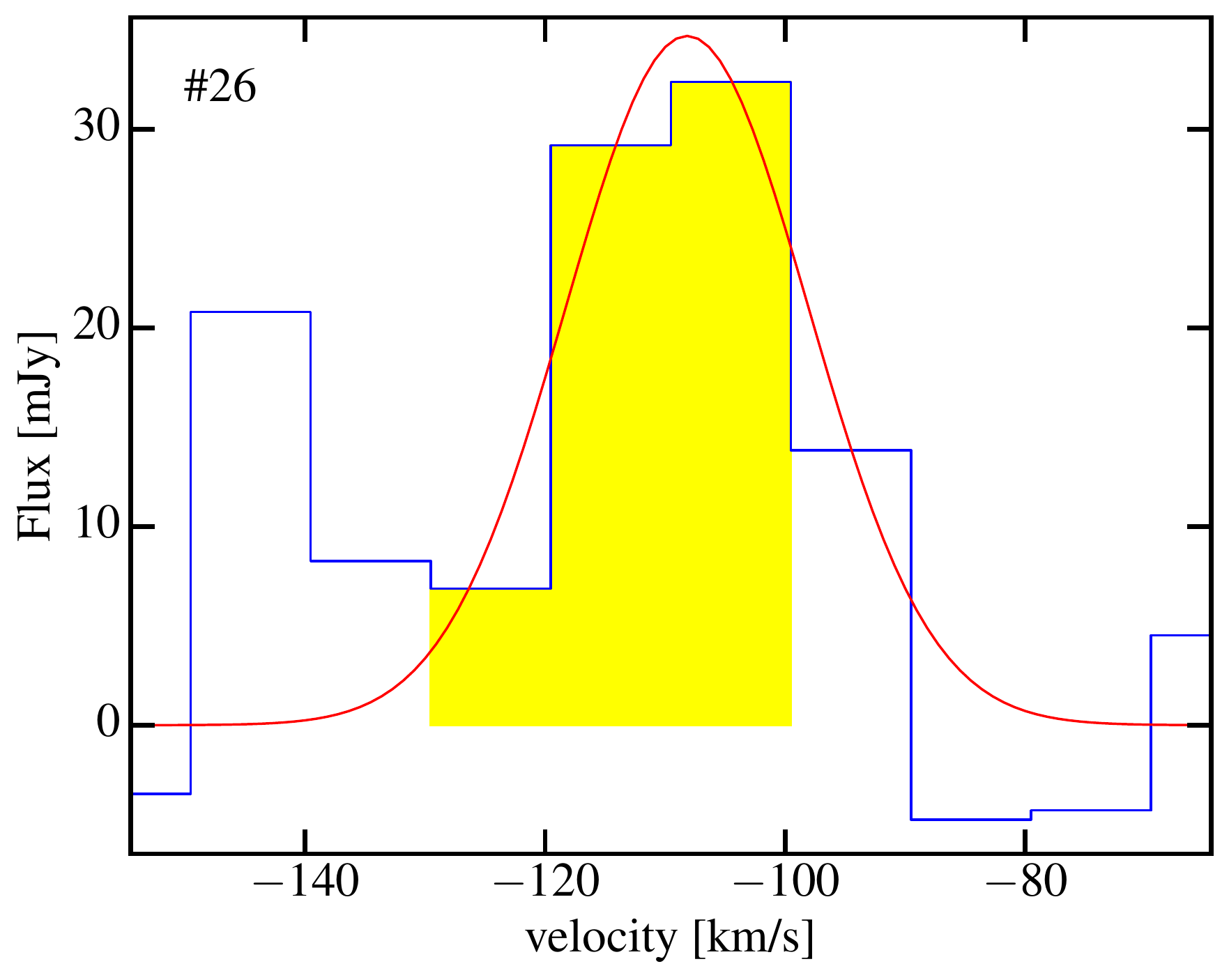}}
 \caption{CO(2-1) spectra of the 17 clouds detected in NGC~5044
 along with a Gaussian fit to the emission line (red curve). The velocity bin size
is 10 km\,s$^{-1}$ except for the cloud \#22 where the bin size is 50 km\,s$^{-1}$.}
\label{fg:clspc5044a}
\end{figure*}

\begin{figure*}[ht]
\centerline{
  \includegraphics[width=5.8cm]{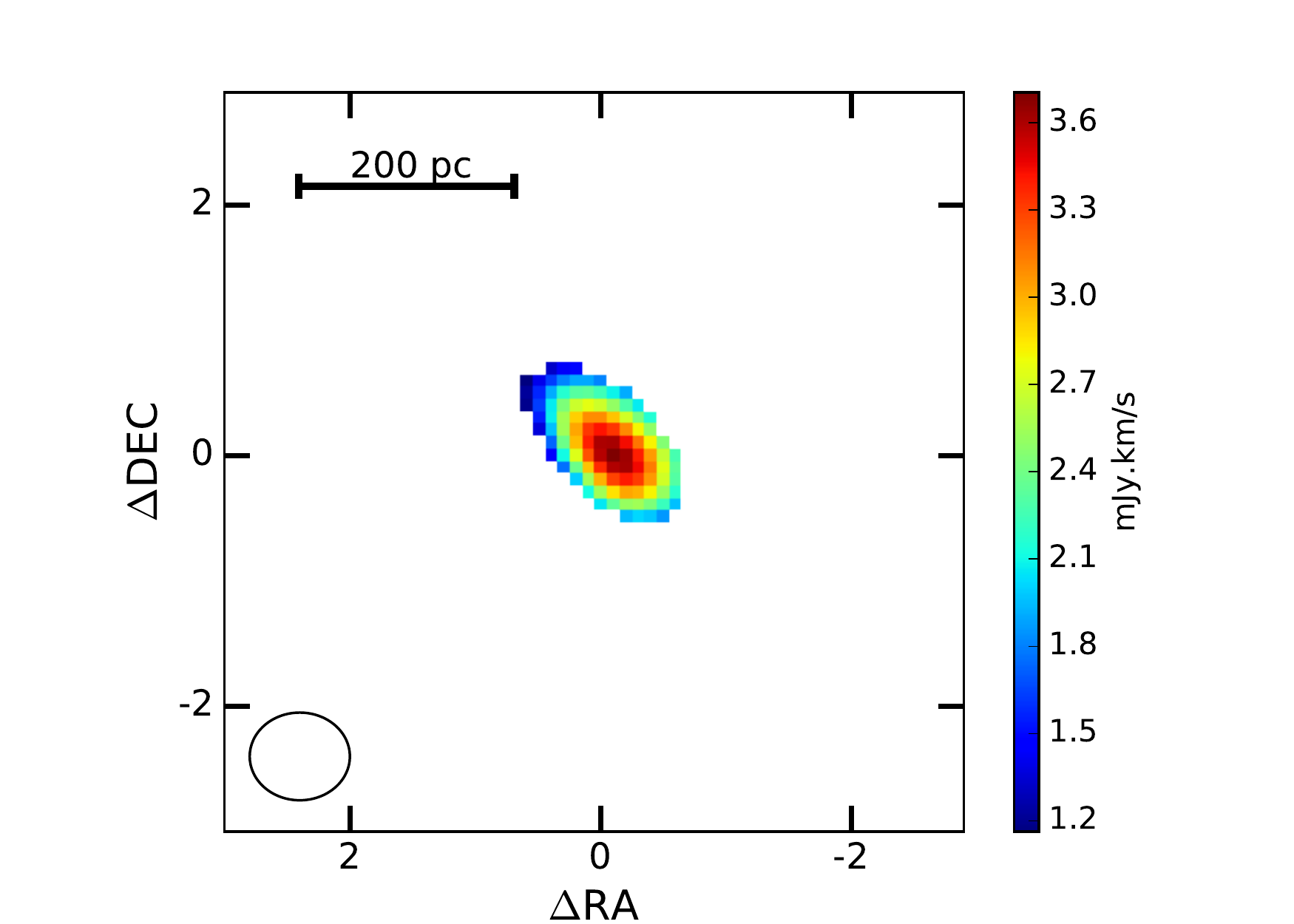}
  \includegraphics[width=5.9cm]{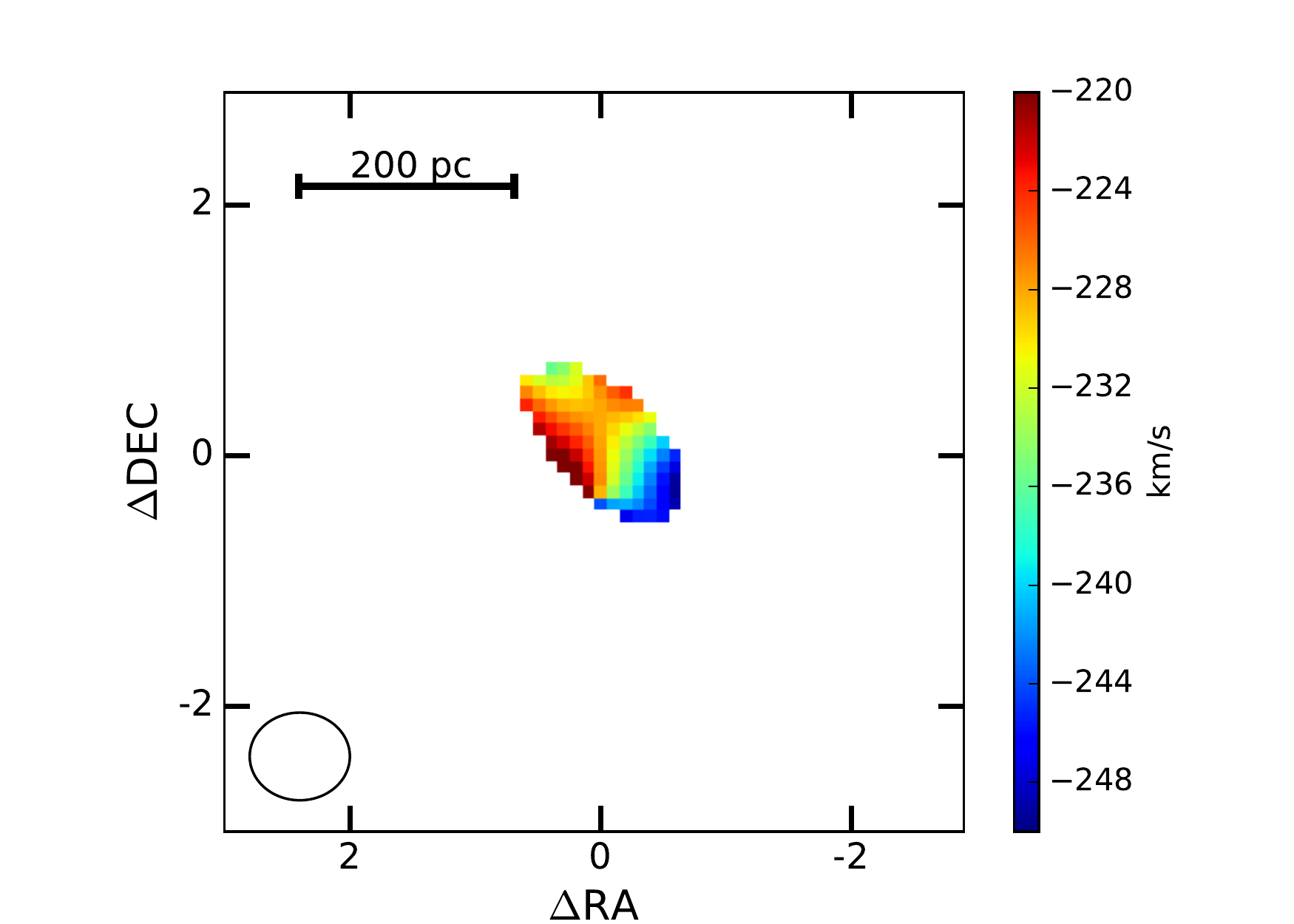}
  \includegraphics[width=5.8cm]{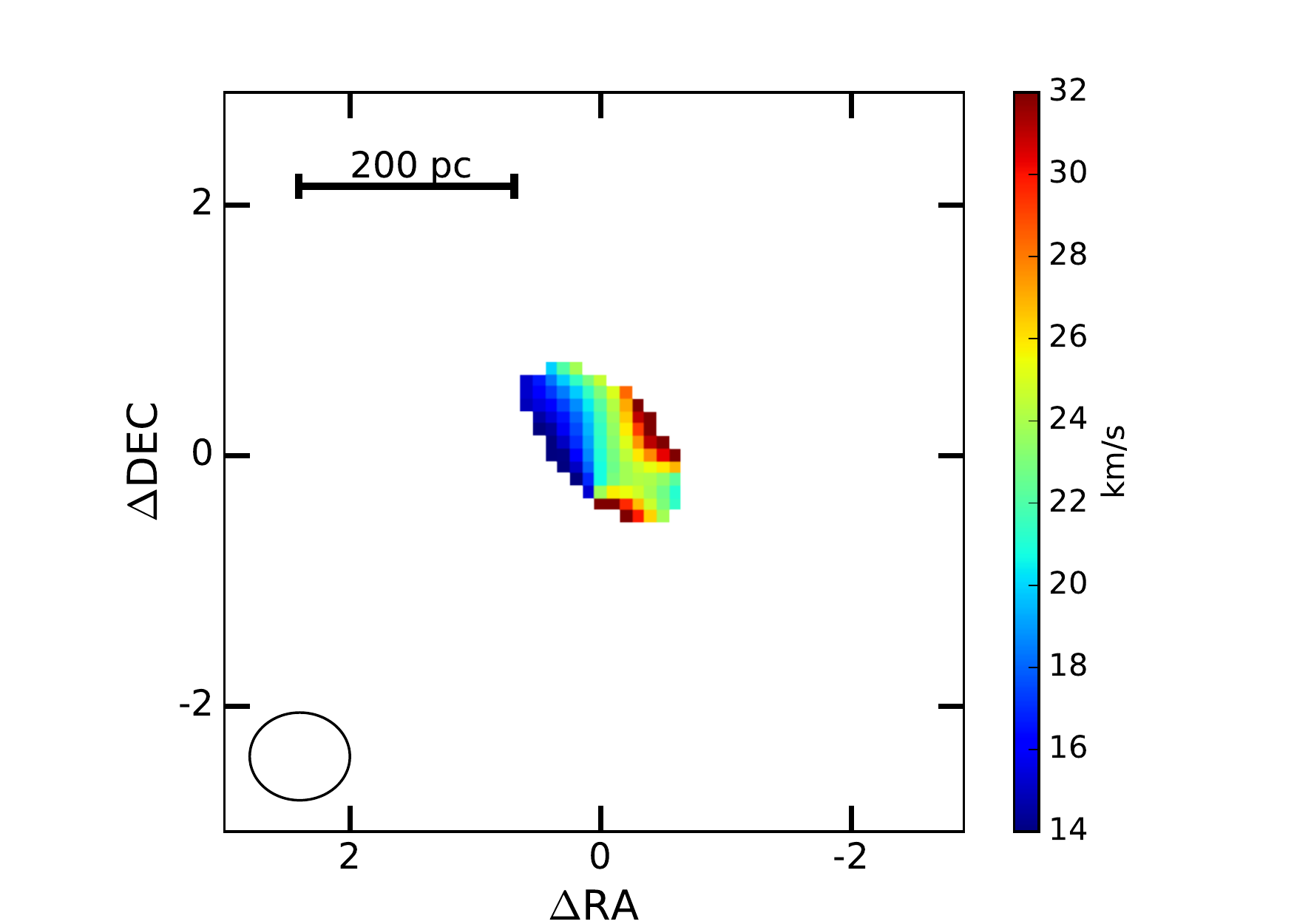}
}
\centerline{
  \includegraphics[width=5.9cm]{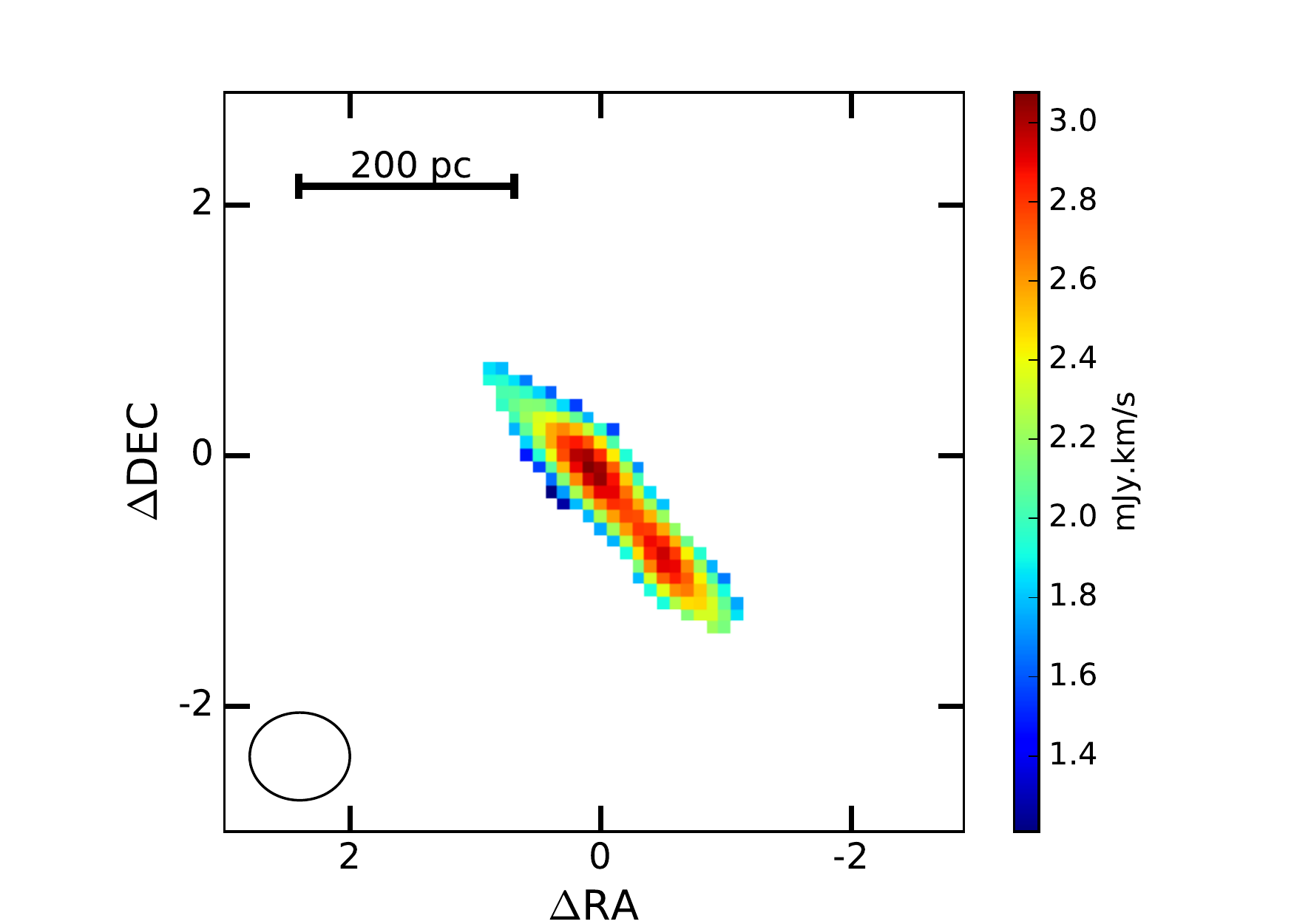}
  \includegraphics[width=6.0cm]{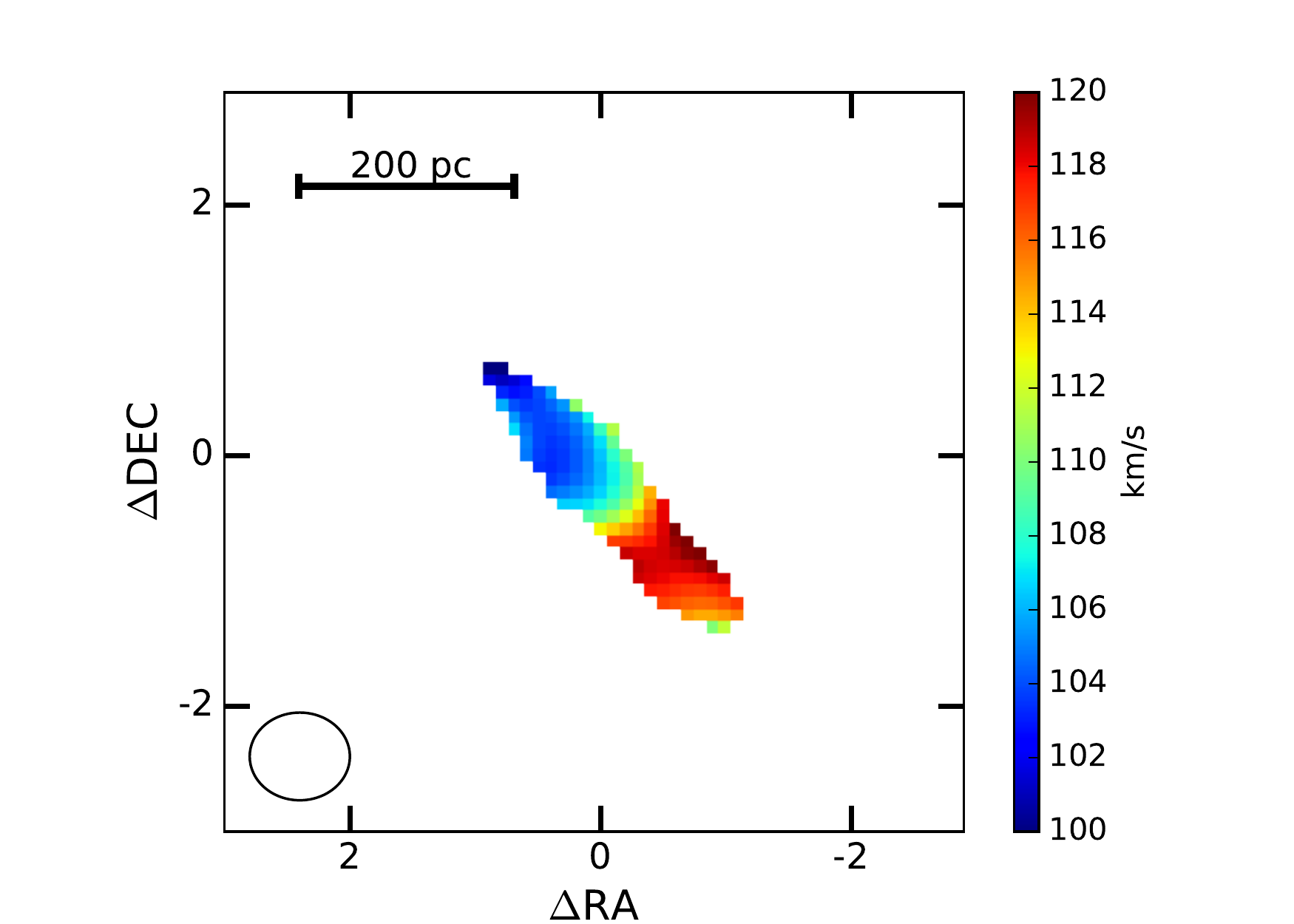}
  \includegraphics[width=5.9cm]{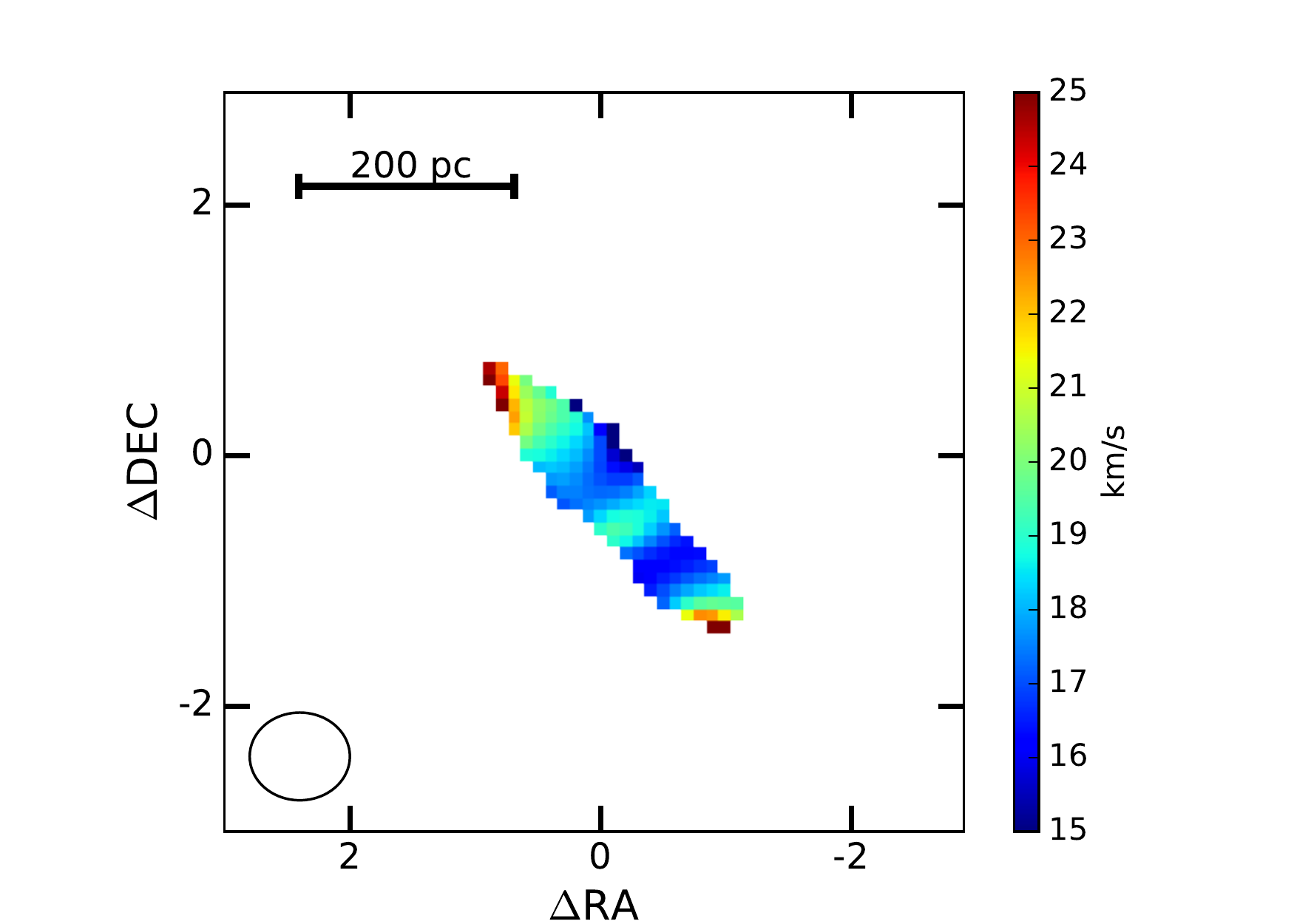}
}
  \caption{Flux (left), velocity (center) and velocity dispersion (right)
    in cloud \#1 (top) and \#3 (bottom) of NGC~5846. The flux image has been
    calculated by integrating the signal from -303 to -163 km\,s$^{-1}$ and 37 to 157
    km\,s$^{-1}$ respectively, flux errors are typically about 0.3 mJy\,km\,s$^{-1}$.
    Velocity and velocity dispersion have been obtained
    by fitting the emission line and have a typical respective
    uncertainty of 4 and 5 km\,s$^{-1}$. The ellipse in the left bottom corner
    represents ALMA beam size. }
 \label{fg:clmom1}
\end{figure*}

\begin{figure*}[ht!]
  \centerline{
    \includegraphics[width=5.9cm]{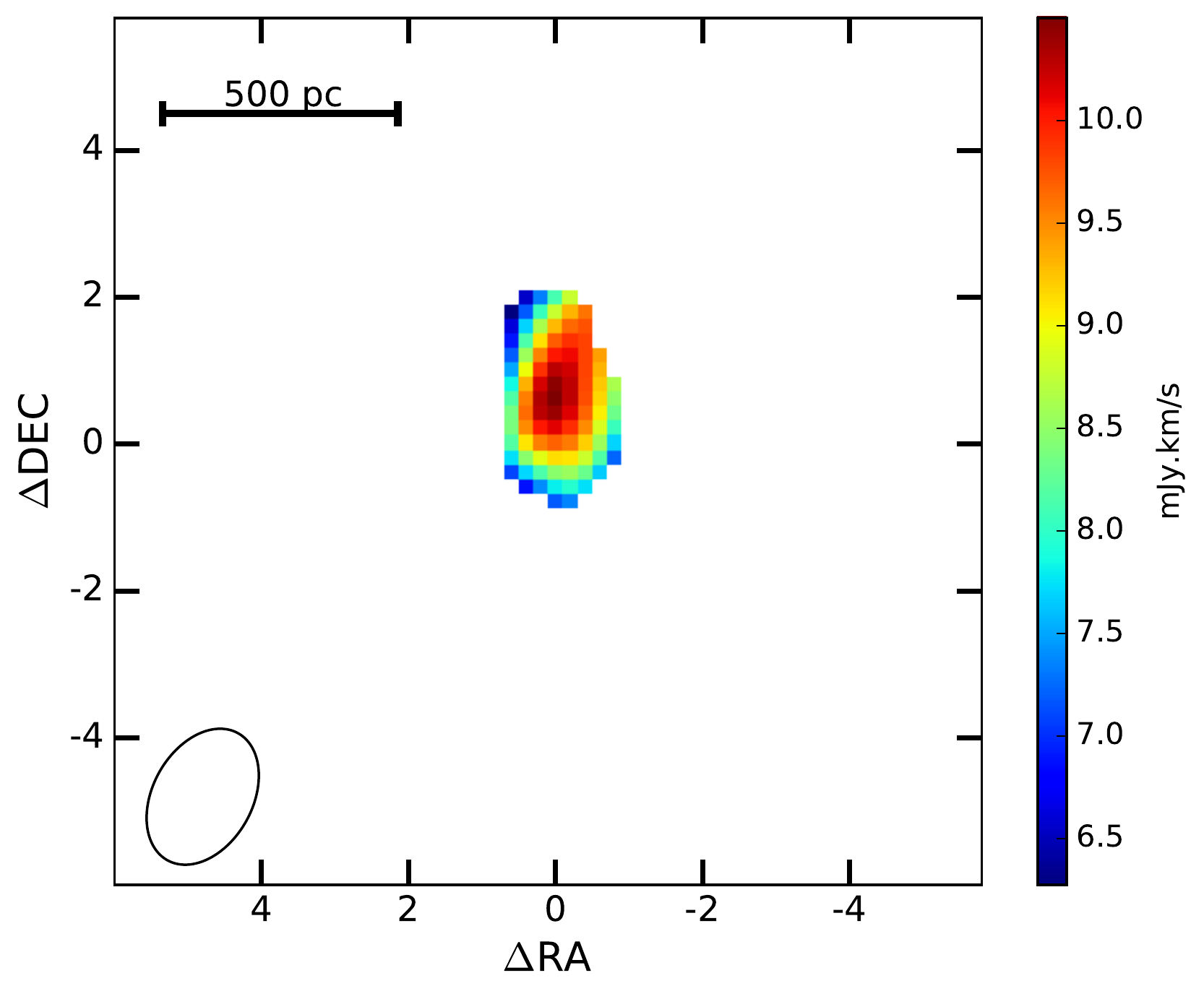}
    \includegraphics[width=6.05cm]{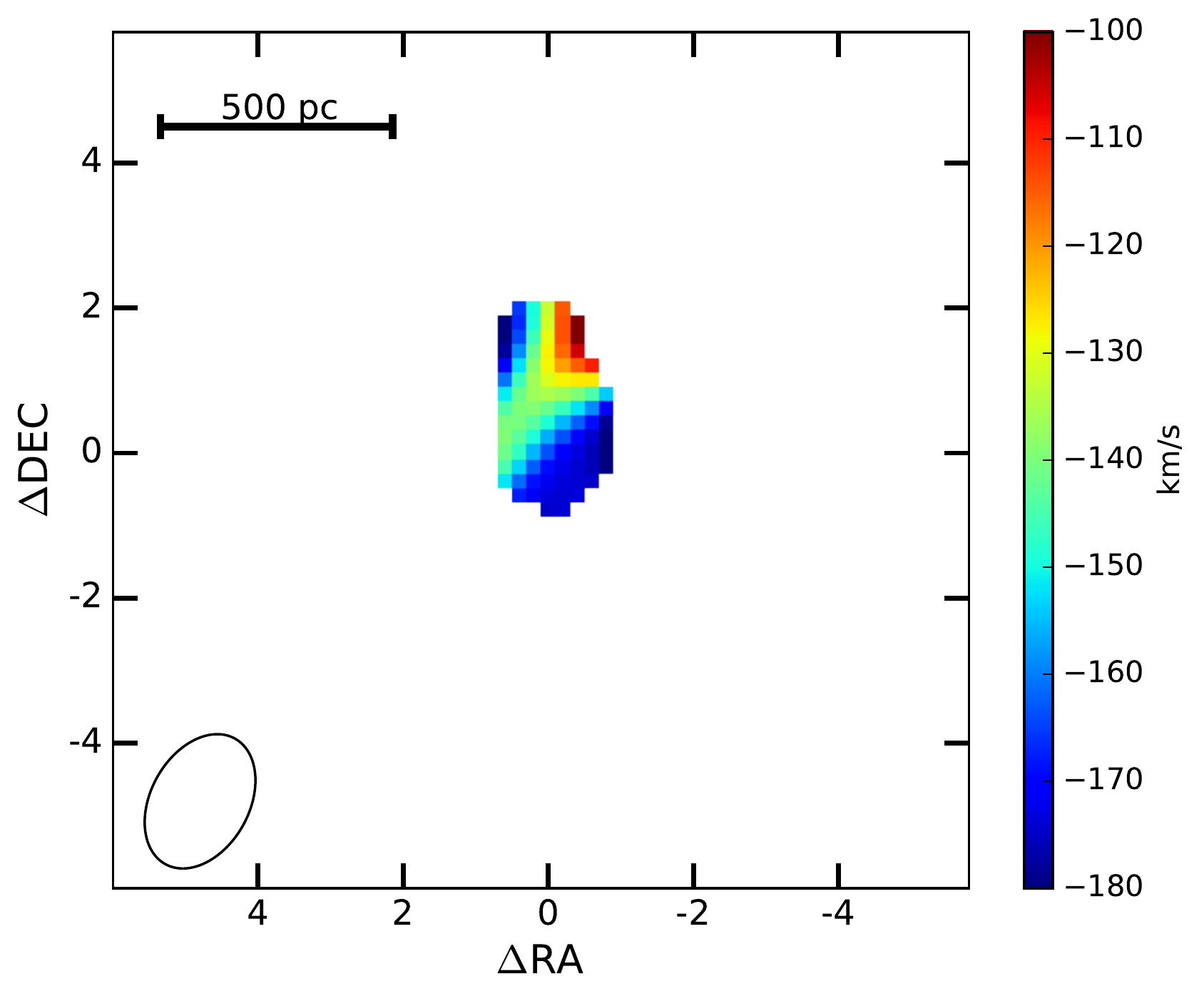}
    \includegraphics[width=5.8cm]{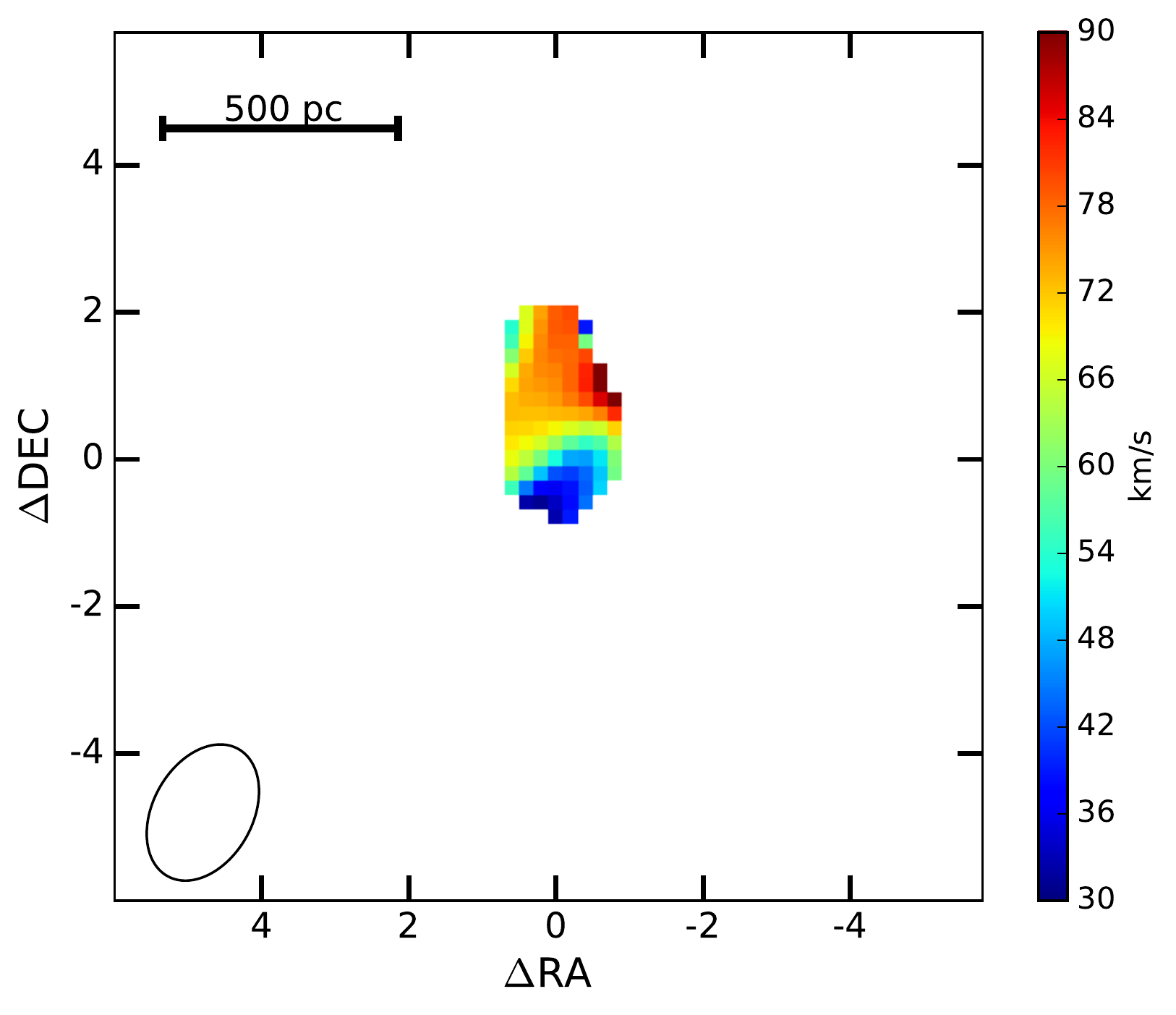}}
  \centerline{
    \includegraphics[width=5.8cm]{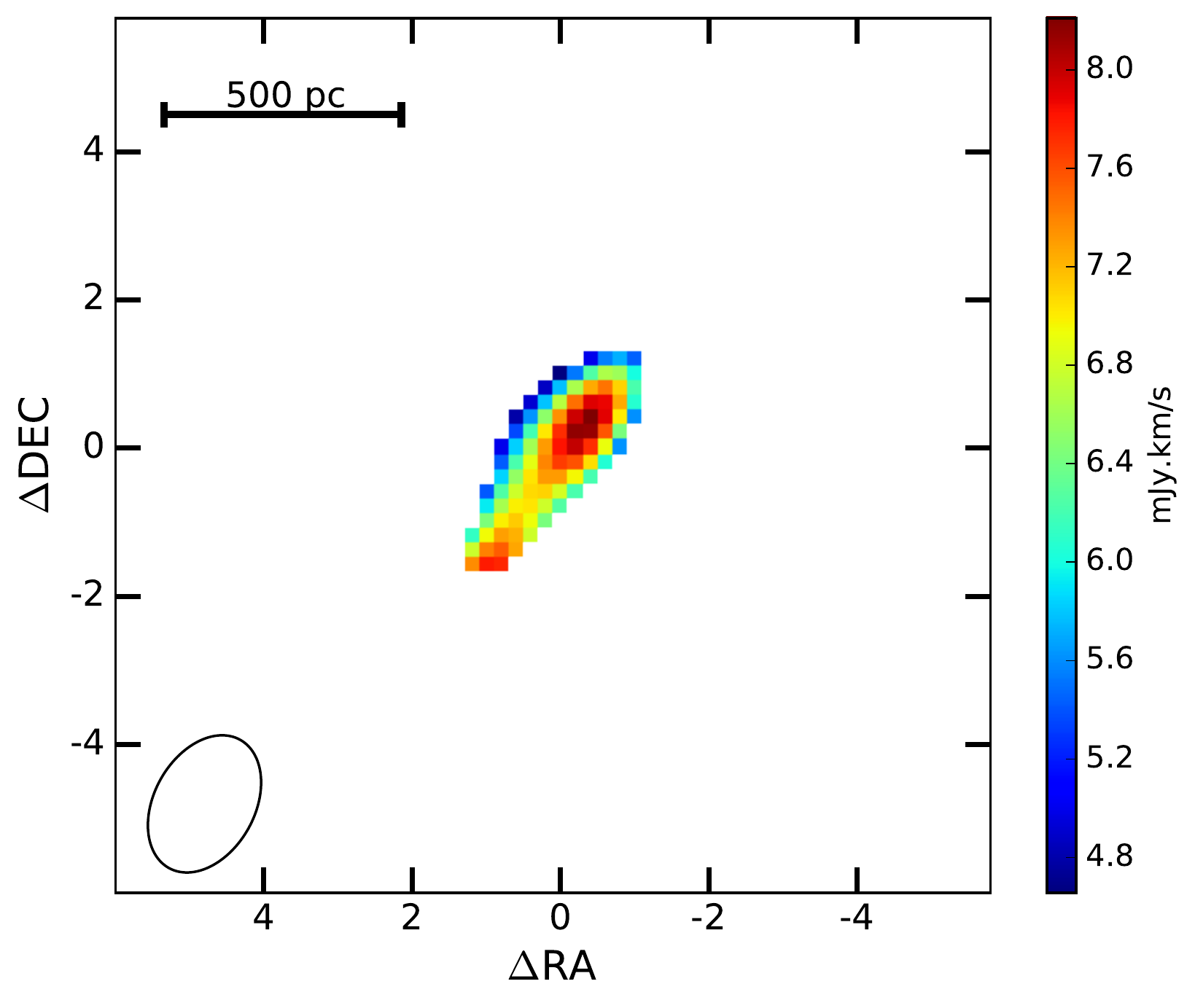}
    \includegraphics[width=6.05cm]{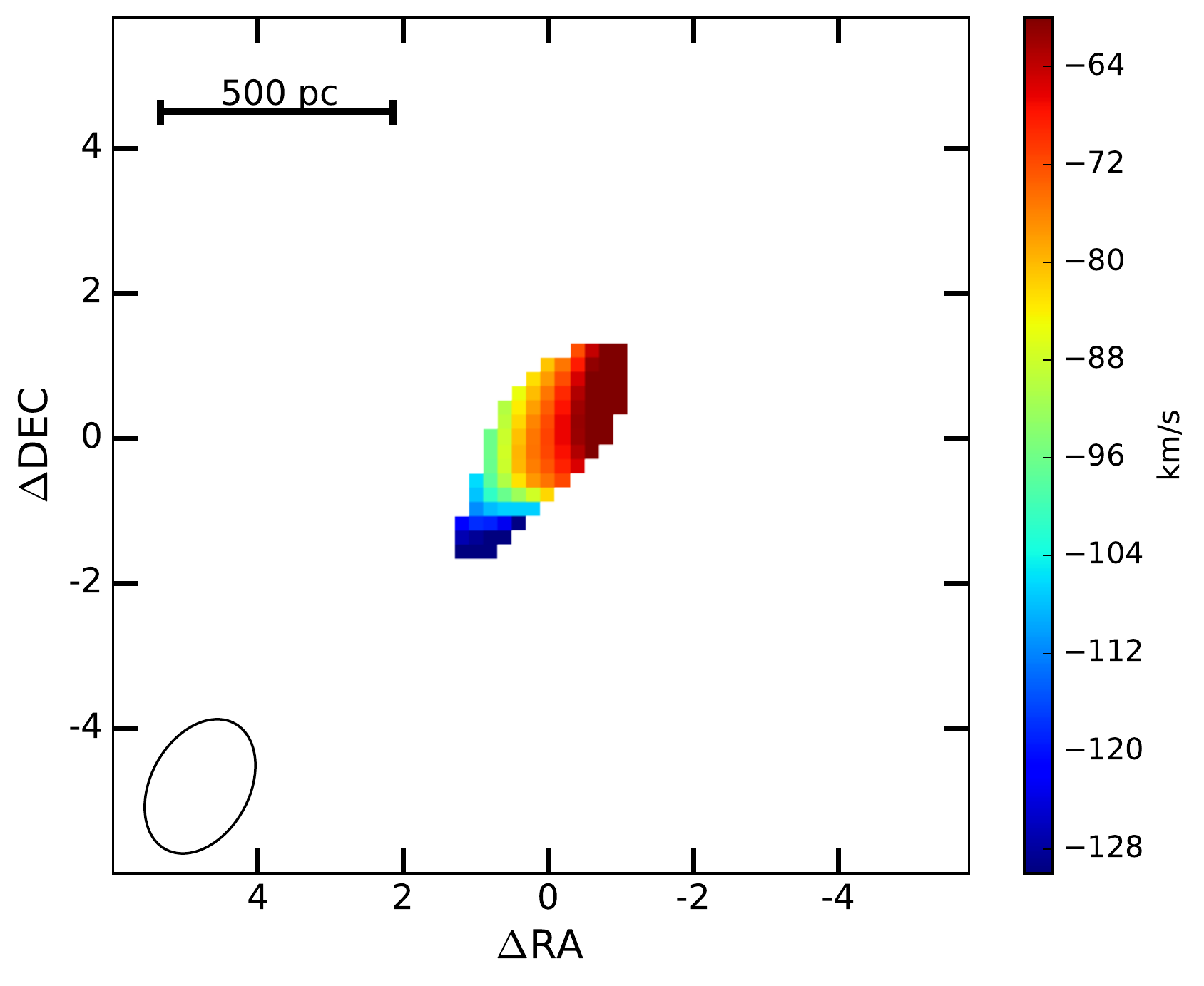}
    \includegraphics[width=5.8cm]{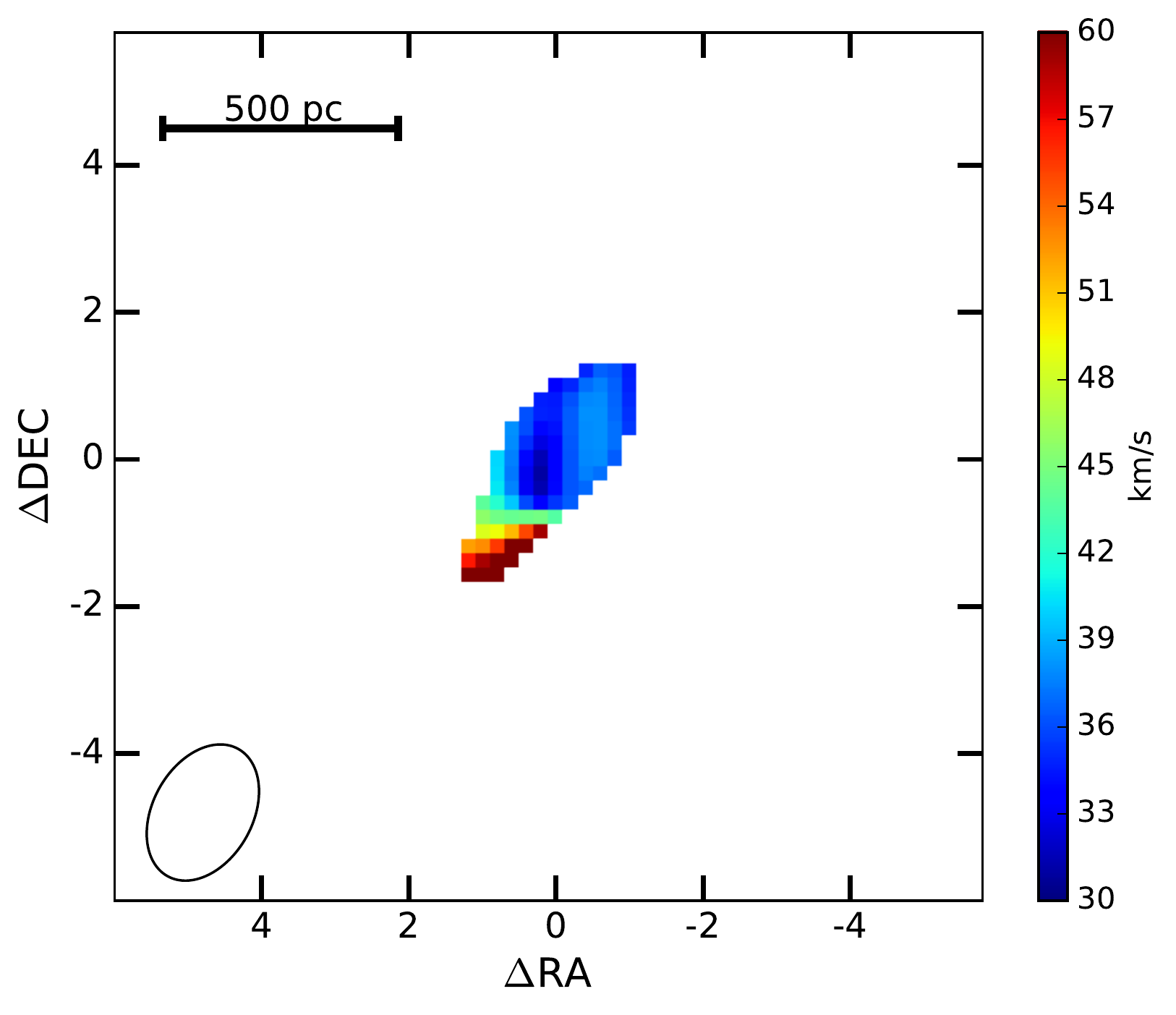}}
  \centerline{
    \includegraphics[width=5.8cm]{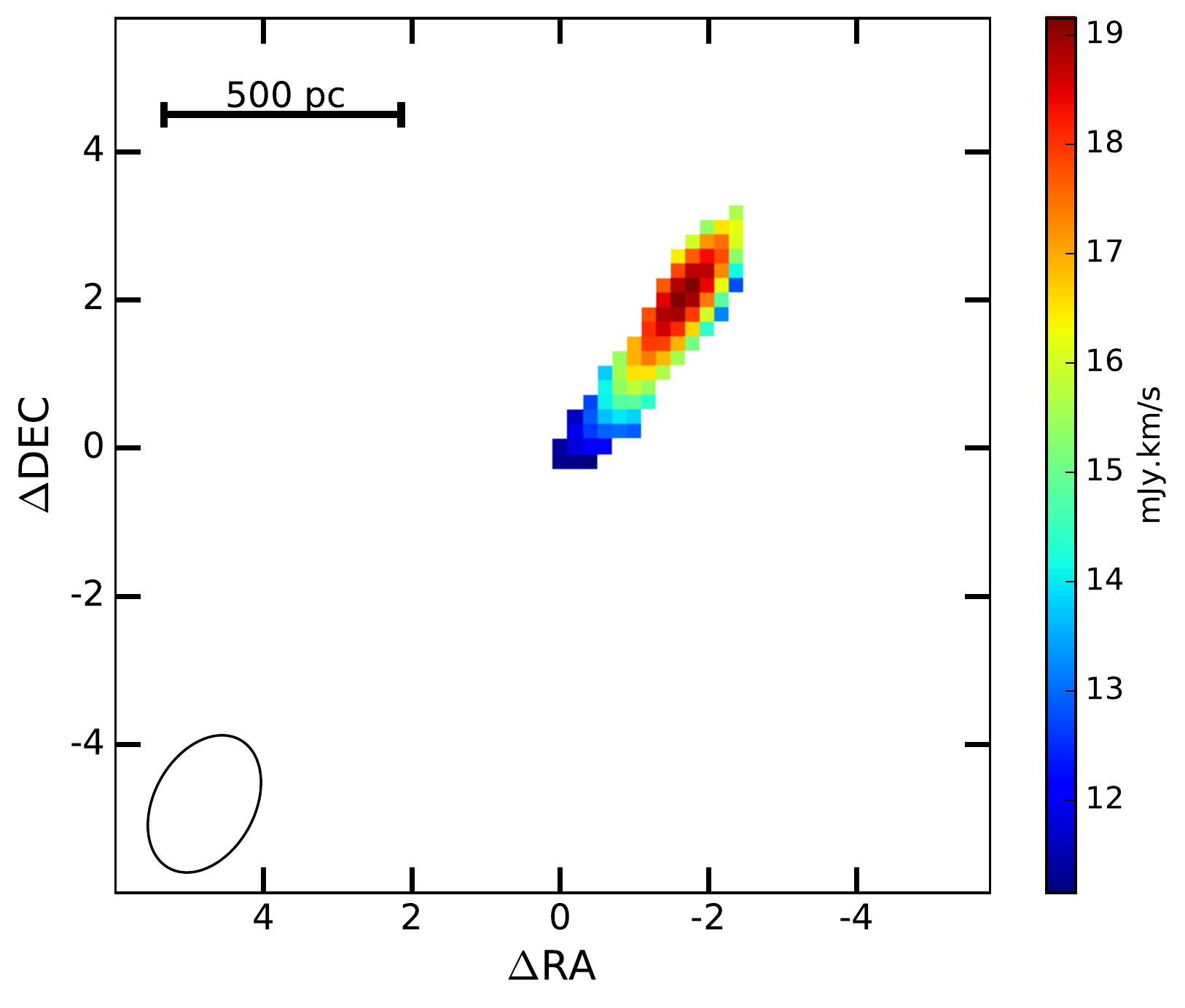}
    \includegraphics[width=5.95cm]{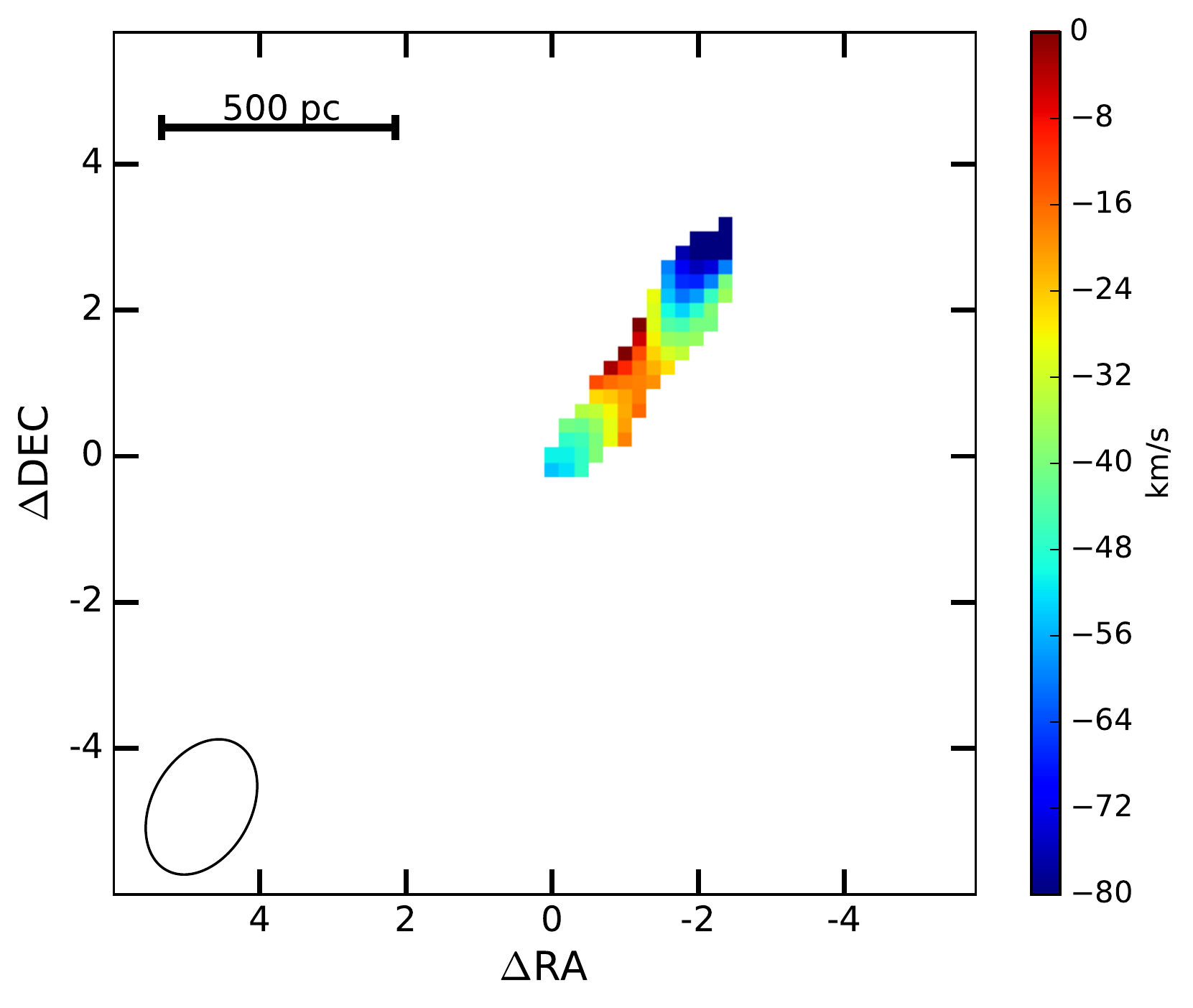}
    \includegraphics[width=5.8cm]{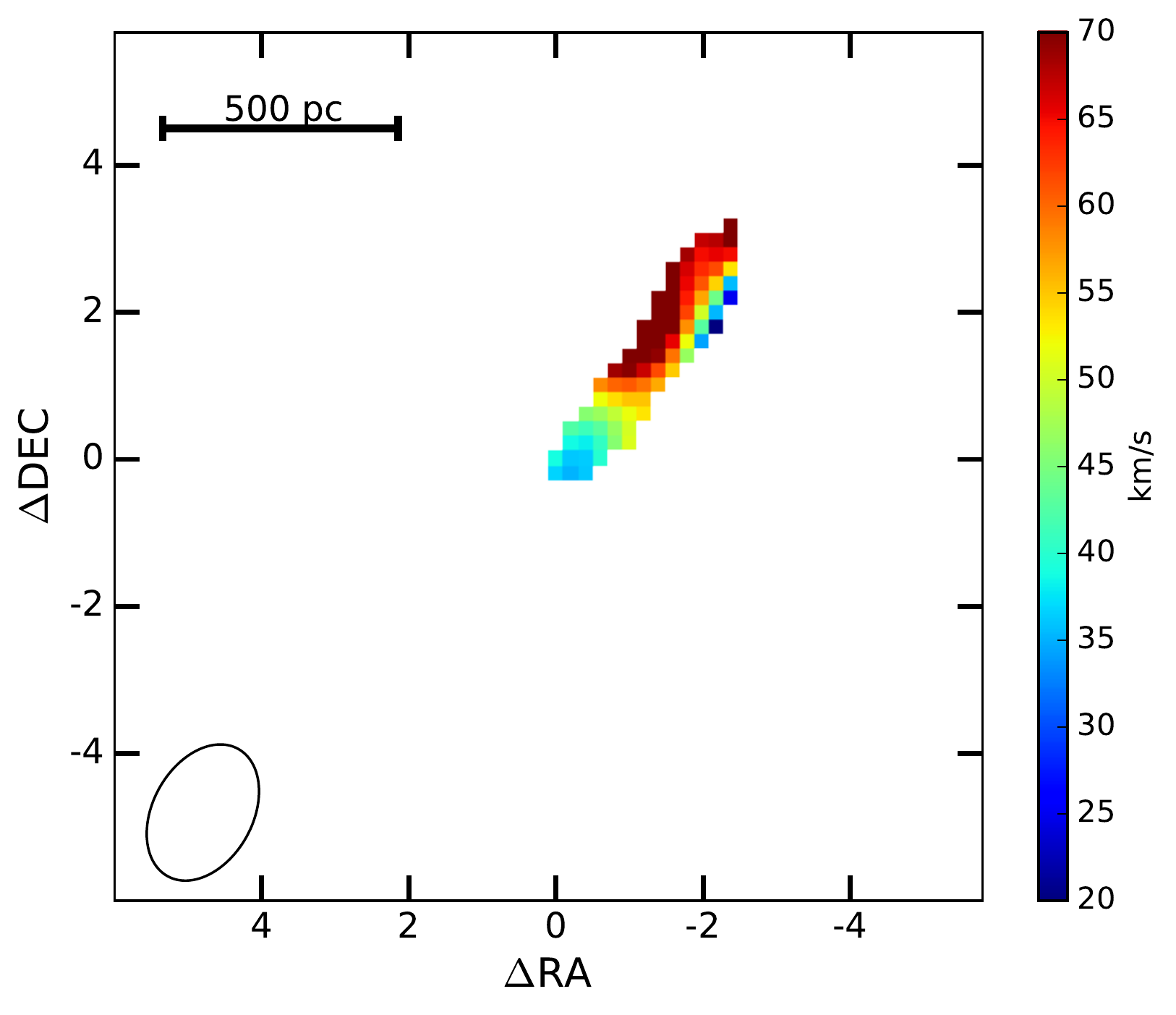}}
  \centerline{
    \includegraphics[width=5.8cm]{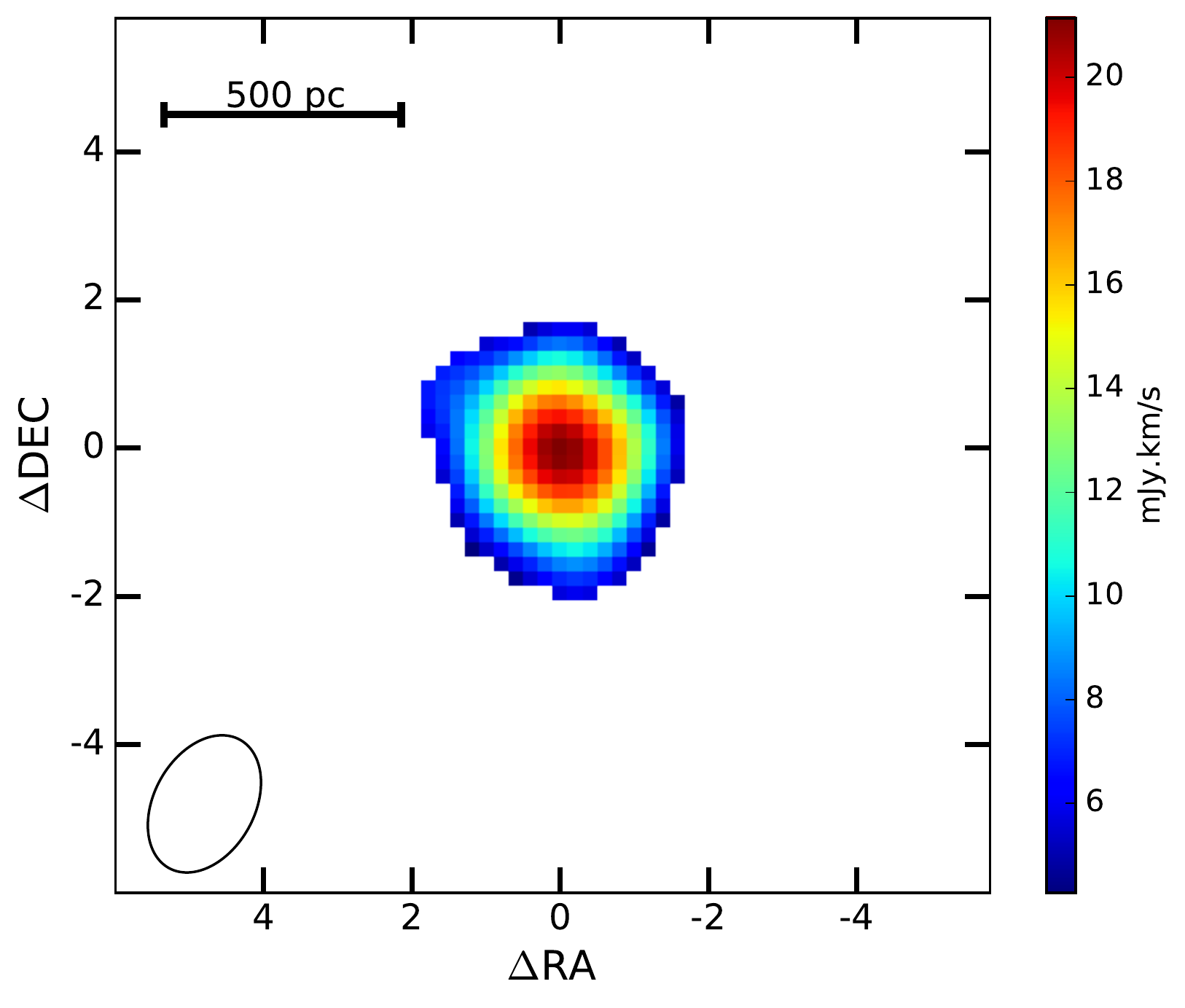}
    \includegraphics[width=5.8cm]{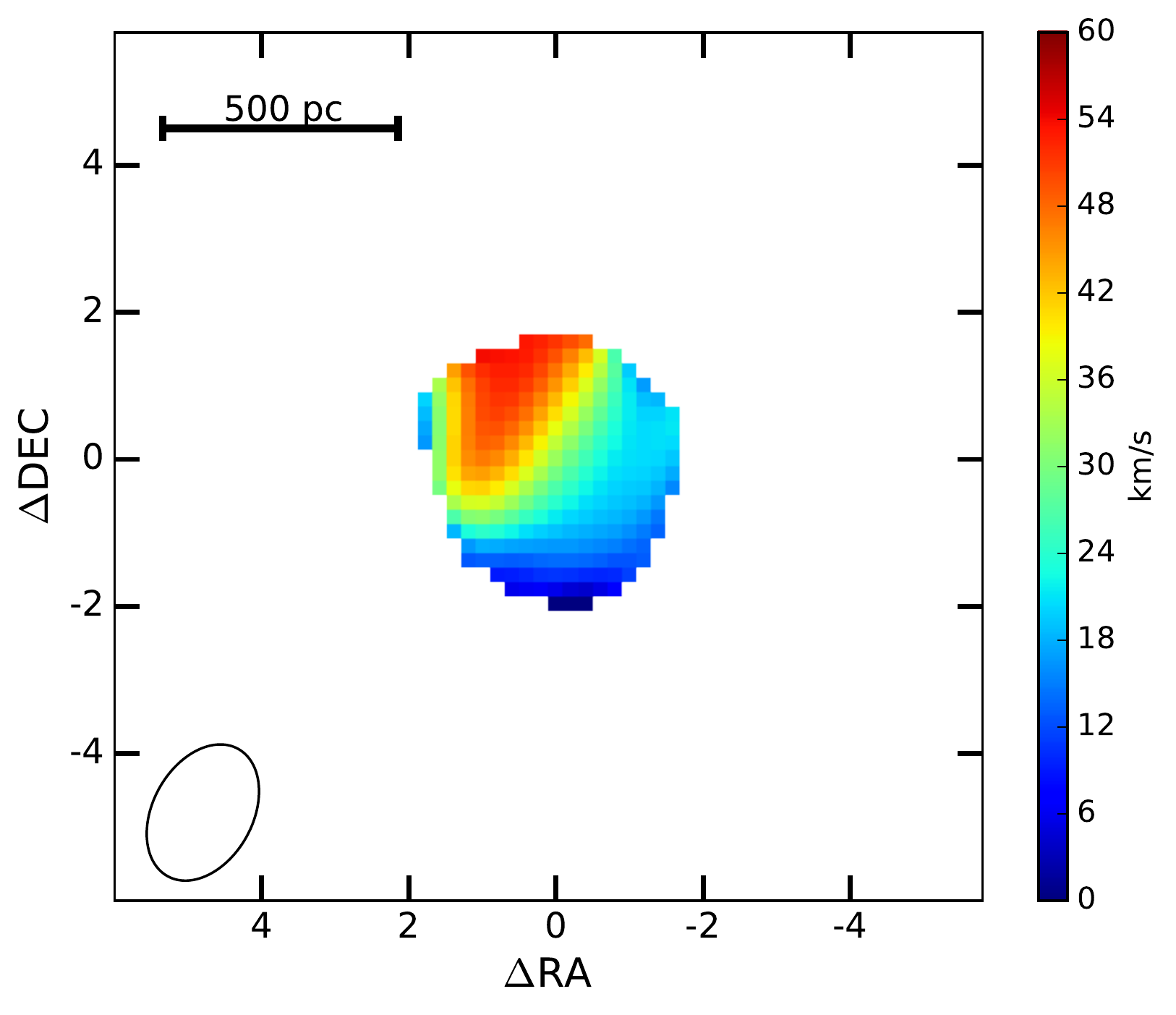}
    \includegraphics[width=5.8cm]{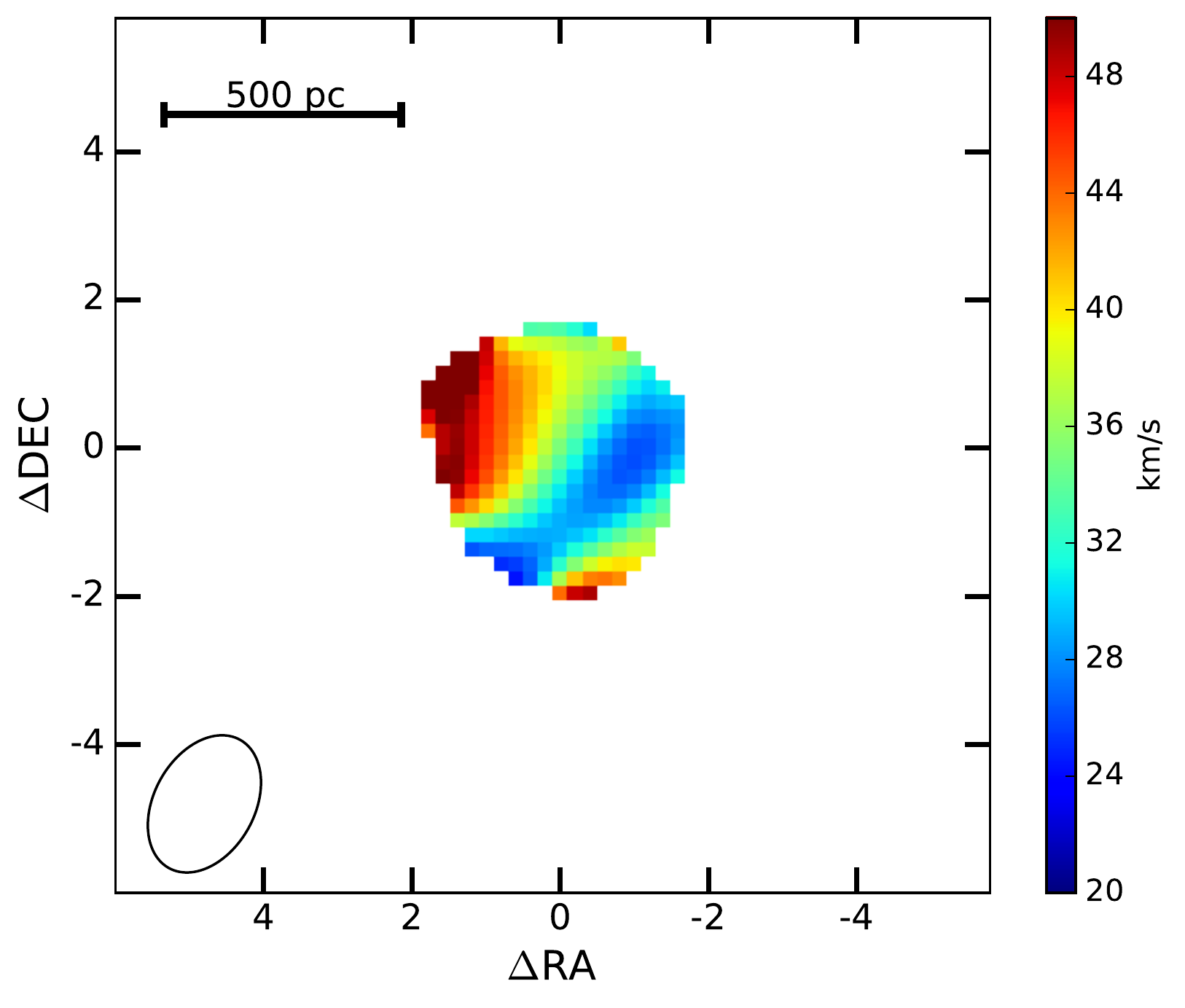}}
 \caption{Flux (left), velocity (center) and velocity dispersion (right)
    in cloud \#8,13,14,18 of  NGC~5044. The flux image has been calculated by
    integrating the signal 3 $\sigma$ around the peak, flux errors are typically
    about 1 mJy\,km\,s$^{-1}$ (except for cloud \#13 where it is 4 mJy\,km\,s$^{-1}$). 
    Velocity and velocity dispersion have been obtained
    by fitting the emission line and have typical uncertainty of km\,s$^{-1}$. The
    ellipse in the left bottom corner represents ALMA beam size.}
\label{fg:clmom2}
\end{figure*}

\subsection {CO Clouds in NGC~5846 and NGC~4636}
\noindent
Figure \ref{fg:clloc1} and Figure \ref{fg:clloc2} show the distribution of the CO(2-1) 
detected clouds in NGC~5846 and NGC~4636 projected against an H$\alpha$+[NII] 
map (left panel) and a HST dust absorption map (right panel). 
The H$\alpha$+[NII] images have been taken with the SOAR telescope with an average seeing 
of $\sim 0.7^{\prime \prime}$. 
The registration of the H$\alpha$+[NII] and CO 
images is correct within an uncertainty
of about $0.2^{\prime \prime}$ due to the astrometry in the SOAR data
\citep{werner14}.
In the figures, the optical dust maps were generated from archival 
HST data recorded with the WFPC2 in the F555W, F547M and F814W filters.

Two of the three CO clouds detected in NGC~5846, cloud \#1 and \#3 are resolved in at least one direction by the 12 meter array observations and extend to 1.2$^{\prime \prime}$ and 2.9$^{\prime \prime}$, respectively.
Black and white contours outline CO clouds, and are defined as the area where the emission line signal-to-noise is greater than 4.
Cloud \#2  in NGC~5846 and \#1 in NGC~4636 are outside the field of view of the high resolution HST dust map. These clouds are also quite close to the edge of the ALMA primary beam and were only detected when the cleaning threshold was set to about 1.5 time the rms noise, therefore they could still be artefacts created by the ``cleaning'' algorithm.

Three out of five clouds detected in these two galaxies are located in a region where the H$\alpha$+[NII] emission is relatively large although not in the strongest emission region and not quite at the center of the galaxy. Clouds \#1 and \#3 in NGC~5846 are about $5.5^{\prime \prime}$ and $8.4^{\prime \prime}$ away (0.6 and 1.0 kpc) from the galaxy center and cloud \#2 in NGC~4636 is about $2.6^{\prime \prime}$ away from the central nucleus (200 pc).

The HST images have a $25^{\prime \prime}$ field of view and only central cloud contours are visible.  There is a good correlation between the cloud position and the dust absorption in both NGC~5846 and NGC~4636, although several heavily obscured parts of each galaxies do not show CO(2-1) surface brightness large enough to be detected in our data.

In NGC~5846, cloud \#3 is aligned almost exactly along a 
dust filament, while cloud \#1 also coincides with small dust extinction structures (about 0.5'' $\sim$ 60 pc wide), but 
other similar dusty regions were not detected in CO(2-1).
Cloud \#2 in NGC~4636 is located in the same area of one of the most obscured region of NGC~4636.
Both clouds in NGC~4636 are not resolved by these observations, implying an angular size $\leq 0.7^{\prime \prime}$ ($\sim 50$ pc).

Figure \ref{fg:clspc} show the spectra of the five CO(2-1) clouds detected. For the two resolved clouds (clouds \#1 \& \#3 in NGC~5846), the spectra have been obtained by summing the signal from the pixels of each cloud (contiguous pixels with a signal-to-noise ratio
greater than 3 times the rms noise within the velocity range of the emission line). 
For unresolved clouds, the spectra is taken from the pixels with the largest signal-to-noise within the velocity range of the emission line. 

A Gaussian fit  to the emission line is presented as a red line in Figure \ref{fg:clspc}. 
Cloud \#1 in NGC~4636 has a double peaked spectrum that cannot be fitted with a single gaussian curve. 
The spectrum could be interpreted as generated by two distinct clouds that have similar projected 
position on the sky, but with a differential peak-to-peak velocity of $\sim$40 km\,s$^{-1}$. 
Unfortunately, since the cloud is not resolved by our current ALMA configuration, we cannot quantify its kinematics properties, including its gas velocity dispersion.

\subsection {CO Clouds in NGC~5044}
\noindent
Our new reduction of the Cycle 0 ALMA data on NGC~5044 confirmed the detection of 17 CO(2-1) clouds.
Figure \ref{fg:clouds3} shows the distribution of the
detected clouds overlapped to a false color H$\alpha$+[NII] 
map (left panel) and a HST dust absorption map (right panel). 
Most of the clouds in NGC~5044 are located within a radius of approximately $5^{\prime \prime}$ around the center of the galaxy. In Figure \ref{fg:clouds3}, unresolved clouds are identified with an ellipse roughly the size of the beam, while contours, at 4 times the rms noise level, are used to represent the extended molecular clouds/associations. 
With a resolution of $2.0^{\prime \prime}\times 1.4^{\prime \prime}$ , only four clouds are resolved by these observations: cloud \#8, \#13, \#14 and \#18.  

The HST dust extinction map presents several compact knotty features in its central region. 
All of the central CO clouds lie on top of these strong dust absorption features, with the larger (resolved) clouds encompassing few clumps of dust. 
In the central $5^{\prime \prime}\times 5^{\prime \prime}$ there is a good correlation between dust features and molecular cloud. However, more distant clouds in HST field of view (clouds \#7,\#12,\#21) are not correlated with dust absorptions. 

The H$\alpha$+[NII] emission map shows a very peaked central emission with the addition of diffuse and filamentary structures that extend up $30^{\prime \prime}$ from the galaxy center.
From the total of 17 clouds, 12 are located where the H$\alpha$+[NII] emission is fairly large and 8 of these clouds are within 4.6$^{\prime \prime}$ from the center of the galaxy. 
In this central $\sim5^{\prime \prime}$ region, the molecular clouds are located where the H$\alpha$+[NII] emission is strongest, broadly providing a correlation between optical line emission and CO gas. On the other hand, the spatial distribution of the 8 central clouds does not rigorously follow the morphology of the H$\alpha$+[NII] emission
(see insert in Figure \ref{fg:clouds3}).
Given the weak angular resolution of NGC~5044 CO(2-1) observation, we cannot exclude that some better correlation exists with a subset of clouds. The remaining 5 clouds (\#5, \#14, \#20, \#25, \#26) are located far from the center, between 12$^{\prime \prime}$ to 22$^{\prime \prime}$. Clouds \#25 \& \# 26 were only detected with the cleaning threshold set to 1.5 times the rms noise, they could be artifacts created by the ``cleaning'' algorithm.

Figure \ref{fg:clspc5044a} show the spectra of the 17 CO(2-1) clouds detected. 
For the four resolved clouds, the spectra have been obtained using a similar procedure described earlier for NGC~5846. Also,
for unresolved clouds, the spectra is taken from the pixels with the largest signal-to-noise within the velocity range of the emission line. 
A Gaussian fit  to the emission line is presented as a red line in Figure \ref{fg:clspc5044a}. The spectra are displayed with a bin size in velocity of 10 km\,s$^{-1}$. Since cloud \#22 has a large velocity dispersion, its spectrum is presented with a velocity bin size of 50 km\,s$^{-1}$.

\section {Analysis}
\noindent
Table 1 contains the basic parameters of the detected clouds, including
the average velocity, their velocity dispersion, and total flux (corrected from the primary beam effect), as well as, the corresponding molecular mass
and cloud dimension.
The velocity and velocity dispersion has been obtained by fitting a Gaussian on the spectra of each clouds. 
For each cloud, the flux has been obtained by summing the emission of the spectra described previously and range from 70 to 3000 mJy\,km\,s$^{-1}$. The molecular mass of each cloud is computed by using (\citealt{bolatto13}) 

\begin{equation}
M_{mol} = 1.05 \times 10^4 \left(\frac{X_{CO}}{2 \times 10^{20}}\right) \; \frac{S_{CO} \Delta\nu \; D_L^2}{ (1+z)} \; M_\odot,
\label{eq:mmol}
\end{equation}
where $S_{CO} \Delta\nu $ is the integrated line flux density in Jy\,km\,s$^{-1}$
in the ground rotational transition $J = 1 \to 0$, $D_L$ is the luminosity distance to the source in Mpc, and {\it z} is the redshift. 
Throughout the paper we use the {\it reference} conversion factor $X_{CO}=2\times 10^{20}$ cm$^{-2}$ (K km s$^{-1}$)$^{-1}$
to evaluate the molecular gas mass in each detected cloud.

To convert the CO(2-1) flux into CO(1-0) flux, we assume that the CO(2-1) to CO(1-0) temperature brightness ratio is 0.8 as found by \cite{braine92} so that the flux density ratio is 3.2 due to the frequency factor \citep{david14}.
The ACA data do not exhibit any detection and the clouds detected in the 12m array data are below the sensitivity of the ACA data. 
Cloud \#3 of NGC~5846 has a peak flux density of 10 mJy in a 10 km\,s$^{-1}$ bin that only represents a 2.5 rms noise signal at the center of the ACA observation. Cloud \#1 of NGC~4636 has a peak flux density of about 5.5 mJy in 10 km\,s$^{-1}$ bin that only represents a 2.3 rms noise signal at the center of the ACA observation.
Using the known location in space and in velocity of cloud \#3 of NGC~5846, we measure a signal with a flux density of 8.1$\pm$2.0 mJy/beam by increasing the bin size to 50 ${\rm km\,s^{-1}}$. However we detect other sources as well. Given the large amount of data, we expect several spurious detections at 4 time the rms noise. We verify that ACA data are consistent with the 12m data, but the ACA data alone are not sufficient to detect even cloud \# 3 of NGC~5846.

Using the reference $X_{CO}$ factor, the derived molecular masses of the detected clouds vary from 1$\times$10$^5$ to 9$\times$10$^6$ M$_\odot$. 
For clouds that have been resolved along at least one of their dimension by ALMA observations, we used the {\it imfit} function from the CASA package to calculate their size. For simplicity, we refer to these clouds as resolved or extended even if one of their dimension is smaller than the beam.{\it imfit} fits an elliptical Gaussian component and returns the size of the major/minor axis and of the position angle, as well as their associated errors \citep{condon97}. This model might not be the best match for the molecular cloud shape, however our measured residuals are consistent with pure noise, which might be due to the fact that none of the clouds is very well resolved.
 The values quoted in Table~\ref{tab:AAAA} correspond to the size deconvolved from the beam, except for cloud \#3 of NGC~5846, because its fitted minor axis is too close to the beam size (within the beam size considering the uncertainties) , so the size of the minor axis can not be deconvolved.
 
Clouds \#1 and \#3 in NGC~5846  measure 1.2$^{\prime \prime}$ and 2.9$^{\prime \prime}$ (major axis) respectively; these scales correspond to a physical size of 143 and 346 pc. Clouds \#8, 13, 14, 18 measure 3.9$^{\prime \prime}$, 
3.8$^{\prime \prime}$, 3.1$^{\prime \prime}$ and 1.9$^{\prime \prime}$ (major axis) corresponding to 590, 575, 469 and 287 pc.
It is not clear whether these clouds are really contiguous or formed of several smaller clouds.
Unresolved clouds are constrained to have a physical size smaller than 82, 50 and 300 pc for NGC~5846, NGC~4636 and NGC~5044 respectively given the angular resolution of our ALMA observations and the assumed distance of these galaxies.

The larger surface brightness and size of clouds \#1 and \#3 in NGC~5846
and clouds \# 8, 13, 14 and 18 in NGC~5044 allow to map out their properties.
Figures \ref{fg:clmom1} \& \ref{fg:clmom2} present the surface brightness, velocity and velocity dispersion maps of NGC~5846 \& NGC~5044 clouds.
The surface brightness has been calculated by integrating between -3$\sigma$ and +3$\sigma$ (the velocity dispersion $\sigma$ is fitted on the spectra) the spectra of each pixel ( -303 to -163 km\,s$^{-1}$  and 37 to 157 km\,s$^{-1}$  for cloud \#1 \& \#3 of NGC~5846 for instance). The velocity and velocity dispersion have been calculated by fitting a Gaussian to each pixel spectra.
The elongation and potential bimodality of the surface brightness of cloud \#3 of NGC~5846 could indicate that it might be composed of 2 nearby (in projection) clouds. The velocity vary slowly along the major axis of the cloud from about 100 km\,s$^{-1}$ to 120 km\,s$^{-1}$ and could either substantiate the existence of 2 clouds or indicate some rotation of a larger cloud. The velocity dispersion of cloud \#3 is fairly uniform between 17 and 21 km\,s$^{-1}$, given the uncertainty of about 5 km\,s$^{-1}$, and only seems to increase to 25 km\,s$^{-1}$ at the edges of the major axis. Our observation angular resolution is not quite good enough to distinguish a two or one cloud scenario.

Cloud \#1 of NGC~5846 has a more unimodal surface brightness shape and its surface brightness peaks roughly at its center. The velocity is varying across the cloud with a $\Delta v = 25\ km\,s^{-1}$, which given the uncertainty of 4 km\,s$^{-1}$, is statistically significant. The shape of the velocity distribution is compatible with tidal disruption.
 The velocity dispersion vary quite significantly within the cloud, from 15 km\,s$^{-1}$ to 32 km\,s$^{-1}$ (error estimate is about 4 km\,s$^{-1}$).\\
All resolved clouds of NGC~5044 are quite unimodal and all clouds but cloud \#15 are fairly elongated. They have a much larger velocity spread along their major axis ($\Delta v \sim $60 to 80 km\,s$^{-1}$) compared to resolved clouds in NGC~5846 ($\Delta v \sim $20 to 30 km\,s$^{-1}$).

\begin{figure*}[ht!]
\centerline{\includegraphics[width=18cm]{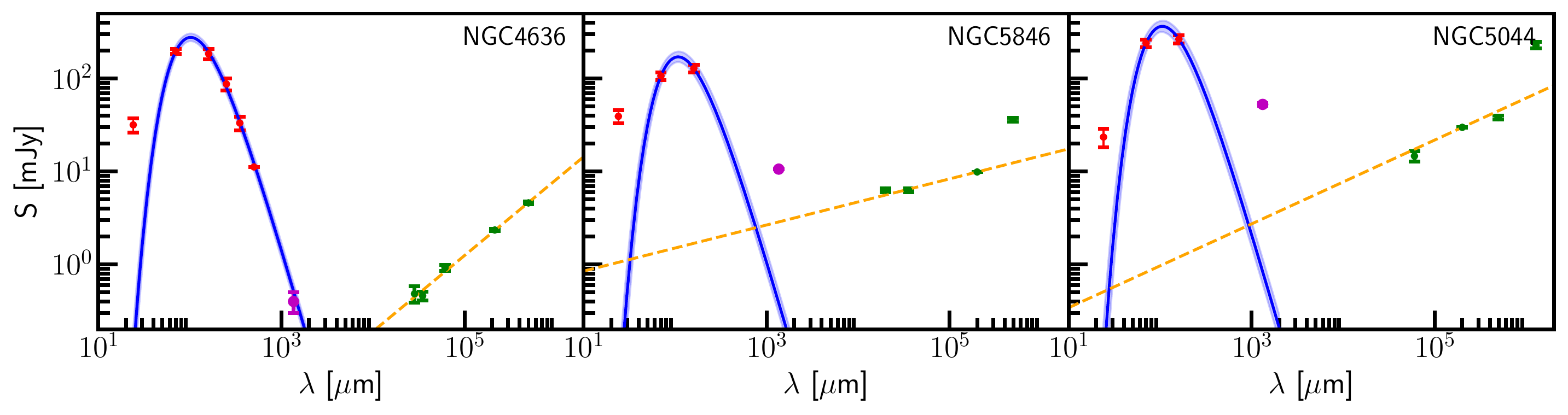}}
\caption{FIR to radio spectra of NGC~4636 (left), NGC~5846 (center) and NGC~5044 (right). Red circles represent
  data from {\it Spitzer}-MIPS and {\it Herschel}-SPIRE (NGC~4636), green circles represent radio measurements, magenta
  circles represent our ALMA continuum measurements. The blue line is a modified black-body (emissivity $\beta$=1.5) fit
  to data in the 70$\mu$m-500$\mu$m range, the light blue shade represents its uncertainty. The orange dashed line is a
fit to the radio data by a power-law spectrum.}

\label{fg:continuum}
\end{figure*}

\subsection{Continuum}

All three galaxies analyzed in this paper have a detected continuum around the CO(2-1) line.
The measured continuum for NGC~4636, NGC~5846, NGC~5044 is respectively 0.4$\pm$0.1, 10.63$\pm$0.03,
and 52.8$\pm$1.5 mJy. Each continuum is detected as a point source with a size limited by the
PSF of the observation (about 0.6'' for NGC~4636 and NGC~5846 and 1.7'' for NGC~5044, corresponding
to 51 and 255 pc).\\
An absorption feature seen in the NGC~5044 continuum spectrum of the central continuum source 
shows that the
emission must be very compact and probably arises from the AGN \citep{david14}.
We have not detected any absorption feature associated with the central continuum source in NGC~5846 or NGC~4636

Using FIR data from {\it Spitzer} and {\it Herschel}, when available, as well as radio data,
it is possible to fit for the FIR and radio spectra to weight the contribution of each to the
continuum at 230 GHz as seen by ALMA. FIR data from {\it Spitzer} and {\it Herschel} were taken from \cite{temi09} and \cite{Amblard14} respectively, in these data all three galaxies are unresolved. The radio data were taken from a variety of observations \cite{vollmer04,nagar05,filho04,giacintucci11}, the fluxes were corrected assuming uniform brightness when the scale at which they were measured differed greatly from ALMA PSF.\\

Figure \ref{fg:continuum} summarizes observations between 10 $\mu$m (30 THz) and 1 m (300 MHz)
for our three galaxies. NGC~4636 has the best coverage in the FIR with data from {\it Herschel}-SPIRE.
The fit of a FIR modified black body spectra (emissivity $\beta$=1.5) returns a temperature of
31.8 $\pm$ 0.9, 30.0$\pm$1.1, 30.7$\pm$1.2 K and a FIR luminosity of 10$^{7.91\pm0.02}$, 10$^{8.13\pm0.03}$,
10$^{8.71\pm0.03}$ L$_\odot$  for NGC~4636, NGC~5846 and NGC~5044 respectively. The
faint NGC~4636 ALMA continuum is in good agreement with the expected emission from cold dust,
which would indicate that the dust content of NGC~4636 is fairly centrally located.
The stronger continuum from NGC~5846 and NGC~5044 is an order of magnitude larger than the expected emission
from cold dust or synchrotron. This strong continuum could be due to free-free emission from HII regions which can dominate in the 30 to 200 GHz range. 
However, these results are somewhat weakened by the fact that some of these observations were performed at difference scales. Given FIR fluxes were measured on a larger scale than ALMA fluxes, cold dust emission could not contribute more to ALMA fluxes for NGC~5846 and NGC~5044, but the synchrotron contribution could be larger or lower depending on the model adopted to match the observed scales. It is also not possible to exclude that NGC~4636 ALMA flux could be a combination of synchrotron and dust emission, even if it matches the cold dust SED. New radio observations matching ALMA PSF would allow to improve this analysis.

\begin{table}[ht!]
  \begin{tabular}{lcccc}
    \hline
    \hline
    ID & $r_c$  & $\Sigma$           & $n_\mathrm{\rm H_2}$  &  $\alpha$ \\ 
       &   (pc)    & (M$_\odot$\,pc$^{-2}$) & (cm$^{-3}$)       &            \\
\hline
\multicolumn{5}{c}{NGC~5846}\\
1 &  24.8 $\pm$  12.6 & 460.2 $\pm$ 331.9 & 274.2 $\pm$ 241.7 &  18.0 $\pm$   9.4 \\
3 &  72.2 $\pm$   4.3 &  39.1 $\pm$   4.1 &   8.0 $\pm$   1.0 &  53.0 $\pm$   7.7 \\
\hline
\multicolumn{5}{c}{NGC~5044}\\
8 & 139.0 $\pm$  41.8 &  55.9 $\pm$  24.3 &   5.9 $\pm$   3.1 & 283.8 $\pm$  96.9 \\
13 & 104.8 $\pm$  54.2 &  75.4 $\pm$  55.6 &  10.6 $\pm$   9.6 &  89.2 $\pm$  52.6 \\
14 & 113.1 $\pm$  41.4 &  76.4 $\pm$  43.4 &  10.0 $\pm$   6.8 &  66.8 $\pm$  35.2 \\
18 &  79.2 $\pm$  25.1 & 476.1 $\pm$ 214.2 &  88.9 $\pm$  48.9 &  14.1 $\pm$   4.6 \\
\hline\hline
  \end{tabular}
  \caption{Cloud radius, surface mass density, volume density and virial factor for
  the 2 resolved clouds of NGC~5846 and the 4 resolved clouds of NGC~5044.}
  \label{tab:dyn}
\end{table}

\subsection{Kinematics interpretation}
\noindent
Cloud \#1 and \#3 of NGC~5846 are resolved and have an average radius ($r_c$ = $\sqrt{\sigma_\mathrm{maj}\sigma_\mathrm{min}}$) of  0.21$^{\prime \prime}\pm$0.11$^{\prime \prime}$ and 0.61$^{\prime \prime}\pm$0.04$^{\prime \prime}$
respectively (see Table \ref{tab:dyn}). For these clouds, we calculate the surface mass density that is respectively 460$\pm$332 and 39$\pm$4 M$_\odot$pc$^{-2}$ and the $n_{\rm H_2}$ volume densities are respectively 274$\pm$240  and  8.0$\pm$1.0 cm$^{-3}$. NGC~5044 clouds have an average radius between 79 and 139 pc, a surface mass density of 56$\pm$24, 75$\pm$56, 76$\pm$43 and 476$\pm$214 M$_\odot$pc$^{-2}$ and a $n_{\rm H_2}$ volume density of 6$\pm$3, 11$\pm$10, 10$\pm$ and 89$\pm$49 cm$^{-3}$ for cloud \#8,12,13,15 respectively.
The large uncertainty of these values is due to the fact that the minor axis of all the clouds is very close to the size of the beam, 
the deconvolved value of the minor axis ($\sigma_\mathrm{min}$) is therefore very small and has a large relative uncertainty. The lower uncertainty on cloud \#3 is due to the fact that we are using the convolved value of its dimension since we could not deconvolve the value of its minor axis.

To derive whether these molecular associations are gravitationally bounded, we can calculate the virial parameters from \cite{bertoldi92} :
\begin{equation}
  \alpha = \frac{5\sigma^2R}{GM}
  \label{eq:vir}
\end{equation}

A virial parameter $\sim 1$ indicates that a cloud is gravitationally bound, while
a virial parameter $\gg 1$ indicates that a cloud is unbound or pressure bound.
All the detected and resolved clouds have a virial parameter much $\gg 1 $ (see Table 2), although with large uncertainties
(due to large errors in the cloud radii). Deeper observations with a better PSF could allow to more properly resolve each clumps, and potentially detect sub-components with thinner CO lines. These sub-components could have a smaller virial parameters.
Indeed the large virial parameters might indicate that the clouds in NGC~5846 and NGC~5044 are unbound giant molecular associations drifting in the turbulent velocity field (dominated by the large eddies at kpc scale) and which may disperse in a relatively short timescale $r_c/\sigma\approx 1-10$ Myr.
Alternatively, the large CO linewidths may arise in molecular gas flowing out from clouds surfaces due to heating by the local hot gas atmosphere. Deeper observations with a smaller PSF are needed to better characterize the CO clumps detected in these observations.

\section{Discussion}
\noindent
Our new observations and detection of CO emitting clouds in NGC~5846 and 
NGC~4636 confirm the presence of molecular gas in group centered galaxies 
in the form of compact clouds. 
A diffuse CO 
component, if present, has not been detected by our ACA observations in 
these two galaxies. Previous attempts of detecting diffuse molecular gas 
with single dish observations have failed as well (Combes, IRAM observations).

Because of the angular sensitivity of interferometric measurements, ALMA observations are sensitive to emission in a selected range of angular scales. At the scale distance of galaxy groups, ALMA observations preferentially detect 
individual clouds, which may comprise a small 
fraction of the total molecular mass.
In more distant clusters, where the ALMA beam size is of the order of few kpc, the 
maximum recoverable size becomes a good match for detecting the diffuse emission \citep{David17}.

It is worth to note that molecular gas masses reported in Table \ref{tab:AAAA} have been evaluated 
assuming the conversion between CO luminosity and $H_2$ mass derived for our Galaxy
or other normal star-forming galaxies.
However, such conversion factor may not be appropriate for 
massive elliptical galaxies at the center of groups and clusters \citep[and references therein]{Lim17}.
The environment in which the molecular gas is immersed in massive ellipticals 
strongly constrains and defines the physical parameters of the molecular gas
and it differs substantially from 
the surrounding ambient of Galactic GMCs. 
The Galactic $X_{CO}$ may overpredict the mass of CO clouds in this
very different group environment where the CO line emissivity may be
unusually large.
Using ALMA observations of CO(3--2) and $^{13}$CO(3--2) in the brightest
cluster galaxy of RXJ0821+0752, \cite{Vantyghem17} have estimated that
the galactic $X_{CO}$ factor overestimates the true molecular mass by a
factor of two.

\begin{figure*}[ht!]
  \centerline{\includegraphics[width=6cm]{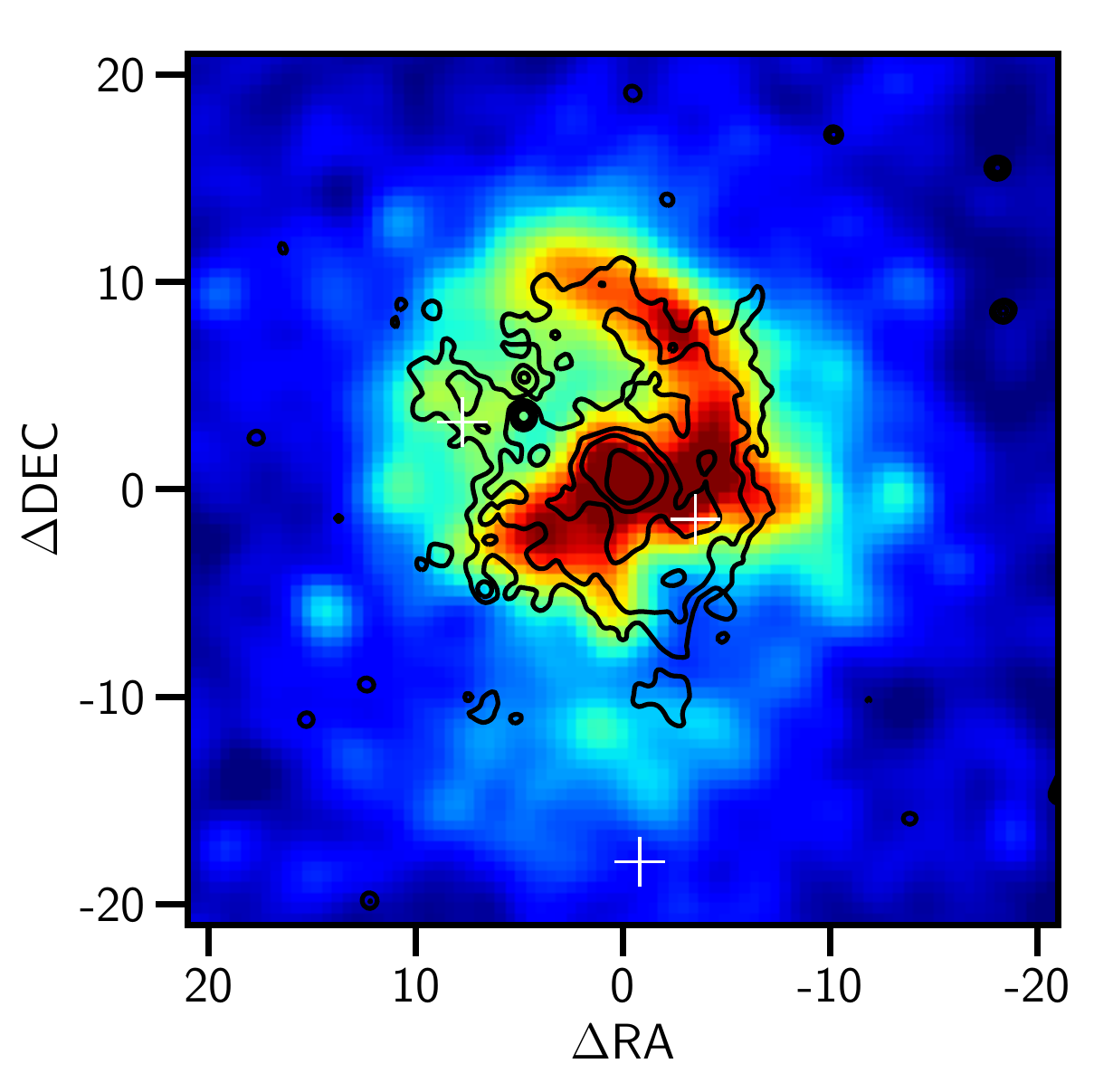}
  \includegraphics[width=6cm]{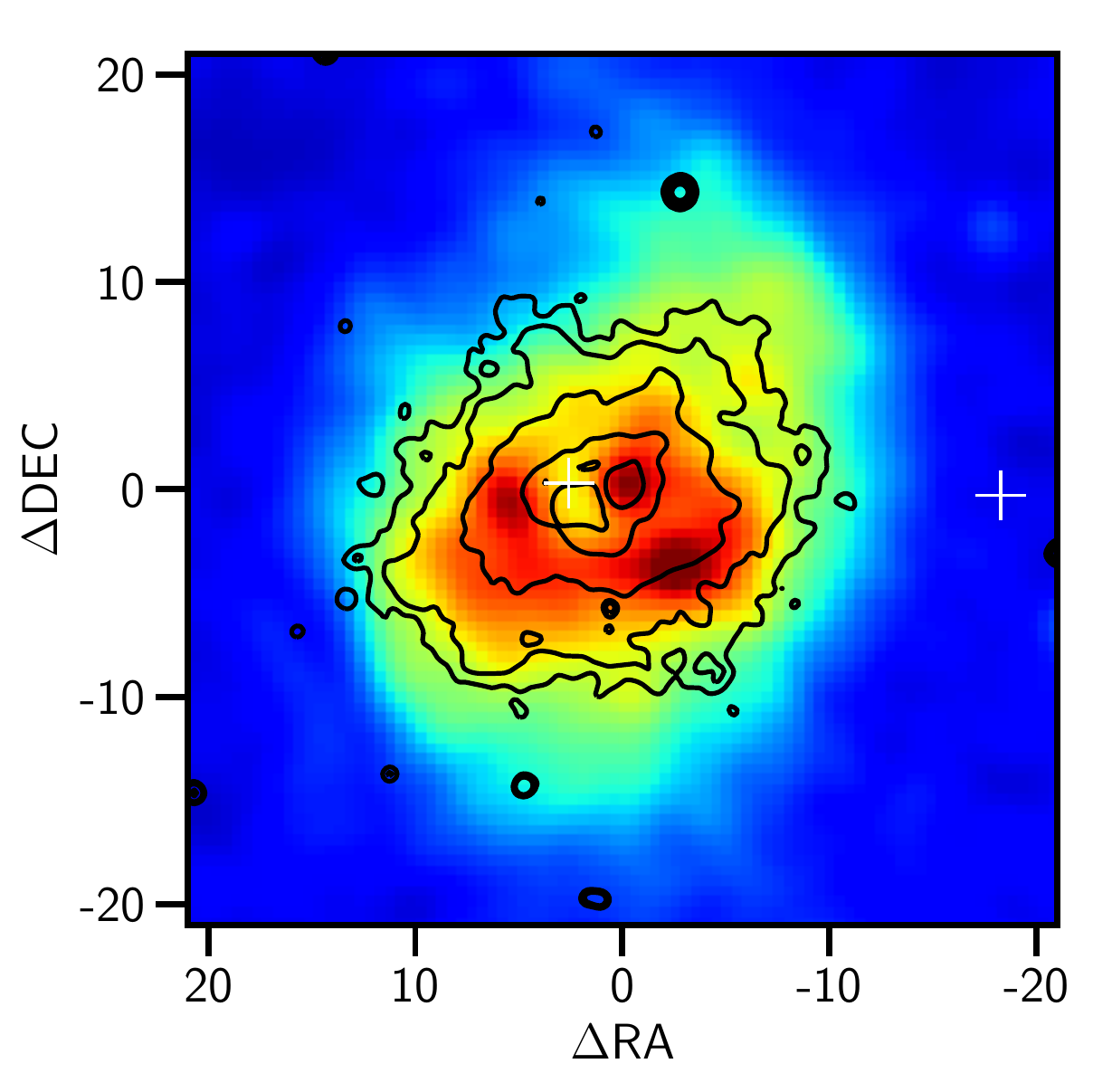}
  \includegraphics[width=5.8cm]{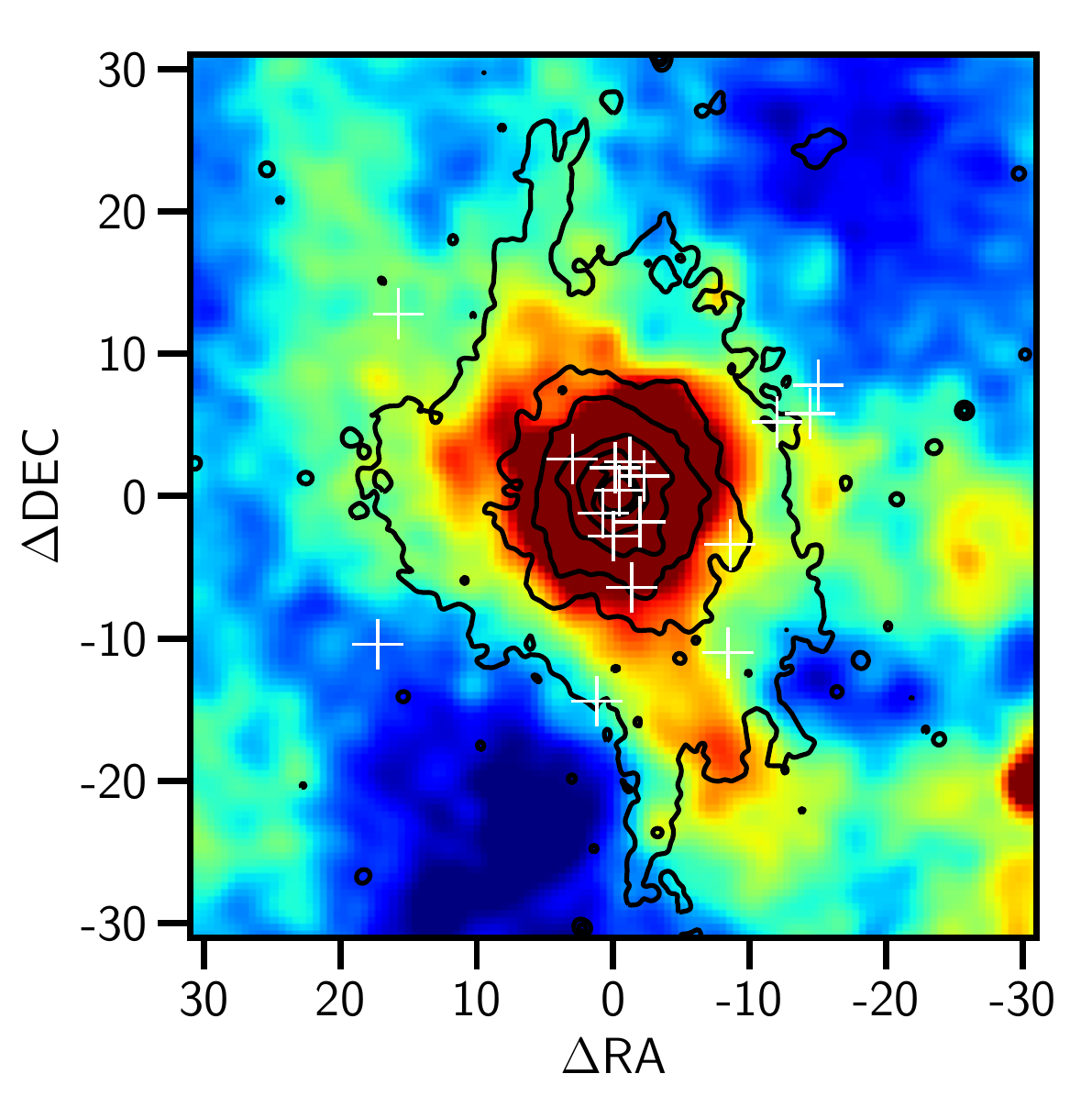}}
  \caption{{\it Chandra} images of NGC~5846, NGC~4636 and NGC~5044 with superimposed 
contours of their respective H$\alpha$+[NII] emission. An excellent correlation is apparent in the
central region as well as in the extended filamentary structures of the three galaxies. 
White crosses mark the detected CO cloud positions.} 
\label{fg:X-H1}
\end{figure*}

There is mounting evidence that the molecular gas in these systems
cooled from the hot interstellar or intragroup medium \citep[e.g.][and references therein]{Hogan17,Pulido17}. 

A remarkable observational result is the cospatiality among the different gas phases that extend to the observed filamentary structures of each gaseous component.
For instance, the X-ray emission is highly cospatial with the optical, infrared, and radio emission of cold/warm gas
\citep[e.g.][]{werner14}.
The [CII] emission, thought to originate in warm gas that roughly moves with the hot gas, correlates well with the H$\alpha$+[NII] emission, and in all three galaxies, the [CII]/H$\alpha$+[NII]) ratio remains relatively constant with total luminosity ratio $\sim$0.6 \citep{werner14}, corroborating the scenario of in-situ gas condensation via nonlinear thermal instability.
Dust absorption features in the optical HST images show a good correlation with the central  H$\alpha$+[NII] emission and filamentary structures  (see Figure ~\ref{fg:clloc1} and ~\ref{fg:clloc2} ).  

In addition
multi-wavelength
observations indicate that the presence of a multiphase ISM strongly
correlates with the hot atmosphere properties, such as short cooling
time ($t_{\rm cool} \lesssim 3\times10^8$ yr), low entropy parameter
($kT/n^{2/3} \lesssim 15$ keV cm$^2$), small cooling time to dynamical
time ratio ($t_{\rm cool}/t_{\rm dyn} \lesssim 20$), and
$t_{\rm cool}/t_{\rm eddy} \approx 1$. Early 2D 
hydrodynamic simulations of AGN heated galaxy group atmospheres
\citep{Brighenti02} suggested that spatially distributed
cooling in localized compressed regions, where nonlinear density
perturbations cools rapidly to $T\lesssim 5\times 10^4$ K. In recent
years much progress has been made in our understanding of the
heating/cooling cycle in galaxies and clusters 
\citep{revaz08,McCourt12, sharma12,gaspari12,gaspari13,gaspari17b,gaspari17_cca,li14,li15,
brighenti15,valentini15,Voit17}. These studies have shown that
cooling perturbations can be generated by several sources, like
turbulence \citep[e.g.][]{gaspari12}, AGN outflows
\citep[e.g.][]{gaspari11}, or buoyant cavities \citep[e.g.][]{brighenti15}. 
The latter two mechanisms are particularly effective in
stimulating cooling because they cause some of the low entropy central
gas to be transported to larger radii (albeit only along a preferential direction), where the dynamical time is
larger \citep[see also][]{McNamara16}.

However, a general prediction for these cooling scenarios is that the
cold gas should be dust poor \citep[e.g.][and references therein]{valentini15},
although additional physics (as dust growth by accretion of gas-phase metals in the cold gas) may alter this outcome \citep{Hirashita17}.  
This prediction seems to be confirmed by our observation, where many 
CO clouds do not extinct background starlight, especially in NGC 5044, although 
given their small angular sizes, dusty clouds may not obscure much starlight.
Thus, the exact alignment of NGC~5846 clouds \#1
and \#3 in Figure \ref{fg:clloc1} with the dust suggests a further
mechanism to form cold gas. 
 
They could have only recently formed from relatively dust-rich gas produced by mixing dusty stellar mass loss or dusty warm gas originally present in the center of the galaxy and uplifted by the AGN outburst with hot ISM. Dust-enhanced cooling can easily cool hot gas on time scales less than the local freefall time \citep{Mathews03}. Another possible origin for orbiting dusty molecular clouds is direct acceleration of molecular gas present in the central region. However, as discussed by \cite{McNamara16} and references therein, is unclear how low-density jets or buoyant cavities can drag very dense and compact clouds to height of several kpc.

More distant CO clouds, like Cloud \#2 in N5846, not associated with 
gas having an enhanced dust/gas ratio, are thought to originate from cooling in the wakes beneath buoyant X-ray cavities where compressions stimulate significant large-scale, coherent radiative cooling  \citep{brighenti15,valentini15}.

The tight spatial correlation between the soft X-ray and warm gas
emission (Figure ~\ref{fg:X-H1}), together with the cold molecular gas, 
is an important clue that the multiphase gas arises from 
the top-down nonlinear condensation process \citep{gaspari17b}, i.e., cascading from the turbulent hot plasma to warm filaments and to the molecular clouds at the overdensity peaks.
The high-resolution 3D simulations in \cite{gaspari17b,gaspari17_cca} allow a
quantitative comparison with the observations presented here.  
\cite{gaspari17b,gaspari17_cca} show that the cold clouds often agglomerate in giant
molecular associations (sometime in projection), which are not
virialized, but rather characterized by significant velocity
dispersion (and thus virial parameter $\alpha \gg 1$).  Typical
properties of simulated clouds (related to the high-end tail of the
molecular gas distribution) are masses in the range
$10^4$\,-\,$10^6\;\msun$ and effective radii 50\,-\,250 pc, which are
well consistent with the observational data.  The simulations also
predict several smaller and less massive clouds, which can not be
detected by our observations.

Analysis of the kinematics of the multiphase gas, in long-term AGN jet
feedback simulations runs \citep{gaspari17b}, reveals that warm and
cold gas are -- as ensemble -- robust kinematic tracers of the turbulent hot gas.
This is substantiated by multi-wavelength observations of NGC~5044. 
The ensemble velocity dispersion for NGC~5044 molecular clouds (which can be approximated via the RMS of the
velocity offsets; Tab.~\ref{tab:AAAA}) is $\sigma_{v, \rm los, ens}\simeq
177\;\kms$. The velocity dispersion from
H$\alpha$+[NII] data results to be $190\;\kms$ \citep{gaspari17b}. The two
are comparable (errors are $\lta10\;\kms$) and very similar to the hot plasma value of $172\;\kms$
(\citealt{ogorzalek17}).
Unfortunately, for NGC~4636 and NGC~5846 the small number of clouds
detected prevent a similar analysis.
\cite{gaspari17b} runs also show that, due to the turbulence cascade ($\sigma_v \propto l^{1/3}$), the small-scale structures typically have low velocity
dispersion ($<100\;\kms$) but large velocity shift, up to several
100\;$\kms$, in agreement with our observational findings 
(Sec.~4-5 and Tab.~\ref{tab:AAAA}).

Overall, the consistent results between the numerical predictions and the observational findings 
-- e.g., the cospatiality among the soft X-ray, optical/IR and radio structures, the correlation with short plasma cooling times, the large virial parameter, the masses and radii of the clouds, the tightly linked ensemble kinematics among all the phases (X-ray, Halpha+[NII], [CII], CO), and the cloud $\sigma_v$ tracing the turbulent cascade -- 
they all corroborate the important role that the the top-down condensation cascade and related CCA
play in shaping the multiphase halos of massive galaxies, groups, and clusters. 

The broad CO line widths of the molecular clouds in these galaxies,
which would imply that are unbound systems, is of particular
interest. 
The large CO linewidths may arise in gas flowing out from clouds surfaces
due to heating by the local hot gas atmosphere.
Broad, rather symmetric CO line profiles are formed by the combined
emission from outflows on both the near and far cloud surfaces. Fast CO outflows also guarantee
that CO lines are absorbed locally; doppler-shifted CO line radiation from distant surfaces will not
be absorbed. Furthermore, extreme CO outflow velocity gradients will sharply reduce optical depths
in all CO lines and CO line emissivities may be greatly increased by
collisional excitation at the relatively high temperature required to drive the observed expansion velocities.
Future ALMA observations of CO at higher J transitions and proper knowledge of excitation mechanisms, 
would provide a better characterization of the physical properties of the molecular gas and the inferred gas mass.

\section{Conclusions}
\noindent
We have presented new CO(2-1) ALMA observations of two group centered elliptical galaxies, NGC~5846 and NGC~4636.
With the addition of the revised cycle 0 observations of NGC~5044, we have confirmation that CO molecular clouds
are common features in these massive ellipticals.
Our main results are summarized as follows.

\begin{itemize}
\item[\textbf{$\bullet$}] The off-center orbiting clouds exhibit
CO linewidths that are $\gtrsim 10$ times
broader than Galactic molecular clouds. The associated total molecular mass
ranges from $2.6\times 10^5 M_\odot$ in NGC~4636 to $6.1\times 10^7 M_\odot$ in
NGC~5044. 
These masses have been estimated assuming the CO--to--H$_2$ conversion factor
calibrated for the Milky Way and nearby spiral galaxies. Since significant deviations from the Galactic
X$_{CO}$ are observed in other galaxies \citep{Vantyghem17}, it is expected that the presented molecular masses are
overestimated. 
A diffuse CO component, if present, has not been detected at the sensitivity level of our ACA 
observations in NGC~5846 and NGC~4636.
It is worth to note that given the angular 
sensitivity of interferometric measurements, ALMA observations at a distance scale of local groups preferentially detect 
individual clouds, which may account for a small fraction of the total molecular mass.
\item[\textbf{$\bullet$}] The origin of the detected molecular
features is still uncertain, but there is evidence that the molecular gas has cooled from the hot gas.
The observed spatial and kinematical correlation among the different phases - hot (soft X-ray), warm (H$\alpha$), cold ([CII]), and molecular (CO) - of the multiphase gas in the atmosphere of these galaxies, support the scenario of in-situ condensation, as opposed to gas stripping from merging galaxies. Also, the hot atmosphere properties - short cooling time, low entropy parameter, and small cooling time to dynamical/eddy time ratio - are consistent with the necessary conditions to promote gas cooling via thermal instabilities, as predicted by hydrodynamic simulations. 

The central CO clouds in strong spatial correlation with dust (e.g. cloud \#1 and \#3 in NGC~5846)
may have recently formed and cooled from the hot gas phase via dust-enhanced cooling.
The global condensation mechanism can be triggered via nonlinear perturbations generated in the chaotic turbulent velocity field or during the bubble uplift.
\item[\textbf{$\bullet$}] The large virial parameter of the molecular structures, their large CO(2-1) line widths, and correlation with the warm/hot phase kinematics provide evidence that they are unbound giant molecular associations drifting in the turbulent field, consistently with numerical predictions of the CCA process.
Alternatively, the
observed large CO line widths may be generated by molecular gas
flowing out from cloud surfaces due to heating by the local hot gas
atmosphere.
\end{itemize}

\vskip0.1cm
Despite the uncertainties listed above, we expect the surfaces of CO-emitting clouds in galaxy group environments to be strongly heated by their environment. This heating is expected to excite
higher rotational J-levels with high CO emissivity. 
Future ALMA observations of CO line fluxes at higher J-levels in these galaxy groups will be critical 
for our understanding of the formation and evolution of such clouds.

\section*{Acknowledgements}
\noindent
This paper makes use of the following ALMA data: ADS/JAO.ALMA\#2015.1.00860.S, ADS/JAO.ALMA\#2011.0.00735.SSB.
ALMA is a partnership of ESO (representing its member states), NSF (USA) and NINS (Japan), together with NRC (Canada), MOST and ASIAA (Taiwan), and KASI (Republic of Korea), in cooperation with the Republic of Chile. The Joint ALMA Observatory is operated by ESO, AUI/NRAO and NAOJ. The National Radio Astronomy Observatory is a facility of the National Science Foundation operated under cooperative agreement by Associated Universities, Inc.
Massimo Gaspari is supported by NASA through Einstein Postdoctoral Fellowship Award Number PF5-160137 issued by the Chandra X-ray Observatory Center, which is operated by the SAO for and on behalf of NASA under contract NAS8-03060. Support for this work was also provided by Chandra grant GO7-18121X.
HPC resources were in part provided by 
the NASA/Ames HEC Program (SMD-16-7320, SMD-16-7321, SMD-16-7305). 
Myriam Gitti thanks the Italian ARC staff in Bologna, in particular E. Liuzzo and M. Massardi, for their helpful advice during the data reduction. Myriam Gitti acknowledges partial support from PRIN-INAF 2014.

\bibliography{alma_bib}

\end{document}